\documentclass[useAMS,usenatbib]{mn2e}
\usepackage[dvips]{graphicx}
\usepackage{longtable}
\usepackage{pdflscape}
\usepackage{url}
\usepackage{amssymb}
\newcommand{\bz}{$\langle B_z \rangle$}
\newcommand{\nz}{$\langle N_z \rangle$}
\newcommand{\bd}{$B_{\rm d}$}
\newcommand{\vsini}{$v \sin i$}
\newcommand{\kms}{km\,s$^{-1}$}

\newcommand{\mdot}{$\dot{M}$}
\newcommand{\vinf}{$v_\infty$}

\newcommand{\msun}{$M_\odot$}

\newcommand{\ra}{$R_{\rm A}$}
\newcommand{\rk}{$R_{\rm K}$}

\newcommand{\lx}{$\log{L_{\rm X}}$}
\newcommand{\teff}{$T_{\rm eff}$}

\title[Magnetic field, rotation, and emission of $\xi^1$~CMa]{The pulsating magnetosphere of the extremely slowly rotating magnetic $\beta$ Cep star $\xi^1$ CMa \thanks{Based on observations obtained using the MuSiCoS spectropolarimeter at Pic du Midi Observatory; ESPaDOnS at the Canada-France-Hawaii Telescope (CFHT), which is operated by the National Research Council of Canada, the Institut National des Sciences de l'Univers of the Centre National de la Recherche Scientifique of France, and the University of Hawaii; on observations collected at the European Organisation for Astronomical Research in the Southern Hemisphere under ESO program 292.D-5028(A) at the Paranal Observatory, ESO Chile, with the NIR interferometers AMBER and PIONIER; on observations obtained at the La Silla Observatory, ESO Chile, with the echelle spectrograph CORALIE at the 1.2m Euler Swiss telescope; and on NEWSIPS data from the IUE satellite.}}
\author[M. Shultz et al.]
{M. Shultz$^{1,2,3,4}$\thanks{E-mail: matthew.shultz@physics.uu.se},
G.A. Wade$^{2}$,
Th. Rivinius$^3$,
C. Neiner$^5$,
H. Henrichs$^{6,7}$,
\newauthor{W. Marcolino$^8$, and the MiMeS Collaboration}\\
$^1$Department of Physics, Engineering Physics \& Astronomy, Queen's University, Kingston, ON Canada, K7L 3N6 \\
$^2$Department of Physics, Royal Military College of Canada, Kingston, Ontario K7K 7B4, Canada\\
$^3$ESO - European Organisation for Astronomical Research in the Southern Hemisphere, Casilla 19001, Santiago 19, Chile\\
$^{4}$Department of Physics and Astronomy, Uppsala University, Box 516, Uppsala 75120 \\
$^5$LESIA, Observatoire de Paris, PSL Research University, CNRS, Sorbonne Universit\'es, UPMC Univ. Paris 06, Univ. Paris Diderot,\\
Sorbonne Paris Cit\'e, 5 place Jules Janssen, 92195 Meudon, France\\
$^6$Anton Pannekoek Institute for Astronomy, University of Amsterdam, Science Park 904, 1098 XH Amsterdam, The Netherlands
\\
$^7$Department of Astrophysics/IMAPP Radboud University Nijmegen, PO Box 9010, 6500 GL Nijmegen, The Netherlands\\
$^8$Universidade Federal do Rio de Janeiro, Observat\'orio do Valongo, Ladeira Pedro AntUnio, 43, CEP 20.080-090, Rio de Janeiro, Brazil
\\
}

\begin{document}

\date{}

\pagerange{\pageref{firstpage}--\pageref{lastpage}} \pubyear{2002}

\maketitle

\label{firstpage}

\begin{abstract}

$\xi^1$ CMa is a monoperiodically pulsating, magnetic $\beta$ Cep star with magnetospheric X-ray emission which, uniquely amongst magnetic stars, is clearly modulated with the star's pulsation period. The rotational period $P_{\rm rot}$ has yet to be identified, with multiple competing claims in the literature. We present an analysis of a large ESPaDOnS dataset with a 9-year baseline. The longitudinal magnetic field \bz~shows a significant annual variation, suggesting that $P_{\rm rot}$ is at least on the order of decades. The possibility that the star's H$\alpha$ emission originates around a classical Be companion star is explored and rejected based upon VLTI AMBER and PIONIER interferometry, indicating that the emission must instead originate in the star's magnetosphere and should therefore also be modulated with $P_{\rm rot}$. Period analysis of H$\alpha$ EWs measured from ESPaDOnS and CORALIE spectra indicates $P_{\rm rot} > 30$~yr. All evidence thus supports that $\xi^1$ CMa is a very slowly rotating magnetic star hosting a dynamical magnetosphere. H$\alpha$ also shows evidence for modulation with the pulsation period, a phenomenon which we show cannot be explained by variability of the underlying photospheric line profile, i.e.\ it may reflect changes in the quantity and distribution of magnetically confined plasma in the circumstellar environment. In comparison to other magnetic stars with similar stellar properties, $\xi^1$ CMa is by far the most slowly rotating magnetic B-type star, is the only slowly rotating B-type star with a magnetosphere detectable in H$\alpha$ (and thus, the coolest star with an optically detectable dynamical magnetosphere), and is the only known early-type magnetic star with H$\alpha$ emission modulated by both pulsation and rotation.
\end{abstract}


\section{Introduction}\label{sec:intro}

The bright ($V = 4.3$ mag), sharp-lined $\beta$ Cep pulsator $\xi^1$ CMa (HD 46328, B0.5 IV) was first reported to be magnetic by \cite{hubrig2006}, based upon FORS observations at the Very Large Telescope (VLT). \cite{silvester2009} confirmed the detection using ESPaDOnS at the Canada-France-Hawaii Telescope (CFHT). 

Despite intensive observation campaigns with different spectropolarimeters, there are conflicting claims regarding the star's rotational period: \cite{hub2011a} derived a rotational period of $\sim$2.18~d based on FORS1/2 and SOFIN data, while \cite{fr2011} found a longer period of $\sim$4.27~d based on ESPaDOnS measurements. Both \citeauthor{hub2011a} and \citeauthor{fr2011} agreed that the star was likely being viewed with a rotational pole close to the line of sight. However, \citeauthor{fr2011} noted that given the small variation in \bz~they could not rule out intrinsically slow rotation. 

\cite{frost1907} was the first to note $\xi^1$ CMa's variable radial velocity. \cite{mcnamara1955} found a pulsation period of $\sim$0.2~d, identifying the star as a $\beta$ Cephei variable. While many $\beta$ Cep stars are multi-periodic, $\xi^1$ CMa appears to be a monoperiodic pulsator \citep{saesen2006} with an essentially constant period \citep{1999NewAR..43..455J}. In their analysis of CORALIE high-resolution spectroscopy, \cite{saesen2006} found a pulsation period of 0.2095764(4) d (where the number in brackets refers to the uncertainty in the final digit). \citeauthor{saesen2006} also reported that $\xi^1$ CMa is one of the few $\beta$ Cep stars that achieves supersonic pulsation velocities. Several spectroscopic mode identification methods revealed that the oscillation frequency most likely corresponds to either a radial $(l,m)=(0,0)$ mode, or a dipolar $(l,m)=(1,0)$ mode, the latter viewed at small inclination. An $l=2$ mode was ruled out via modelling of the velocity moments. Photometric mode identification indicates that only the radial mode agrees with the frequency \citep{1994AAS..105..447H}.

Like many magnetic early-type stars, $\xi^1$ CMa displays the emission signatures of a magnetized stellar wind in optical, ultraviolet, and X-ray spectra. The star's X-ray emission spectrum is very hard, and the brightest amongst all magnetic $\beta$ Cep stars, as demonstrated with {\em Einstein} \citep{cassinelli1994}, ROSAT \citep{cassinelli1994}, and {\em XMM-Newton} \citep{oskinova2011}. The star's X-ray emission has recently been shown to be modulated with the pulsation period \citep{2014NatCo...5E4024O}, a so-far unique discovery amongst massive magnetic pulsators \citep{2009A&A...495..217F,2015A&A...577A..32O}\footnote{Short-term X-ray variability has also been reported for the magnetic $\beta$ Cep star $\beta$ Cen \citep{1997ApJ...487..867C,alecian2011}, and for the non-magnetic (Jason Grunhut, priv. comm.) $\beta$ Cep star $\beta$ Cru \citep{2008MNRAS.386.1855C}. However, while subsequent analyses of the X-ray light curves of these stars have confirmed variability, they have not confirmed those variations to be coherent with the primary pulsation frequencies \citep{2005A&A...437..599R,2015A&A...577A..32O}, thus $\xi^1$ CMa is the only $\beta$ Cep star for which X-ray variation is unambiguously coherent with the pulsation period.}. Various wind-sensitive ultraviolet lines show strong emission \citep{schnerr2008}, which is typical for magnetic stars. $\xi^1$ CMa has also been reported to display H$\alpha$ emission \citep{fr2011}, although the properties of this emission have not yet been investigated in detail. 

Our goals in this paper are, first, to determine the rotational period based upon an expanded ESPaDOnS dataset, and second, to examine the star's H$\alpha$ emission properties. 

The observations are presented in \S~2. In \S~3 we refine the pulsation period using radial velocity measurements. In \S~4 we analyze the line broadening, determine the stellar parameters, and investigate the relationship between the star's pulsations and its effective temperature. In \S~5 we examine the possibility that the H$\alpha$ emission may be a consequence of an undetected binary companion, constraining the brightness of such a companion using both interferometry and radial velocity measurements. The magnetic measurements and magnetic period analysis are presented in \S~6. We examine the long- and short-term variability of the H$\alpha$ emission in \S~7, and investigate the influence of pulsation on wind-sensitive UV resonance lines. Magnetic and magnetospheric parameters are determined in \S~8, \S~9 presents a discussion of the paper's results, and the conclusions are summarized in \S~10. 

\section{Observations}\label{sec:observation}

\subsection{ESPaDOnS spectropolarimetry}

Under the auspices of the Magnetism in Massive Stars (MiMeS) CFHT Large Program \citep{2016MNRAS.456....2W}, 29 ESPaDOnS Stokes $V$ spectra were acquired between 2008/01 and 2013/02. One additional observation was already published by \cite{silvester2009}. A further 4 ESPaDOnS observations were acquired by a PI program in 2014, and another 22 by a separate PI program in 2017\footnote{Program codes 14AC010 and 17AC16, PI M. Shultz.}. ESPaDOnS is a fibre-fed echelle spectropolarimeter, with a spectral resolution $\lambda/\Delta\lambda \sim 65,000$, and a spectral range from 370 to 1050 nm over 40 spectral orders. Each observation consists of 4 polarimetric sub-exposures, between which the orientation of the instrument's Fresnel rhombs are changed, yielding 4 intensity (Stokes $I$) spectra, 1 circularly polarized (Stokes $V$) spectrum, and 2 null polarization ($N$) spectra, the latter obtained in such a way as to cancel out the intrinsic polarization of the source. The majority of the data were acquired using a sub-exposure time of 60 s, with the exception of the first observation (75 s), and the data acquired in 2017, for which a 72 s sub-exposure time was used to compensate for degradation of the coating of CFHT's mirror. The data were reduced using CFHT's Upena reduction pipeline, which incorporates Libre-ESPRIT, a descendent of the ESPRIT code described by \cite{d1997}. \cite{2016MNRAS.456....2W} describe the reduction and analysis of MiMeS ESPaDOnS data in detail. 

The log of ESPaDOnS observations is provided in an online Appendix in Table \ref{esp_obs_log}. The quality of the data is excellent, with a median peak signal-to-noise ratio (S/N) per spectral pixel of 828 in the combined Stokes $V$ spectrum. The 2 observations acquired on 2017/02/11 had a peak S/N below 500, and were discarded from the magnetic analysis. The log of sub-exposures, which were used for spectroscopic analysis, is provided in an online Appendix in Table \ref{esp_rv_ew}. Note that while there are 56 full polarization sequences, due to one incomplete polarization sequence there are 227 rather than 224 sub-exposures listed in Table \ref{esp_rv_ew}. 

\subsection{MuSiCoS spectropolarimetry}

Three Stokes $V$ spectra were obtained in 2000/02 with the MuSiCoS spectropolarimeter on the Bernard Lyot Telescope (TBL) at the Pic du Midi Observatory. This instrument, one of several similar fibre-fed \'echelle multi-site continuous spectroscopy (hence MuSiCoS) instruments \citep{1992A&A...259..711B} constructed at various observatories was uniquely coupled to a polarimeter \citep{1992A&A...259..711B,1999A&AS..134..149D}. It had a spectral resolution of 35,000 and covered the wavelength range 450--660 nm, across 40 spectral orders. As with ESPaDOnS, each polarimetric sequence consisted of 4 polarized subexposures, from which Stokes $I$ and $V$ spectra, as well as a diagnostic null $N$ spectrum, were extracted. The log of MuSiCoS data is provided in an online Appendix in Table \ref{esp_obs_log}. One of the MuSiCoS observations has a low S/N (below 100), and was discarded from the analysis. The data were reduced using ESPRIT \citep{d1997}.

Normal operation of the instrument was verified by observation of magnetic standard stars in the context of other observing programs \citep{2002A&A...392..637S,2004A&A...414..613K,2006A&A...458..569W}. In addition to this, we utilize the MuSiCoS spectra of 36 Lyn presented by \cite{2006A&A...458..569W}, one of which was obtained during the same observing run as those of $\xi^1$ CMa, in order to compare them to observations of the same star acquired recently with Narval, a clone of ESPaDOnS which replaced MuSiCoS at TBL, and which achieves essentially identical results \citep{2016MNRAS.456....2W}. This comparison is provided in Appendix \ref{append:musicos}. 

\subsection{CORALIE optical spectroscopy}

A large dataset (401 spectra) was obtained between 2000/02 and 2004/10 with the CORALIE fibre-fed echelle spectrograph installed at the Nasmyth focus of the Swiss 1.2~m Leonard Euler telescope at the European Southern Observatory's (ESO) La Silla facility \citep{2000A&A...354...99Q,2001Msngr.105....1Q}. The spectrograph has a spectral resolving power of $\sim$100,000, and covers the wavelength range 387--680 nm across 68 spectral orders. The data were reduced with TACOS \citep{1996A&AS..119..373B}.  The first analysis of these data was presented by \cite{saesen2006}.

The log of CORALIE observations is provided in an online Appendix in Table \ref{cor_obs_log}. The median peak S/N is 205. One observation, acquired on 10/12/2003, was discarded as it had a S/N of 16, too low for useful measurements. 

\subsection{IUE ultraviolet spectroscopy}

$\xi^1$ CMa was observed numerous times with the International Ultraviolet Explorer (IUE). The IUE could operate in two modes, high-dispersion ($R\sim2000$) or low-dispersion ($R\sim300$), with two cameras, the Short Wavelength (SW) from 115 to 200~nm and the Long Wavelength (LW) from 185 to 330~nm. We retrieved the data from the MAST archive\footnote{Available at https://archive.stsci.edu/iue/}. The data were reduced with the New Spectral Image Processing System (NEWSIPS). There are two simultaneous low-resolution spectra obtained with the Short Wavelength Prime (SWP) and Long Wavelength Redundant (LWR) cameras, along with 13 high-resolution spectra obtained with the SWP camera. The high-resolution data are of uniform quality as measured by the S/N, which is approximately 20 in all cases. Twelve of the spectra were obtained in close temporal proximity, covering a single pulsation cycle in approximately even phase intervals. The first observation was obtained 154 days previously. We used the absolute calibrated flux, discarded all pixels flagged as anomalous, and merged the various spectral orders.  

\subsection{VLTI near infrared interferometry}

While the angular radii of $\xi^1$ CMa and its circumstellar material are certainly too small to be resolved interferometrically, the star is bright enough that the data can be used to search for a high-contrast binary companion.

We have acquired four Very Large Telescope Interferometer (VLTI) observations: low-resolution AMBER H and K photometry, a high-resolution AMBER NIR spectro-interferogram, and one low-resolution H band PIONIER observation. AMBER (Astronomical Multi-Beam Recombiner) offers three baselines and can operate in either low-resolution photometric mode or high-resolution ($R\sim12000$) spectro-interferometric mode \citep{2007A&A...464....1P}. PIONIER (Precision Integrated-Optics Near-infrared Imaging ExpeRiment) combines light from 4 telescopes, offering visibilities across 6 baselines together with 4 closure phase measuremnts \citep{2011A&A...535A..67L}. All data were obtained using the 1.8~m Auxilliary Telescopes (ATs), which have longer available baselines, and hence better angular resolution than the 8~m Unit Telescopes (UTs). AMBER observations were obtained with baselines ranging from 80 to 129~m. PIONIER was configured in the large quadruplet, with a longest baseline of 140 m. The nearby star $\xi^2$ CMa (A0 III, $H = 4.63$) was used as a standard star for calibration. 

The observing log is given in Table \ref{vlti_obs_log}. As the execution time for a full measurement ($\sim$1 hr) is a significant fraction of the pulsation period, pulsation phases are not given. 

Owing to the relatively faint magnitude, the wavelength edges of the AMBER H and K profiles were particularly noisy. These data points were edited out by hand before commencing the analysis. No such procedure was necessary for the PIONIER observation.

\section{Pulsation Period}\label{sec:pulsation}

\begin{figure}
\centering
\includegraphics[width=\hsize]{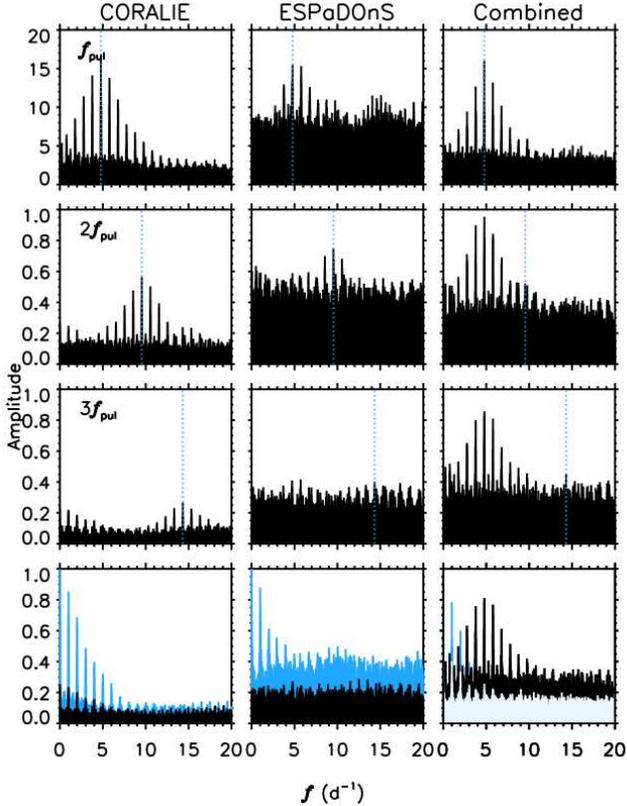}
\caption{Frequency spectra for radial velocities (RVs) measured from CORALIE data (left), ESPaDOnS data (middle), and the combined (right) dataset, for the RVs (top row), and after pre-whitening with significant frequencies (subsequent rows). Dotted light blue lines indicate $f_{\rm pul} \sim 4.77153~{\rm d}^{-1}$ and its first two harmonics. In the bottom panels, the window function is overplotted in light blue. In the bottom right panel, the white frequency spectrum shows the results of prewhitening the combined data with a pulsation period increasing at a linear rate of $0.0096(5) {\rm s}~{\rm yr}^{-1}$. By contrast, using a constant period to prewhiten the combined dataset (black frequency spectrum) does not fully remove the peak at $f_{\rm pul}$.}
\label{periods}
\end{figure}

\begin{figure}
\centering
\includegraphics[width=\hsize]{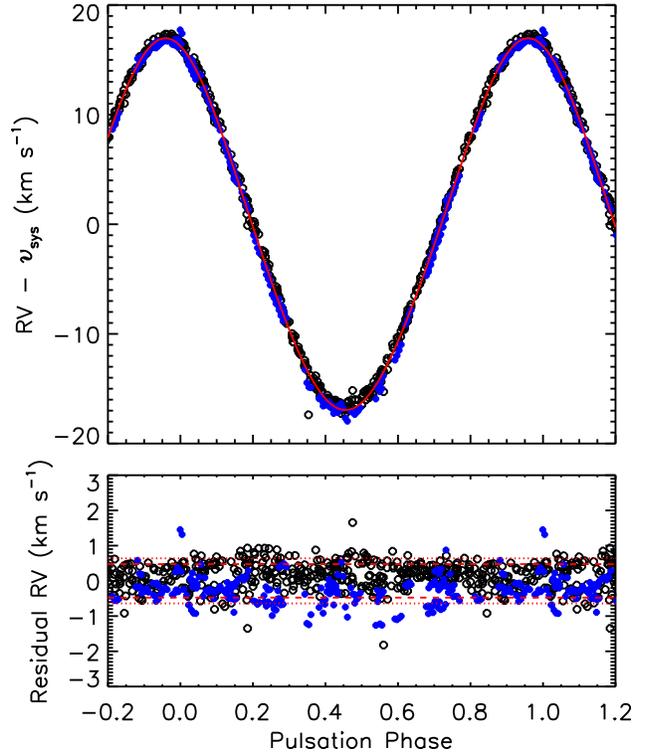}
\caption{{\em Top panel}: Composite radial velocities (using N {\sc II} 404.4 and 422.8 nm; N {\sc III} 463.4 nm; O {\sc II} 407.2, 407.9, 418.5, and 445.2 nm; Ne {\sc II} 439.2 nm; Al {\sc III} 451.3 nm; Si~{\sc iii} 455.3 nm; and Si {\sc IV} 411.6 nm) as a function of pulsation phase, using the nonlinear ephemeris defined in Eqn.\ \ref{ephem}. Open black circles are CORALIE measurements, filled blue circles ESPaDOnS measurements. The solid red line indicates a $3^{rd}$-order sinusoidal fit. {\em Bottom panel}: residual radial velocities. Dashed red lines indicate $\pm$1 standard deviation, about 0.48~\kms, and the dotted red lines show the median RV uncertainty.}
\label{rvs}
\end{figure}

\subsection{Radial Velocities}

We measured radial velocities (RVs) from the ESPaDOnS and CORALIE spectra using the centre-of-gravity method \citep{2010aste.book.....A}. In addition to using the same line (Si {\sc iii} 455.3 nm) used by \cite{saesen2006}, we have used: N {\sc ii} 404.4 and 422.8 nm; N {\sc iii} 463.4 nm; O {\sc ii} 407.2, 407.9, 418.5, and 445.2 nm; Ne {\sc ii} 439.2 nm; Al {\sc iii} 451.3 nm; and Si {\sc iv} 411.6 nm. For ESPaDOnS data we used individual sub-exposures rather than the Stokes $I$ profiles corresponding to the full polarization sequences: each polarization sequence encompasses about 2.3\% of a pulsation cycle (4$\times$60~s sub-exposures + 3$\times$60~s chip readouts), as compared to about 0.33\% for the sub-exposures (although the 2017 data had slightly longer subexposure times, this was compensated for by shorter readout times). The CORALIE data are not as uniform as the ESPaDOnS data in this regard: the median exposure time corresponds to about 2.8\% of a pulsation cycle, and some are up to about 10\% (all CORALIE measurements were retained, however HJDs were calculated at the middle rather than the beginning of each exposure). 

Since centre-of-gravity measurements can be biased by inclusion of too much continuum, it is important to choose integration limits with care. Thus RVs were measured iteratively. A first set of measurements was conducted using a wide integration range ($\pm 50$ \kms~about the systemic velocity), chosen to encompass the full range of variation. These initial RVs were then used as the central velocities for a second set of measurements, with an integration range of $\pm 30$ \kms. These were in turn used a final time as the central velocities to refine the RVs with a third iteration using the same integration range. The error bar weighted mean RV across all lines was then taken as the final RV measurement for each observation, with the standard deviation across all lines as the uncertainty in the final RV.  ESPaDOnS and CORALIE RV measurements are tabulated in an online Appendix in Tables \ref{esp_rv_ew} and \ref{cor_obs_log}, respectively. We determined the systemic velocity from the mean RV across all observations to be $22.5~\pm~1$~\kms. 

\begin{figure*}
\centering
\includegraphics[width=\hsize]{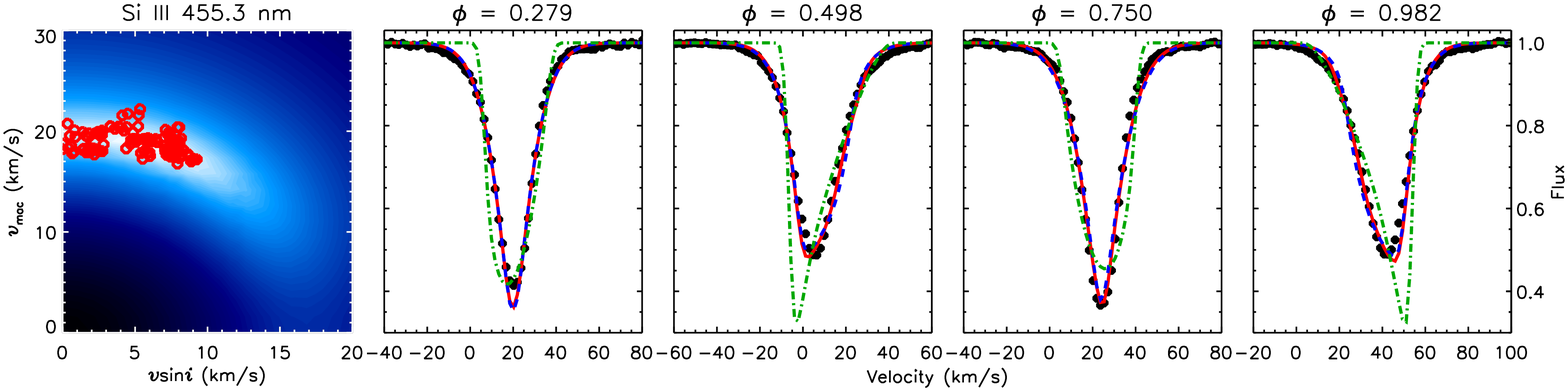} 
\includegraphics[width=\hsize]{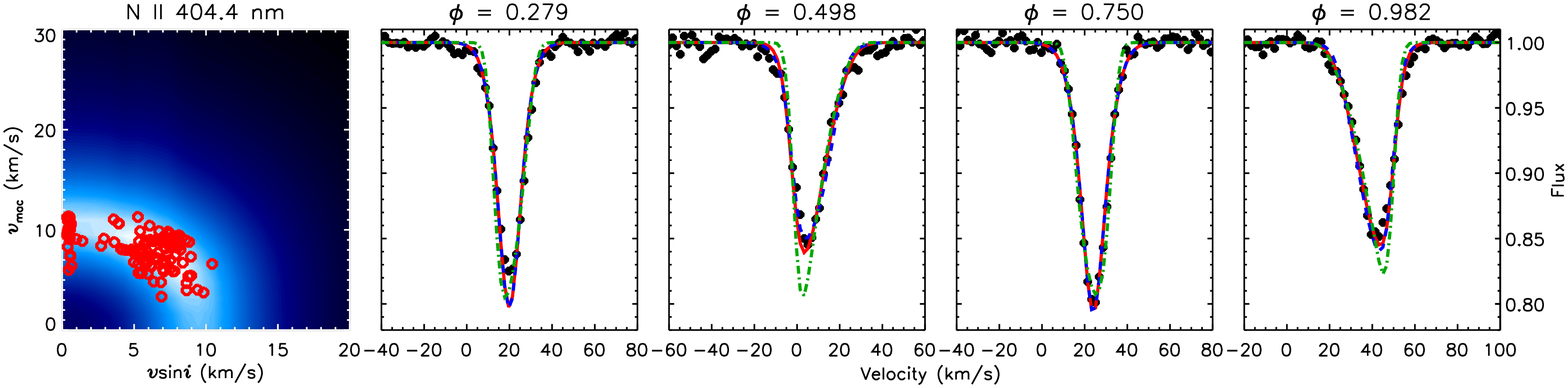} 
\caption{Line profile fitting to determine \vsini~and $v_{\rm mac}$, using Si~{\sc iii} 455.3 nm ({\em top}) and N II 404.4 nm ({\em bottom}). {\em Left}: $\chi^2$ landscapes (high values are dark, low values are light, where the reduced $\chi^2$ goes from approximately 1 to $2\times 10^5$), corresponding to the weighted mean $\chi^2$ across all observations. The best-fit solutions for individual observations are indicated by open red circles. {\em Right}: line profile fits at pulsation phases close to the RV minima and maxima ($\phi=0.498, 0.982$) and $RV=0$ \kms ($\phi=0.279, 0.750$). Observations are shown by black circles. The solid red line shows the best overall fit for each observation. The dashed blue line shows the best fit model with $v\sin{i}=0$~\kms. The dash-dotted green line indicates the best fit model with $v_{\rm mac}=0$ \kms. While the overall best-fit has a non-zero \vsini~of a few \kms, excluding $v_{\rm mac}$ results in a noticeably worse fit, but excluding \vsini~does not substantially worsen the fit. The weaker N~{\sc ii} line yields a lower $v_{\rm mac}$, but does not change the result that $v_{\rm mac}>v\sin{i}$, or that equivalently good fits can be achieved with or without \vsini.}
\label{vsini_vmac}
\end{figure*}

\subsection{Frequency Analysis}

We analyed the RVs using {\sc period04} \citep{2005CoAst.146...53L}. Frequency spectra are shown in Fig.\ \ref{periods}, where the top panels show the frequency spectra for the original dataset, and the panels below show frequency spectra after pre-whitening with the most significant frequencies from previous frequency spectra. A S/N threshold of 4 was adopted as the minimum S/N for significance \citep{1993A&A...271..482B,1997A&A...328..544K}. The uncertainty in each frequency was determined using the formula from \cite{1976fats.book.....B}, $\sigma_{\rm F}=\sqrt{6}\sigma_{\rm obs}/(\pi\sqrt{N_{\rm obs}}A\Delta T)$, where $\sigma_{\rm obs}$ is the mean uncertainty in the RV measurements, $N_{\rm obs}$ is the number of measurements, $A$ is the amplitude of the RV curve, and $\Delta T$ is the timespan of observations.

Period analysis of the CORALIE dataset (left panels in Fig.\ \ref{periods}) yielded the same results as those reported by \cite{saesen2006}, with significant frequencies at $f_{\rm pul,COR} = 4.7715297(5)~{\rm d}^{-1}$ with an amplitude of 16.6~\kms, and at the harmonics 2$f_{\rm pul,COR}$ and 3$f_{\rm pul,COR}$ with amplitudes of 0.6~\kms~and 0.3~\kms, respectively. After pre-whitening with these frequencies, no significant frequencies remain (bottom left panel of Fig.\ \ref{periods}), and all peaks are at the 1$~{\rm d}^{-1}$ aliases of the spectral window. The same analysis of the ESPaDOnS data (middle panels in Fig.\ \ref{periods}) finds maximum power at $f_{\rm pul,ESP} = 4.7715007(4)~{\rm d}^{-1}$, and at 2$f_{\rm pul,ESP}$. 

The combined dataset (right panels in Fig.\ \ref{periods}) yields the strongest signal at $f_{\rm pul,Comb} = 4.7715121(3)~{\rm d}^{-1}$, along with significant peaks at 2$f_{\rm pul,Comb}$ and 3$f_{\rm pul,Comb}$. In this case pre-whitening does not completely remove the peak corresponding to $f_{\rm pul,Comb}$. This could indicate that the pulsation frequency is not the same between the datasets. The difference in frequencies between the CORALIE and ESPaDOnS datasets, $2.9\times 10^{-5}~{\rm d^{-1}}$, is about 30 times larger than the formal uncertainties in either frequency. This difference corresponds to an increase in the pulsation period of 0.1~s, approximately compatible with the increase of $0.063 \pm 0.009~{\rm s}$ expected from the constant period change of $+0.0037(5)~{\rm s}~{\rm yr}^{-1}$ reported by \cite{1999NewAR..43..455J}. 

To explore this hypothesis, phases were calculated assuming a constant rate of period change $-0.02 < \dot{P} < +0.02~{\rm s}~{\rm yr}^{-1}$, with $f_0 = 4.771529(7)~{\rm d}^{-1}$ taken as the frequency inferred from the CORALIE data acquired in 2000, and allowed to vary within the uncertainty in this frequency. The phases $\phi$ were calculated as 

\begin{equation}\label{ephem}
\phi = \frac{{\rm HJD} - (T_0 + NP_0 + 0.5\dot{P}N^2)}{P_0 + N\dot{P}},
\end{equation}

\noindent where $P_0$ is the initial period, and $N$ is the number of pulsation cycles elapsed between the observation and the reference epoch $T_0 = 2451591.42576$, defined as the time of the first RV maximum one cycle before the first observation in the dataset. The goodness-of-fit $\chi^2$ statistic was calculated for each combination of $P_0$ and $\dot{P}$ using a 3$^{rd}$-order least-squares sinusoidal fit in order to account for the pulsation frequency and its first two harmonics. The minimum $\chi^2$ solution was found for $P_0 = 0.2095763(1)~{\rm d}$ and $\dot{P}=+0.0096(5)~{\rm s}~{\rm yr}^{-1}$, where the uncertainties were determined from the range over which $\chi^2$ does not change appreciably. The RVs are shown phased with Eqn.\ \ref{ephem} using these parameters in Fig.\ \ref{rvs}. The bottom panel of Fig.\ \ref{rvs} shows the residual RVs after subtraction of the 3$^{rd}$-order sinusoidal fit. The standard deviation of the residuals is 0.48~\kms, an improvement over the standard deviation of 0.94~\kms~obtained using the constant period determined from the full dataset. The bottom right panel of Fig.\ \ref{periods} shows the frequency spectrum obtained for the residual RVs in Fig.\ \ref{rvs}: in contrast to the results obtained via pre-whitening using a constant pulsation frequency, all power at $f_{\rm pul}$ and its harmonics is removed, with no significant frequencies remainining. 

\subsection{Comparison to previous results}

Our analysis of the CORALIE dataset recovers the same pulsation frequency as that found by the analysis performed by \cite{saesen2006} of the same data. However, we find $\dot{P}$ to be almost 3 times larger than the rate found by \cite{1999NewAR..43..455J}. Whether this reflects an acceleration in the rate of period change will need to be explored in the future when larger datasets with a longer temporal baseline are available. \cite{2015A&A...584A..58N} noted that $\xi^1$ CMa is one of the few $\beta$ Cep stars for which $\dot{P}$ is low enough to be consistent with stellar evolutionary models. The higher $\dot{P}$ suggested by our results brings $\xi^1$ CMa closer to the range observed for other $\beta$ Cep stars, which is to say, slightly above the $\dot{P}$ predicted by evolutionary models for a star of $\xi^1$ CMa's luminosity.

\section{Stellar Parameters}\label{sec:stellar_pars}

\subsection{Line Broadening}\label{subsec:vsini}

The high S/N, high spectral resolution, and wide wavelength coverage of the ESPaDOnS data, combined with the sharp spectral lines of this star, present numerous opportunities for constraining the line broadening mechanisms. As with measurement of RVs, in order to minimize the impact of RV variation, Stokes $I$ spectra from individual sub-exposures were used rather than the spectra computed from the combined exposures. We selected two lines for analysis: the Si~{\sc iii}~455.3~nm line, which is sensitive to the pulsational properties of $\beta$ Cep stars \citep{2003SSRv..105..453A}, and for comparison N~{\sc ii}~404.4~nm, a weaker line with lower sensitivity to pulsation.

We applied a model incorporating the projected rotation velocity \vsini, radial/tangential macroturbulence $v_{\rm mac}$, and assumed radial pulsations, performing a goodness-of-fit (GOF) test on a grid spanning 0--20 \kms~in \vsini~and 0--30 \kms~in $v_{\rm mac}$, in 1 \kms~increments. The disk integration model is essentially as described by \cite{petit2012a}, with the exception of radial pulsations. Pulsations were modeled as a uniform velocity component normal to the photosphere, with the local line profiles in each surface area element shifted by the projected line-of-sight component of the pulsation velocity. A limb darkening coefficient of $\epsilon=0.36$ was used, obtained from the tables calculated by \cite{diazcordoves1995} for a star with \teff$=27$ kK and $\log{g}=3.75$. Local profiles were broadened with a thermal velocity component of 4~\kms~for Si~{\sc iii}, and 5.6 \kms~for N~{\sc ii}, and the disk integrated profile was convolved according to the resolving power of ESPaDOnS. Line strength was set by normalizing the EWs of the synthetic line profiles to match the EWs of the observed line profiles. In the course of this analysis we found a Baade-Wesselink projection factor of measured-to-disk-centre radial velocity of $p=1.45\pm0.02$, which is consistent with determinations for other $\beta$ Cep variables (e.g., \citealt{2013AA...553A.112N}).

Fig.\ \ref{vsini_vmac} shows the resulting best-fit models. Taking the mean and standard deviation across all observations, both Si~{\sc iii} 455.3 nm and N~{\sc ii} 404.4 nm yield $v\sin{i}=5\pm3$~\kms. The two lines give different values for $v_{\rm mac}$, however: 19$\pm$1~\kms~and 8$\pm$2~\kms, respectively. The much lower value of $v_{\rm mac}$ determined from N~{\sc ii} 404.4 nm may indicate that the extended wings of Si~{\sc iii} 455.3 nm, which require a higher value of $v_{\rm mac}$ to fit, are at least partly a consequence of pulsations. 

As demonstrated in Fig.\ \ref{vsini_vmac}, there is essentially no difference in quality of fit between models with nonzero \vsini~and $v_{\rm mac}$, and zero \vsini; conversely, a much worse fit is obtained by setting $v_{\rm mac}=0$ \kms. For Si~{\sc iii} 455.3 nm, the $\chi^2$ of the model without $v_{\rm mac}$ is 13 times higher than the $\chi^2$ of the model without \vsini, while the $\chi^2$ of the model without \vsini~is only 1.4 times higher than the model with both turbulent and rotational broadening. The difference is not as great for N~{\sc ii} 404.4 nm, although the $\chi^2$ of the model with rotational broadening only is still higher than either the model with turbulent broadening only, or both rotational and turbulent broadening. \cite{2013AA...559L..10S} used magnetic O stars with rotational periods known to be extremely long (i.e.\ with \vsini~$\sim 0$~\kms) and found that both the goodness-of-fit test and the Fourier transform methods severely over-estimate \vsini~when $v_{\rm mac} > v\sin{i}$, as is the case for both of the lines we have examined. We conclude that the star's line profiles are consistent with \vsini~$= 0$~\kms.


\begin{figure}
\centering
\includegraphics[width=\hsize]{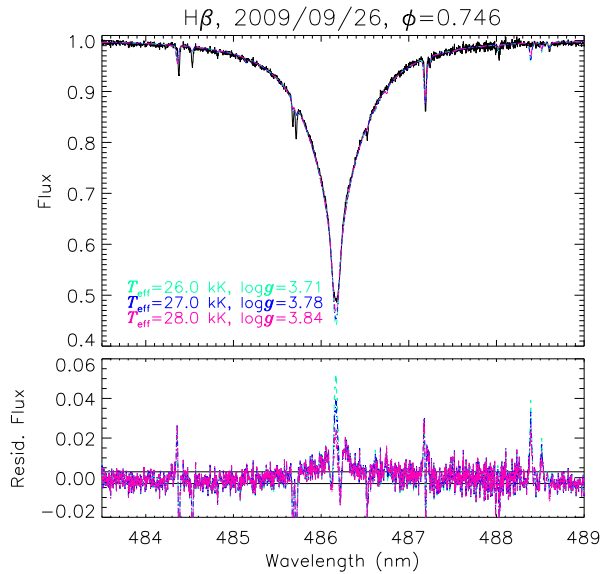}
\caption{Surface gravity determination using H$\beta$. {\em Top}: model fits for \teff$=26,27$, and 28 kK (cyan, blue, and pink, respectively) to the observed profile (black). The observation nearest $\phi=0.75$ (where the RV$=0$ \kms) was used. {\em Bottom}: residuals. The horizontal lines indicate the mean flux uncertainty. As explored in \S~\ref{subsec:halpha}, there is emission near the line core; this is more prominent in H$\alpha$. This region was not included in the fit.}
\label{logg}
\end{figure}

\begin{figure*}
\centering
\includegraphics[width=18cm]{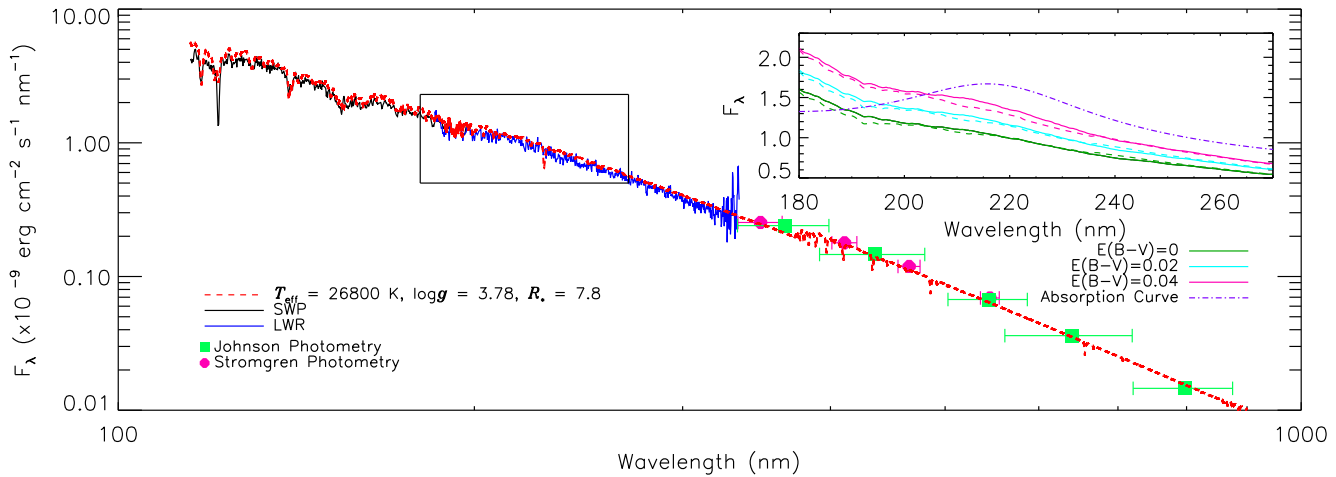}
\caption{Low-resolution IUE spectroscopy (solid lines) with the SWP (black) and LWR (blue) cameras, and photometry (filled symbols) compared to a synthetic photospheric spectrum (red dashed line). No dereddening correction has been applied to the data. {\em Inset}: Comparison between synthetic spectra (dashed lines) and observed spectra (solid lines) deredded using three values of $E(B-V)$: 0 (green), 0.02 (cyan), and 0.04 (pink). Both the observed spectrum and the best-fit synthetic spectra have been smoothed with a boxcar algorithm. The dot-dashed (purple) line is the reddening curve in arbitrary units. Even mild dereddening produces a bulge in the vicinity of the 220 nm~silicate absorption feature.}
\label{iue_lum_teff}
\end{figure*}

\subsection{Surface Gravity}\label{subsec:logg}

The surface gravity $\log{g}$ was determined by fitting {\sc tlusty} BSTAR2006 synthetic spectra \citep{lanzhubeny2007} to the H$\beta$ and H$\gamma$ lines of the ESPaDOnS spectrum acquired at pulsation phases at which the RV was closest to the systemic velocity of 22.5 \kms~(i.e.\ at pulsation phases 0.25 or 0.75). As both of these lines are close to the edges of their respective orders, in order to avoid warping the line the orders were merged from un-normalized spectra, then normalized using a linear fit to nearby continuum regions. Five surface gravities were tested between $\log{g}=3.25$ and $\log{g}=4.25$, and a low-order polynomial was fit to the resulting $\chi^2$ in order to identify the lowest $\chi^2$ solution. Uncertainties were determined by fitting models with \teff~of 26, 27, and 28 kK, spanning the approximate range in the uncertainty in \teff~(see below). Fig.\ \ref{logg} shows the best-fit models, $\log{g} = 3.78 \pm 0.07$, compared to the observed H$\beta$ line. Fitting to H$\gamma$ yielded identical results. 

\subsection{Effective Temperature}\label{subsec:teff}

The effective temperature \teff~was determined using photometry, spectrophotometry, and spectroscopy. Using the Str\"omgren $uvby\beta$ photometric indices obtained by \cite{1998AAS..129..431H}, and the {\sc idl} program {\sc uvbybeta.pro}\footnote{Available at \url{http://idlastro.gsfc.nasa.gov/ftp/pro/astro/uvbybeta.pro}.} which implements the calibrations determined by \cite{1985Ap&SS.117..261M}, yields \teff$=26.2$ kK. Using the Johnson $UBVRIJHK$ photometry collected by \cite{2002yCat.2237....0D} and the colour-temperature calibration presented by \cite{2011ApJS..193....1W} yields \teff$=25\pm3$ kK, where the uncertainty was determined from the range of \teff~across the $(U-B), (V-K), (B-V), (V-R)$, and $(J-K)$ colours.

The \teff~was measured using spectrophotometric data by fitting synthetic BSTAR2006 spectra to the IUE low-dispersion spectra, Johnson UBVRI photometry, and Str\"omgren $uvby\beta$ photometry. The Johnson photometry was converted into absolute flux units using the calibration provided by \cite{1998AA...333..231B}, and the Str\"omgren photometry converted using the calibration determined by \cite{gray1998}. The synthetic spectra were scaled by $1/d^2$ and $4\pi R_*^2$, where the distance $d$ was fixed by the Hipparcos parallax and the stellar radius $R_*$ was left as a free parameter. The resulting fit is shown in Fig.\ \ref{iue_lum_teff}. The best-fit model yields \teff$=26.8 \pm 0.4$ kK and $R_*=7.8~R_\odot$. The uncertainty comes from the spread in \teff~over the range of $\log{g}=3.78\pm0.07$, and for two values of $E(B-V)$, 0 and 0.02. Fluxes were dereddened using the {\sc idl} routine {\sc fm\textunderscore unred}\footnote{Available at 
\url{http://idlastro.gsfc.nasa.gov/ftp/pro/astro/fm_unred.pro}.}, which utilizes the \cite{1990ApJS...72..163F} extinction curve parameterization. As demonstrated by the inset in Fig.\ \ref{iue_lum_teff}, there is a clear absence of interstellar silicate absorption near 220~nm. The de-reddened spectra in the inset in Fig.\ \ref{iue_lum_teff} assume the strength of the absorption bump is $c_3=2.1$, the minimum Galactic value \citep{2003ApJ...588..871C}. Even with this low value of $c_3$, using $E(B-V)=0.04$ results in an increase of the flux near the silicate absorption bump which is not observed. Using the average Galactic value of $c_3=3.23$ instead would require that $E(B-V)$ be even lower. 

\subsection{Comparison to previous results}

Previous measurements of \vsini~have in general yielded somewhat higher, non-zero results than those obtained here ($15\pm 1.5$~\kms, \citealt{saesen2006}; 9$\pm$2~\kms, \citealt{2010AA...515A..74L}; 14$\pm$5~\kms, \citealt{2014AA...569A.118A}). \citeauthor{saesen2006} did not include $v_{\rm mac}$ in their profile fit, but did include pulsation and thermal broadening; \citeauthor{2010AA...515A..74L} did not consider pulsations; and \citeauthor{2014AA...569A.118A} included $v_{\rm mac}$, but neither pulsation velocity nor thermal broadening. Our inclusion of thermal, macroturbulent, pulsational, and rotational velocity fields likely explains why our best-fit value of \vsini, $5 \pm 3$~\kms, is somewhat lower than the values found previously, since all four sources of line broadening are of a similar magnitude in this star. 


Stellar parameters for $\xi^1$ CMa have been determined numerous times using a variety of different spectral modelling methodologies, e.g., DETAIL/SURFACE \citep{2008AA...481..453M}, FASTWIND \citep{2010AA...515A..74L}, and PoWR \citep{oskinova2011}. \cite{2005AA...433..659N} employed an algorithmic method to simultaneously determine reddening, \teff, metallicities, and surface gravities for $\beta$ Cep stars observed with IUE, and for $\xi^1$ CMa found \teff~$=24 \pm 1$~kK, $\log{g}=3.89$, and $E(B-V) < 0.01$, i.e. their analysis favoured a cooler, slightly less evolved star, but with the same very low level of reddening. \citeauthor{2005AA...433..659N} derived $\log{g}$ from photometric calibrations, rather than line profile fitting, which likely explains the different values of $\log{g}$. \cite{2008AA...481..453M} found \teff~$=27.5$~kK and $\log{g}=3.75$, \cite{2010AA...515A..74L} derived \teff~$=27 \pm 1$~kK and $\log{g}=3.80 \pm 0.15$, and \cite{oskinova2011} found \teff~$=27.5 \pm 0.5$~kK: all are in agreement with the results found here. The slightly higher \teff~found in these works is likely a consequence of the higher value of reddening which, as is shown in \S~\ref{subsec:teff} and Fig.\ \ref{iue_lum_teff}, is not consistent with the absence of silicate absorption. While reddening has an effect on the spectrophotometric method applied in \S~\ref{subsec:teff}, we obtained essentially identical results using line strength ratios and ionization balances below in \S~\ref{subsec:teffvar}, which should not be affected by reddening. 

\subsection{Effective Temperature Variation}\label{subsec:teffvar}

\begin{figure}
\centering
\includegraphics[width=8.5cm]{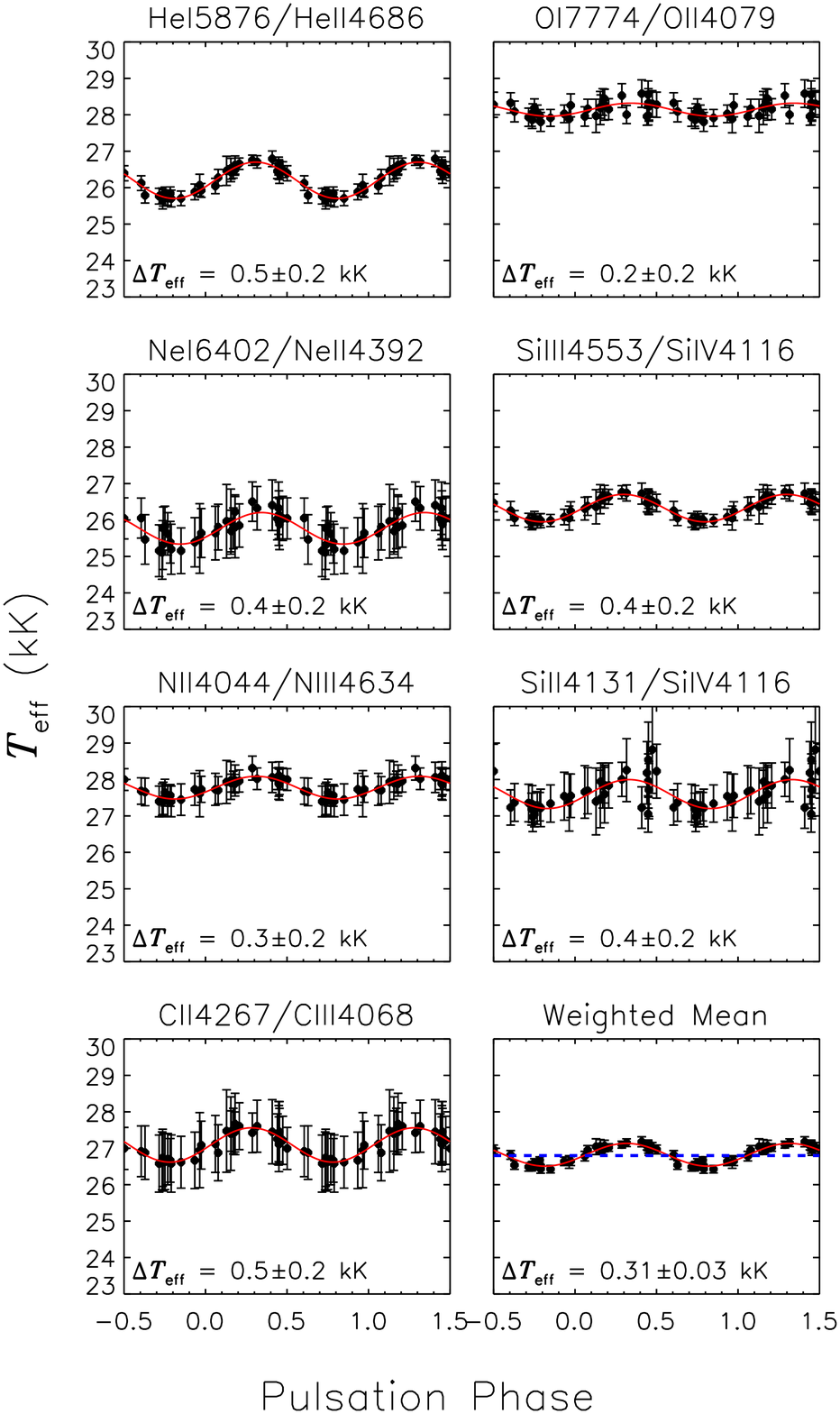}
\caption{\teff~determined from EW ratios of various ion pairs as a function of pulsation phase. While the mean \teff~differs by $\sim$2 kK between different ion pairs, a coherent and identical modulation with the pulsation period is seen in all cases. The semi-amplitude of the \teff~variation, as determined from the least-squares sinusoidal fit to the data, is given at the bottom of each panel. The bottom right panel shows the weighted mean \teff~across all ion pairs, with the dashed blue line indicating the \teff~determined from SED fitting (Fig.\ \ref{iue_lum_teff}).}
\label{ion_teff}
\end{figure}

\begin{figure}
\centering
\includegraphics[width=8cm]{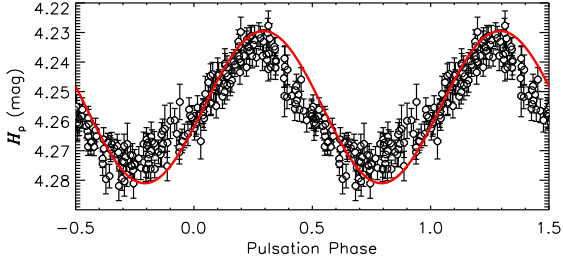}
\caption{Predicted variation in visual magnitude (red line) as determined from the variation in $R_*$ from RVs (Fig.\ \ref{rvs}) and in \teff~from EW ratios (Fig.\ \ref{ion_teff}) as compared to the Hipparcos photometry (black circles). Pulsation phases were calculated using eqn.\ \ref{ephem}.}
\label{phot_mod}
\end{figure}

The photospheric temperature of a $\beta$ Cephei star is variable, a property which can be explored both photometrically and spectroscopically. The only public two-colour photometry for $\xi^1$ CMa are the Tycho $BV_{\rm T}$ observations. While pulsational modulation is visible in both passbands, the precision of the data is not sufficient to detect a coherent signal in the colours. This is not surprising, as $B-V$ colours are poorly suited to determining \teff~for such hot stars. We therefore attempted to constrain the \teff~change via the ionization balances of various spectral lines, under the assumption that all equivalent width (EW) changes are due to temperature variation. 

We first measured EWs from the ESPaDOnS observations, and from appropriate synthetic spectra obtained from the {\sc tlusty} BSTAR2006 library \citep{lanzhubeny2007}. We used the ESPaDOnS Stokes $I$ spectra obtained from the combined sub-exposures, in order to maximize the S/N. The following spectral lines were used: He~{\sc i} 587.6 nm, He {\sc ii} 468.6 nm, C {\sc ii} 426.7 nm, C {\sc iii} 406.8 nm, N~{\sc ii} 404.4 nm, N~{\sc iii} 463.4 nm, O {\sc i} 777.4 nm, O {\sc ii} 407.9 nm, Ne~{\sc i} 640.2 nm, Ne~{\sc ii} 439.2 nm, Si {\sc ii} 413.1 nm, Si~{\sc iii} 455.3 nm, and Si~{\sc iv} 411.6 nm. Since the EW can depend on $\log{g}$ as well as \teff, the model EW ratios were linearly interpolated to $\log{g}=3.78$ between the results for $\log{g}=3.75$ and $\log{g}=4.0$. We then determined \teff~by interpolating through the resulting grid to the EW ratios measured from the observed spectra. The results are illustrated in Fig.~\ref{ion_teff}. While there is a considerable spread in values, from approximately 25 to 28 kK, the mean value of 26.8 kK agrees well with the spectrophotometric determination. A coherent variation with the pulsation period is seen for all ion strength ratios. Individual atomic species yield variations with semi-amplitudes up to $\Delta T_{\rm eff} = \pm 500$ K. Taking the weighted mean across the effective temperatures from all ion strength ratios at each phase yields a variation of $\Delta T_{\rm eff} = \pm 310 \pm 30$ K. If the zero-point of each \teff~variation is first subtracted, the resulting mean \teff~semi-amplitude is $330 \pm 30$~K, slightly larger but overlapping within the uncertainty.

To check the validity of the temperature variation, we used it to model the Hipparcos light curve, as demonstrated in Fig.\ \ref{phot_mod}. With the uncertainty in the pulsation period and rate of period change determined in \S~\ref{sec:pulsation}, the accumulated uncertainty in pulsation phase for the Hipparcos photometry is between 0.009 and 0.012 cycles. We began by determining the mean radius $R_*$ from the mean \teff~and the luminosity $\log{L}$. $\log{L}$ was obtained from the Hipparcos parallax distance $d=424\pm35$~pc, the apparent $V$ magnitude 4.06, and the bolometric correction $BC=-2.61 \pm 0.09$~mag obtained by linear interpolation through the theoretical BSTAR2006 grid according to \teff~and $\log{g}$ \citep{lanzhubeny2007}. The absolute $V$ magnitude $m_V = V - A_V - \mu = -3.86 \pm 0.18$~mag, where $\mu = 5\log{d} - 5 = 8.13 \pm 0.18$~mag is the distance modulus and $A_V = E(B-V)/3.1 < 0.06$~mag is the extinction. The bolometric magnitude is then $M_{\rm bol} = m_V + BC = -6.47 \pm 0.27$~mag, yielding $\log{(L/L_\odot)} = (M_{\rm bol,\odot}-M_{\rm bol})/2.5=4.49 \pm 0.11$ where $M_{\rm bol,\odot} = 4.74$~mag. This yields $R_* = \sqrt{(L/L_\odot)/(T_{\rm eff}/T_{\rm eff,\odot})^4} = 7.9 \pm 0.6~R_\odot$, where $T_{\rm eff,\odot}=5.78$ kK. This radius agrees well with the value found via spectrophotometric modelling, $R_*=7.8~R_\odot$, in which the radius was left as a free parameter (Fig.\ \ref{iue_lum_teff}).

Integrating the radial velocity curve (Fig.\ \ref{rvs}) in order to obtain the absolute change in radius, and assuming radial pulsation, yields a relatively small change in radius of $\pm 6.58 \times 10^4$ km, corresponding to between 1.0 and 1.5\% of the stellar radius. The radius variation was then combined with the \teff~variation from Fig.\ \ref{ion_teff} to obtain $\log{L}$ and the $BC$ at each phase. The apparent $V$ magnitude was then obtained by reversing the calculations in the previous paragraph. The solid red line in Fig.\ \ref{phot_mod} shows the resulting model light curve compared to the Hipparcos photometry, where we made the assumption that $V$ and $H_{\rm p}$ are approximately equivalent. The larger 500 K \teff~semi-amplitude measured from individual pairs of ions is not consistent with the light curve, yielding a photometric variation larger than observed, with a semi-amplitude of 0.037 mag as compared to the observed semi-amplitude of $\sim$0.021 mag. The apparent phase offset of $\sim$0.05 cycles between the predicted and observed photometric extrema may suggest that $\dot{P}$ may not in fact be constant; alternatively, there may simply be too few measurements to constrain the \teff~variation well enough to obtain a close fit to the photometric data, as the latter is clearly not a perfect sinusoid. 

\begin{figure}
\centering
\includegraphics[width=8cm]{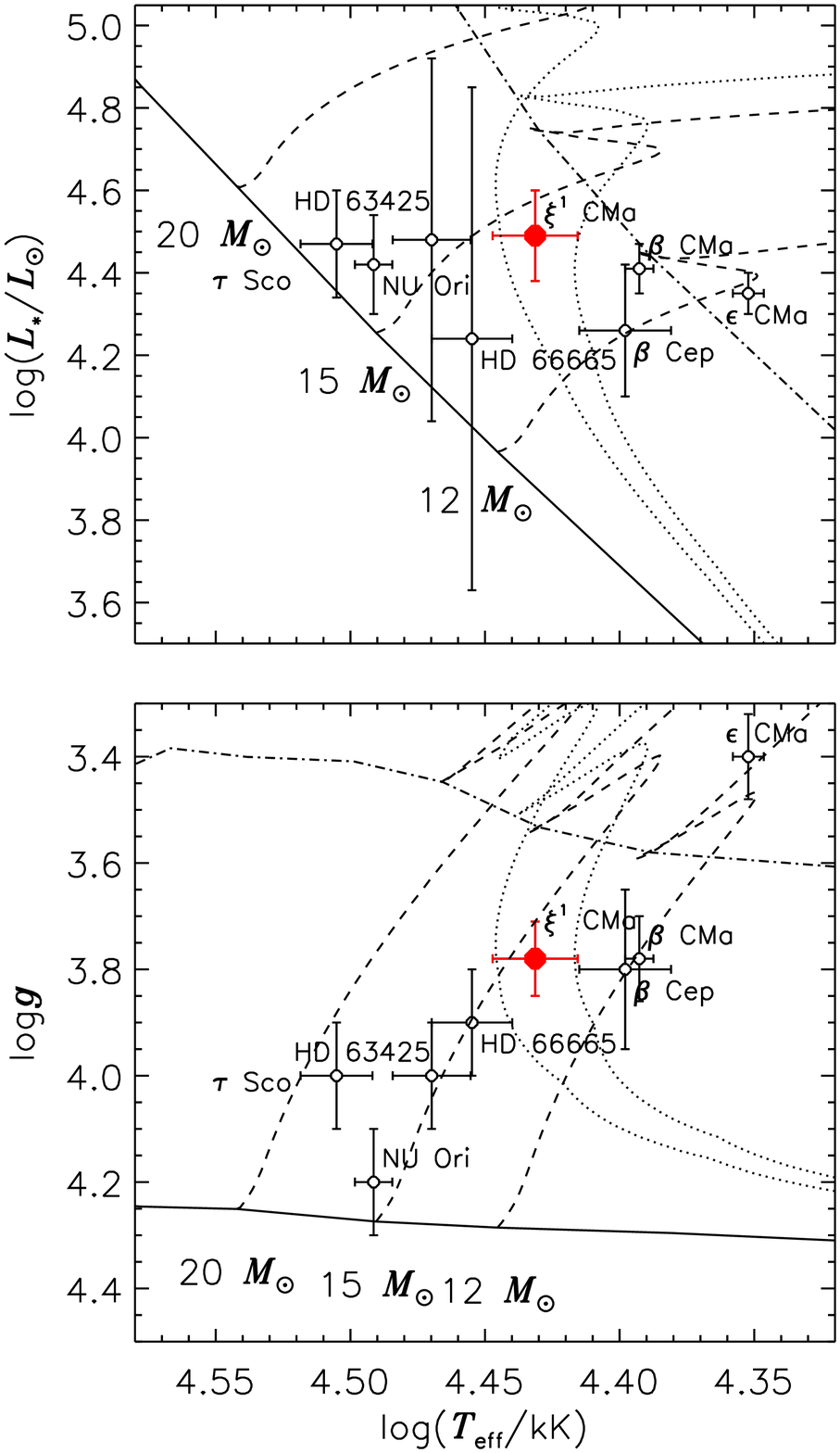}
\caption{$\xi^1$ CMa's position on the HRD (top) and on the \teff-$\log{g}$ diagram (bottom), relative to other magnetic stars with similar stellar parameters. The solid line indicates the Zero-Age Main Sequence (ZAMS), and the dot-dashed line the Terminal Age Main Sequence (TAMS). Dashed lines show evolutionary tracks from the (rotating, non-magnetic) models by \protect\cite{ekstrom2012}, and dotted lines isochrones from the same models for $\log{(t/{\rm yr})}=7.0$ (left) and 7.1 (right). The star's fundamental parameters indicate $M_*=14.2\pm0.4 M_\odot$ and $t=11\pm0.7$ Myr. $\xi^1$ CMa is the most evolved magnetic B-type star above about 12 $M_\odot$.}
\label{hrd}
\end{figure}

\subsection{Mass and Age}\label{subsec:massage}

Fig.\ \ref{hrd} shows $\xi^1$ CMa's position on the Hertzsprung-Russell diagram (HRD) (top) and the \teff-$\log{g}$ diagram (bottom), where the mean \teff~was used. Comparison to the evolutionary models calculated by \cite{ekstrom2012} (which assume an initial rotational velocity of 40\% of the critical velocity) indicates that the stellar mass $M_*=14.2\pm0.4~M_\odot$, and that the absolute stellar age is $t=11\pm0.7$~Myr. The \cite{ekstrom2012} models do not include the effects of magnetic fields, however, grids of self-consistent evolutionary models including these effects in a realistic fashion are not yet available.  

The positions of the star on the two diagrams are mutually consistent. The other known magnetic B-type stars with similar stellar parameters are also shown in Fig.\ \ref{hrd}. The stellar parameters of the majority of the other stars were obtained from the catalogue of magnetic hot stars published by \cite{petit2013} and references therein; those of $\beta$ CMa and $\epsilon$ CMa were obtained from \cite{2015AA...574A..20F}. $\xi^1$ CMa is one of the most evolved stars in the ensemble, and is the most evolved star with a mass above about 12 \msun. Its position on the \teff-$\log{g}$ diagram indicates it may be a more evolved analogue of NU Ori, HD 63425, and HD 66665. 

\begin{table}
\caption{Observational and physical properties. The third column gives the reference, where numerical references indicate the section within this paper where the parameter is determined.}
\begin{center}
\begin{tabular}{lrl}
\hline
\hline
$v_{\rm sys}$~(\kms) & 22.5$\pm$1 & \ref{sec:pulsation} \\
\vsini~(\kms) & $\le 8$ & \ref{subsec:vsini} \\
$v_{\rm mac}$~(\kms) & 8$\pm$2 & \ref{subsec:vsini} \\
\hline
Sp. Type & B0.5 IV & \cite{hubrig2006} \\
$V$ (mag) & 4.3 & \cite{2002yCat.2237....0D} \\
$\pi$ (mas) & 2.36$\pm$0.2 & \cite{vanleeuwen2007} \\
$d$ (pc) & 424$\pm$35 &  \cite{vanleeuwen2007} \\
$BC$ (mag) & -2.61$\pm$0.09 & \ref{subsec:teffvar} \\
$A_{\rm V}$ (mag) & $<0.06$ & \ref{subsec:teffvar} \\
$m_{\rm V}$ (mag) & -3.86$\pm$0.18 & \ref{subsec:teffvar}  \\
$M_{\rm bol}$ (mag) & -6.47$\pm$0.27 & \ref{subsec:teffvar} \\
\hline
$\log{(L/L_\odot)}$ & 4.49$\pm$0.11 & \ref{subsec:teffvar} \\
\teff~(kK) & 27$\pm$1 & \ref{subsec:teff} \\
$\log{g}$ & 3.78$\pm$0.07 & \ref{subsec:logg} \\
$R_*~(R_\odot)$ & 7.9$\pm$0.6 & \ref{subsec:teff} \\
$M_*~(M_\odot)$ & 14.2$\pm$0.4 & \ref{subsec:massage} \\
Age (Myr) & 11.1$\pm$0.7 & \ref{subsec:massage} \\
$\tau_{\rm MS}$ & 0.77$\pm$0.04 & \ref{subsec:massage} \\
$R_*/R_{\rm ZAMS}$ & 1.8$\pm$0.1 & \ref{subsec:massage} \\
\hline
\hline
\hline
\end{tabular}
\end{center}
\label{params}
\end{table}

As an additional check on the stellar mass and radius, we utilized the pulsation period and the relation $P_{\rm fun} = \frac{Q}{\sqrt{\rho}}$, where $P_{\rm fun}$ is the fundamental pulsation period, $\rho$ is the mean density, and $Q$ is the pulsation constant. With $M_* = 14.2\pm0.4~M_\odot$, $R_* = 7.9 \pm 0.6~R_\odot$, and the theoretical value $Q = 0.035$ \citep{1973ApJ...179..235D,stothers1976}, this yields $P_{\rm fun} = 0.18 \pm 0.02$ d, very close to the true pulsation period $P \sim 0.21$~d. If the actual pulsation period is used to calculate $Q$, we obtain $Q = P_{\rm fun} \sqrt{\rho} = 0.041 \pm 0.003$. 

\section{Binarity}\label{sec:interferometry}

\subsection{Interferometry}

\cite{fr2011} reported that $\xi^1$ CMa hosts weak H$\alpha$ emission, which is atypical for a star of $\xi^1$ CMa's stellar properties. There are two possibilities to explain this: first, that it originates in a stellar magnetosphere, and second, that it originates in the decretion disk of a heretofore undetected Be companion star. It seems reasonable to expect that $\xi^1$ CMa has a binary companion, as the binary fraction is $\sim$65\% for B0 stars \citep{2012MNRAS.424.1925C}. Furthermore, early claims that the H$\alpha$ emission of $\beta$ Cep, a similar magnetic early-B type pulsator, originated in its magnetosphere \citep{donati2001}, proved unfounded following the detection of a Be companion star by \cite{2006AA...459L..21S}. 

The peak H$\alpha$ emission has a maximum strength of 28\% of the continuum (see \S~\ref{subsec:halpha}). Be star emission lines can range up to $\sim 10\times$ of the continuum, although the emission is typically much less than this. We therefore expect that the putative companion must have a luminosity of $ \ge 2.8\%$ of that of $\xi^1$ CMa, yielding $\log{(L/L_\odot)} \ge 2.9$ or $V < 7.6$. Assuming the pair to be coeval, and therefore locating the star on the same isochrone as $\xi^1$ CMa, the companion's mass should be near $5~M_\odot$, compatible with a classical Be star. 

$\xi^1$ CMa is listed in the Washington Double Star Catalogue as possessing a binary companion \citep{2001AJ....122.3466M}. However, the reported companion is both too dim to account for the H$\alpha$ emission ($V = 14$) and too far away, located approximately 28" from the primary, i.e.\ well outside the 1.6" ESPaDOnS aperture.

\begin{figure}
\centering
\includegraphics[width=8cm]{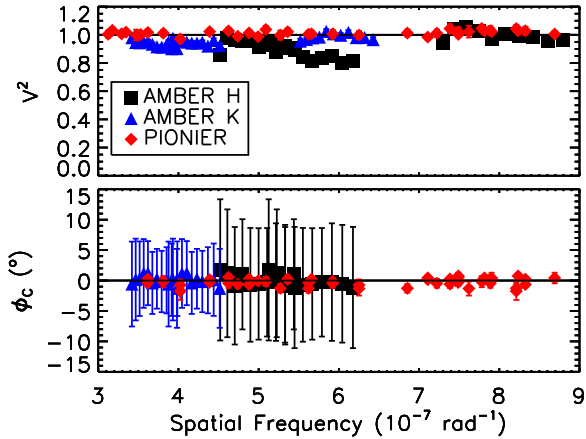}
\caption{Low-resolution VLTI-AMBER H and K data and VLTI-PIONIER data, showing squared visibilities ({\em top}) and closure phases ({\em bottom}).}
\label{amber_pionier}
\end{figure}

We acquired H and K low-resolution VLTI-AMBER data, together with VLTI-PIONIER data, in order to search for the presence of a binary star. The squared visibilities $V^2$ and closure phases $\phi_{\rm C}$ of these data are shown as functions of spatial frequency in Fig.\ \ref{amber_pionier}. Both the AMBER K band and PIONIER $V^2$ measurements are entirely flat, compatible with a single unresolved source. There is furthermore no signal in the PIONIER $\phi_{\rm C}$ measurements, which are more precise than those available from AMBER. 

We analyzed the data using the standard {\sc litpro} package\footnote{{\sc litpro} software is available at http://www.jmmc.fr/litpro} \citep{2008SPIE.7013E..1JT}, fitting a two-point model, with one point fixed in the centre of the map and the (x,y) coordinates of the second free to move. The overall upper limit on the flux of a secondary component is 1.7\% of the primary star's flux. In order to constrain the maximum flux of a binary companion at different distances from the primary, we repeated the two-point model fit in successively wider boxes (as the current version of {\sc litpro} does not support polar coordinates, a Cartesian approximation to annuli was used). As these flux ratios are in the $H$ and $K$ bands, but our estimated minimum flux ratio is in the $V$ band, we used synthetic {\sc tlusty} SEDs \citep{lanzhubeny2007} to convert the $H$ and $K$ band flux ratios to $V$ band flux ratios. In this step we assumed that the secondary's \teff~is near 20 kK, appropriate to a 5-6 $M_\odot$ star near the main sequence. A companion of the required brightness is ruled out beyond $\sim 40$ AU. 

Close binaries containing magnetic, hot stars are extremely rare, with $<2$\% of close binary systems containing a magnetic companion earlier than F0 \citep{2015IAUS..307..330A}. This does not mean that a close companion can be ruled out {\em a priori}. Such a companion may be detectable via radial velocities. As found in \S~\ref{sec:pulsation}, the RV curve of $\xi^1$ CMa is extremely stable, with a standard deviation of the residual RVs of $\sim$0.5~\kms. To determine if a Be companion should have been detected, we computed radial velocities across a grid of models with secondary masses $1~M_\odot < M_{\rm S} < 10~M_\odot$, semi-major axes $0.05~{\rm AU} < a < 500~{\rm AU}$, eccentricities $0 < e < 0.9$, and inclinations of the orbital axis from the line of sight $1^\circ < i < 90^\circ$. We then phased the JDs of the observations with the orbital periods $P_{\rm orb}$, using a single zero-point if $P_{\rm orb}$ was less than the time-span of the observations and multiple, evenly-spaced zero-points if $P_{\rm orb}$ was greater than this span. We then calculated the expected RV of the primary at each orbital phase, and compared the standard deviation $\sigma_{\rm orb}$ of these RVs with the observed standard deviation $\sigma_{\rm obs}$ (RV amplitudes were not used as, for orbits with periods longer than the timespan of observations, the full RV variation would not have been sampled). A given orbit was considered detectable if $\sigma_{\rm orb} > 3\sigma_{\rm obs}$. Numerical experiments with synthetic RV curves including gaussian noise with a standard deviation of 0.5~\kms~indicate that this criterion is likely conservative: with 624 RV measurements, input periods can generally be recovered even when the semi-amplitude of the RV curve is similar to the noise level. 

Each orbit was assigned a probability $P(i)$ assuming a random distribution of $i$ over $4\pi$ steradians (i.e., $P(i) = 1-\cos{i}$), and a flat probability distribution for $M_{\rm S}$, $e$, and $a$. From this we obtained the 1, 2, and 3$\sigma$ upper limits on a companion's mass as a function of $a$. This mass upper limit was then transformed into a $V$ magnitude lower limit by interpolating along the 10 Myr isochrone in Fig.\ \ref{hrd}. Within the 40 AU inner boundary of the interferometric constraints, a binary companion of sufficient brightness to host the H$\alpha$ emission can be ruled out entirely at 1$\sigma$ confidence, and almost entirely at 2$\sigma$. 

We have proceeded under the very conservative assumption of a Be star with H$\alpha$ emission of 10$\times$ its continuum level, however the majority of Be stars have emission of at most a few times the continuum, and many have much less than this. Thus, the upper limits on companion mass and brightness determined above rule out all but the most exceptional of classical Be stars as a possible origin for the H$\alpha$ emission. We conclude that there is unlikely to be an undetected binary classical Be companion star that is sufficiently bright to be the source of the star's H$\alpha$ emission.

\subsection{Spectro-interferometry}

\begin{figure}
\centering
\includegraphics[width=9cm]{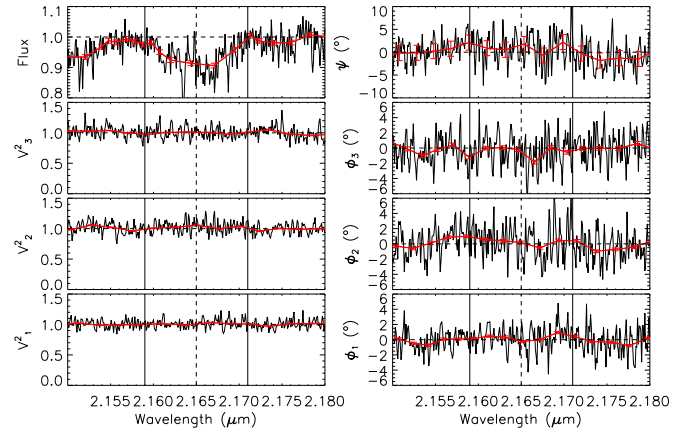}
\caption{High-resolution VLTI/AMBER data showing (clockwise from top left), flux, closure phase $\psi$, differential phases $\phi_i$, and squared visibilities $V^2_i$. Black shows the raw data. Red lines show data binned to 20\% of the mean error bar in the raw data. Vertical dashed lines indicate the central wavelength of the Br$\gamma$ line, solid vertical lines denote the extent of the line.}
\label{amber_hires}
\end{figure}

The formation of the line emission around the $\beta$ Cep star itself is supported by the absence of any spectrointerferometric signal across the Br$\gamma$ line. Such observations were taken with AMBER, and no signature in phase was detected on the level of 2$^\circ$ (Fig.\ \ref{amber_hires}). This means that the photometric position as a function of wavelength remained stable on the level of 60 $\mu$as \citep{2003A&A...400..795L,2009AA...504..929S}. An emission component of about 30\% of the strength of the continuum (Br$\gamma$ emission typically being comparable in strength to that of H$\alpha$ in Be stars) must therefore have its photocenter within 60 $\mu$as from the photocenter of the nearby continuum, as otherwise it would have produced a detectable offset of the phase signal. This not only excludes a general offset from the central star, i.e.\ formation around a companion, but also an extended orbiting structure, in which the blue emission would be formed at an offset opposite to  the red emission.

\section{Magnetic Field}

\subsection{Least Squares Deconvolution}\label{subsec:lsd}

\begin{figure}
\centering
\includegraphics[width=8cm]{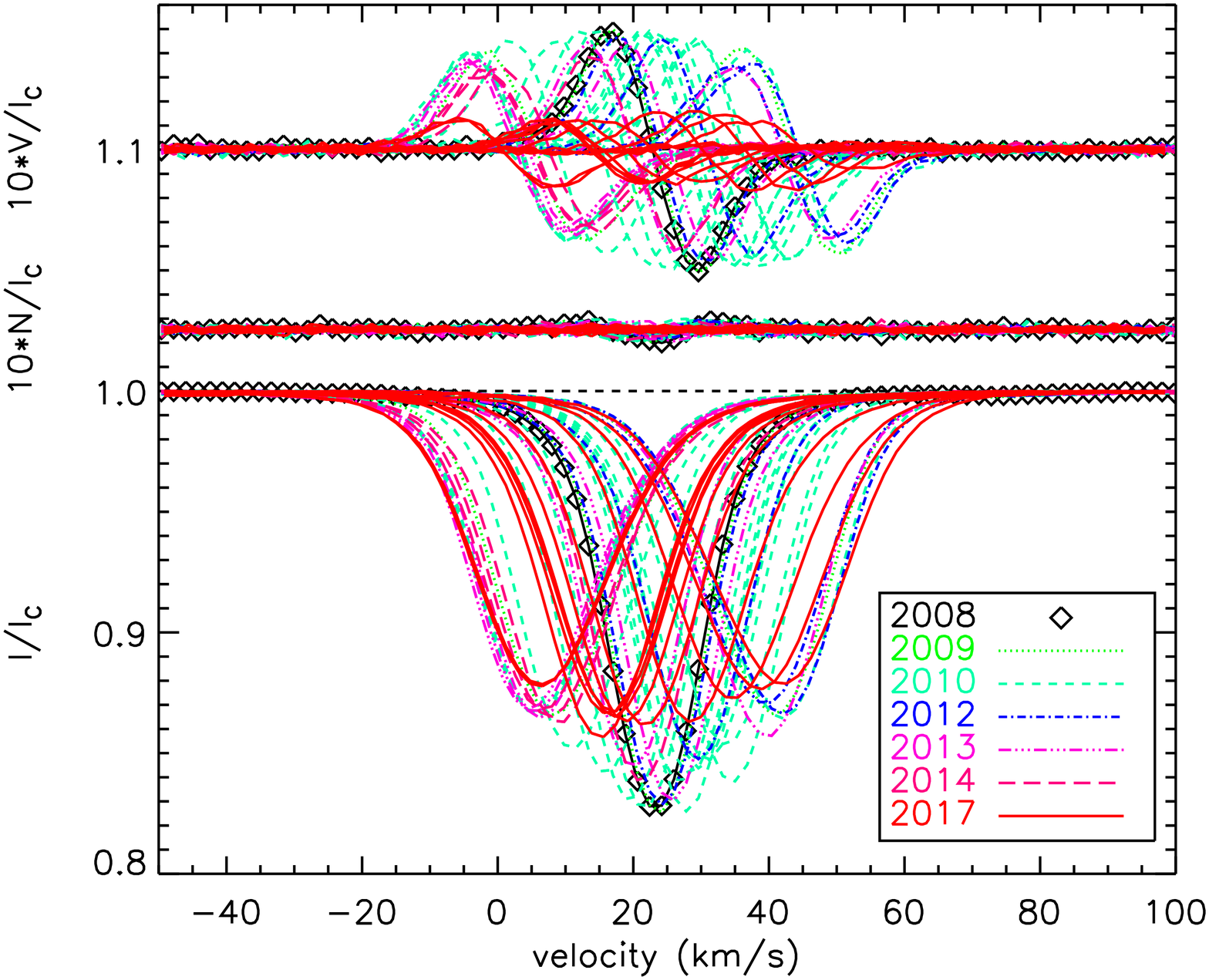}
\includegraphics[width=8cm]{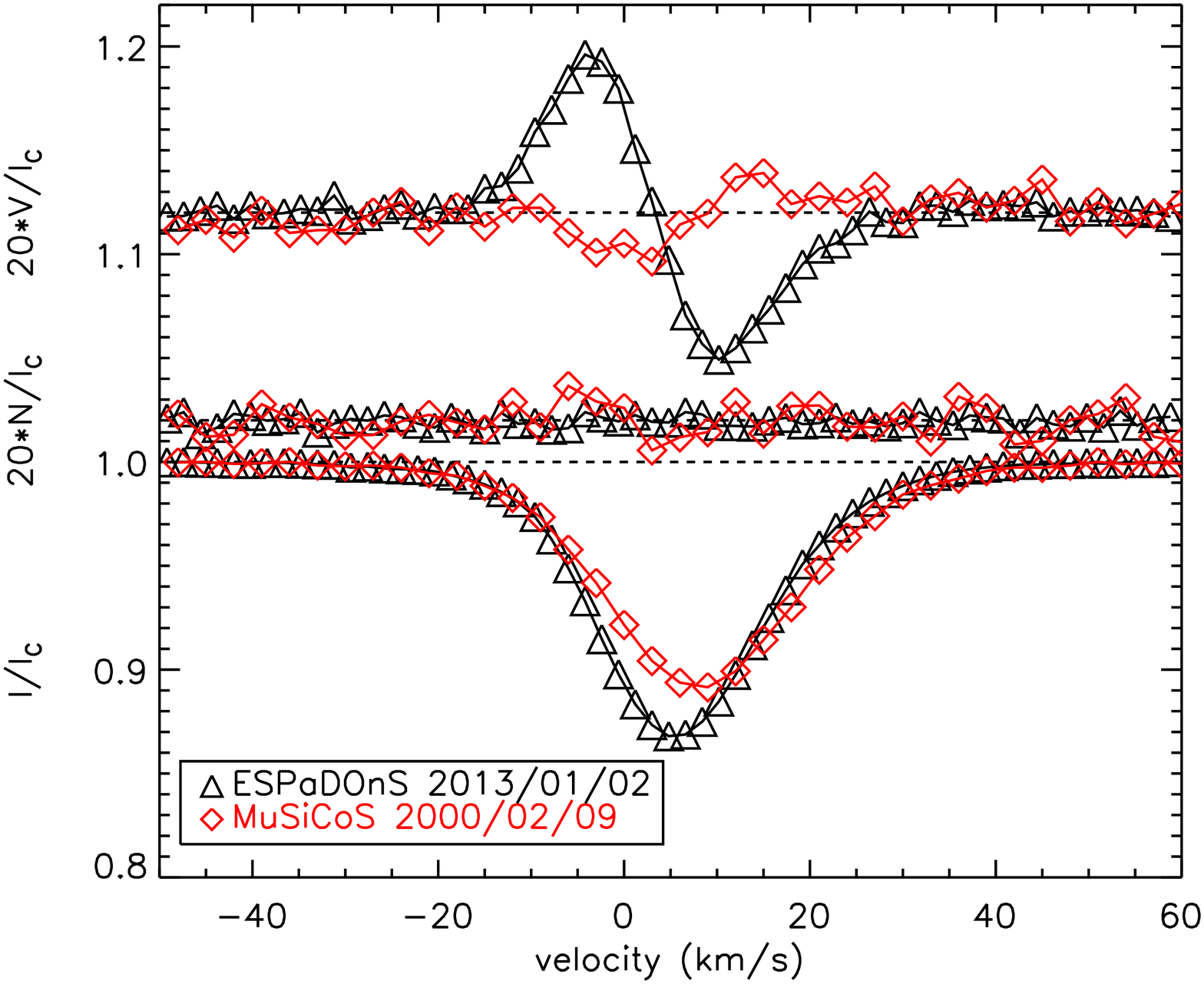}
\caption{{\em Top}: ESPaDOnS LSD profiles: Stokes $I$ (bottom), $N$ (middle), Stokes $V$ (top). {\em Bottom}: ESPaDOnS (black triangles) and MuSiCoS (red diamonds) LSD profiles, extracted with identical line masks, from observations acquired at similar pulsation phases ($\sim$0.44). The broader and shallower MuSiCoS profile is a consequence of the much longer exposure times of the MuSiCoS observations, the instrument's lower spectral resolution, and the intrinsic variability of the star. While the MuSiCoS LSD profile is obviously noisier, the Stokes $V$ signature is clearly of the opposite polarity of the ESPaDOnS observation.}
\label{esp_mus_lsd}
\end{figure}

While $\xi^1$ CMa's sharp spectral lines and the high S/N of the ESPaDOnS observations mean that Zeeman signatures are visible in numerous individual spectral lines, in order to maximize the S/N we employed the usual Least Squares Deconvolution (LSD; \citealt{d1997}) multiline analysis procedure. In particular we utilized the `improved LSD' (iLSD) package described by \cite{koch2010}. iLSD enables LSD profiles to be extracted with two complementary line masks, improving the reproduction of Stokes $I$ and $V$ obtained from masks limited to a select number of lines. 

We used a line mask obtained from the Vienna Atomic Line Database (VALD3; \citealt{piskunov1995, ryabchikova1997, kupka1999, kupka2000}) for a solar metallicity star with \teff~$= 27$~kK. While magnetic early-type stars are often chemically peculiar, $\xi^1$ CMa is of essentially solar composition \citep{2005AA...433..659N}, albeit with a mild N enhancement \citep{2008AA...481..453M}, thus a solar metallicity mask is appropriate. The mask was cleaned and tweaked as per the usual procedure, described in detail by \cite{2012PhDT.......265G}, such that only metallic lines unblended with H, He, interstellar, or telluric lines remained, with the strengths of the remaining lines adjusted to match as closely as possible the observed line depths. Of the initial 578 lines in the mask, 338 remained following the cleaning/tweaking procedure. The resulting LSD profiles are shown in Fig.\ \ref{esp_mus_lsd} (top panel). Note that, due to intrinsic line profile variability and the longer subexposure times used for the 2017 data, the line profiles of the most recent data are slightly broader than the  data acquired previously.

In order to directly compare MuSiCoS and ESPaDOnS results, a second line mask was employed with all lines outside of the MuSiCoS spectral range removed, leaving 139 lines. The bottom panel of Fig.\ \ref{esp_mus_lsd} shows a comparison between the highest S/N MuSiCoS LSD profile and the ESPaDOnS LSD profile with the closest pulsation phase (0.4377 vs.\ 0.4425). The Stokes $I$ profiles from the 2 instruments agree well, considering both the lower spectral resolution and the much longer exposure times of MuSiCoS (corresponding to $\sim$15\% of a pulsation period, as compared to $\sim$2.3\% of a pulsation period for ESPaDOnS). The key point of interest is in Stokes $V$, which is clearly negative in the MuSiCoS LSD profile, but positive in all ESPaDOnS LSD profiles. The second MuSiCoS observation also yields a negative Stokes $V$ profile. The reliability of these MuSiCoS observations is evaluated in detail  in Appendix \ref{append:musicos}. 

False Alarm Probabilities (FAPs) were calculated inside and outside of the line profile, and classified as Definite Detections (DDs), Marginal Detections (MDs), or Non-Detections (NDs) according the methodology and criteria described by \cite{1992AA...265..669D,d1997}. Detection flags for Stokes $V$ and diagnostic null $N$ are given in Table \ref{esp_obs_log}. All MuSiCoS observations are formal non-detections. All ESPaDOnS observations are DDs  (with the exception of the two discarded measurements from 2017/02/11, both of which yielded NDs due to their low S/N). However, in numerous ESPaDOnS observations, $N$ also yields a DD. This phenomenon is considered in greater detail in \S~\ref{subsec:pulsation_impact}. 

\subsection{Longitudinal Magnetic Field}\label{subsec:bz}

The longitudinal magnetic field \bz~was measured from the LSD profiles by taking the first-order moment of the Stokes $V$ profile normalized by the equivalent width of Stokes $I$ \citep{mat1989}. The same measurement using $N$ yields the null measurement \nz, which should be consistent with 0~G. In order to ensure a homogeneous analysis, the LSD profiles were first shifted by their measured RVs to zero velocity, and an identical integration range of $\pm 30$ \kms~around line centre employed. \bz~and \nz~measurements are reported in Table \ref{esp_obs_log}. As expected from the LSD profiles, \bz$< 0$ for both MuSiCoS measurements, while \bz$> 0$ for all ESPaDOnS measurements. The ESPaDOnS data are of much higher quality, with a median $\sigma_B = 6$~G, as compared to 57 G for the MuSiCoS data. In contrast to the FAPs, in which many ESPaDOnS $N$ profiles yield definite detections, \nz$/\sigma_N$ is typically very low, with a maximum of 2.4 and a median of 0.6, i.e.\ statistically identical to zero. 

\begin{figure}
\centering
\includegraphics[width=8cm]{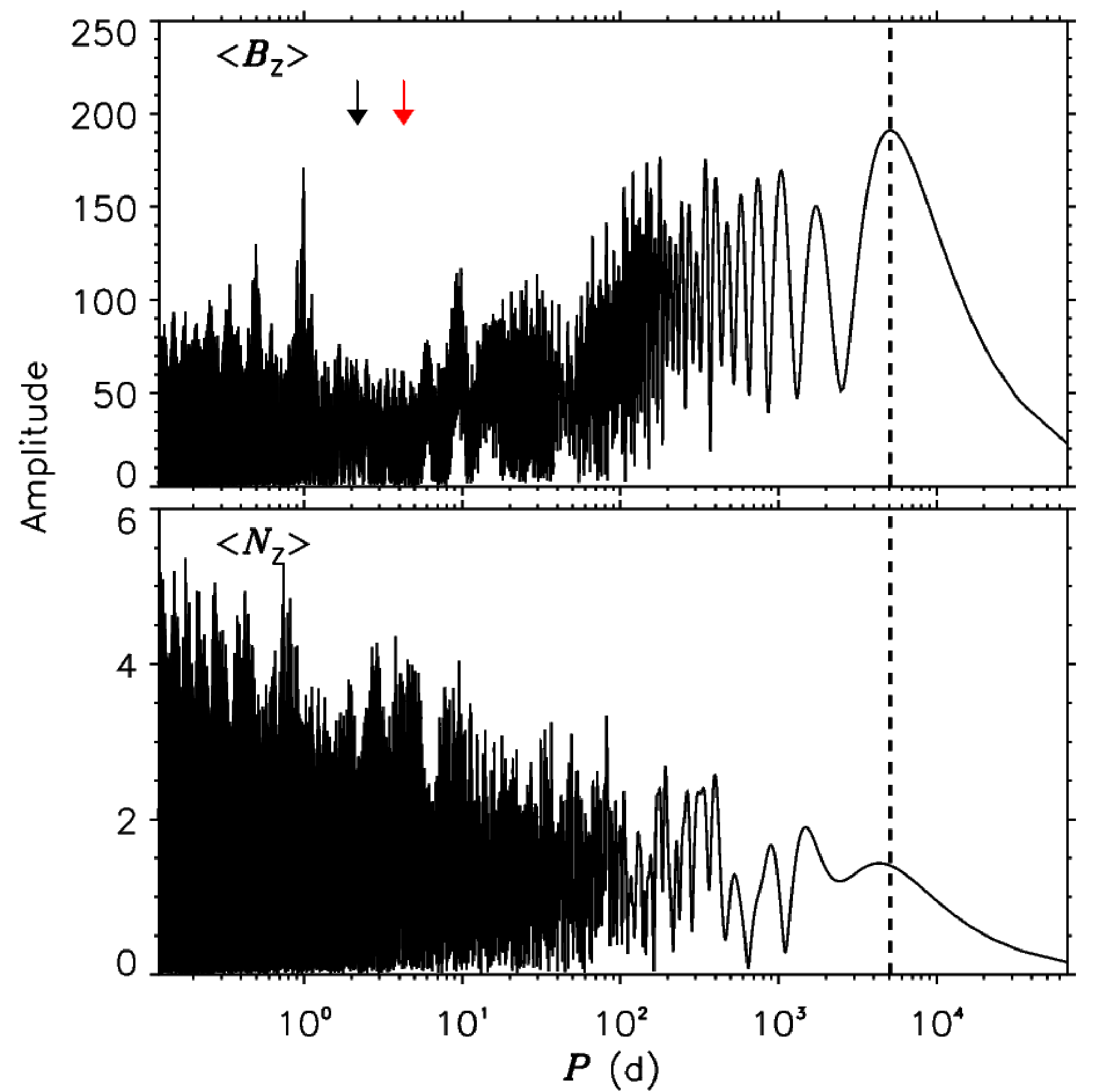}
\caption{Periodograms for ESPaDOnS \bz~({\em top}) and \nz~({\em bottom}). Many of the peaks in the \bz~periodogram also appear in the \nz~periodogram, with the exception of the maximum amplitude peak at $\sim$5100~d (vertical dashed line). The period determined from FORS1/2 data by \protect\cite{hub2011a} is indicated by the black arrow; the period determined by \protect\cite{fr2011} from an earlier, smaller ESPaDOnS dataset is indicated by the red arrow. There is very little amplitude at either period.}
\label{bz_periods}
\end{figure}

\begin{figure}
\centering
\includegraphics[width=\hsize]{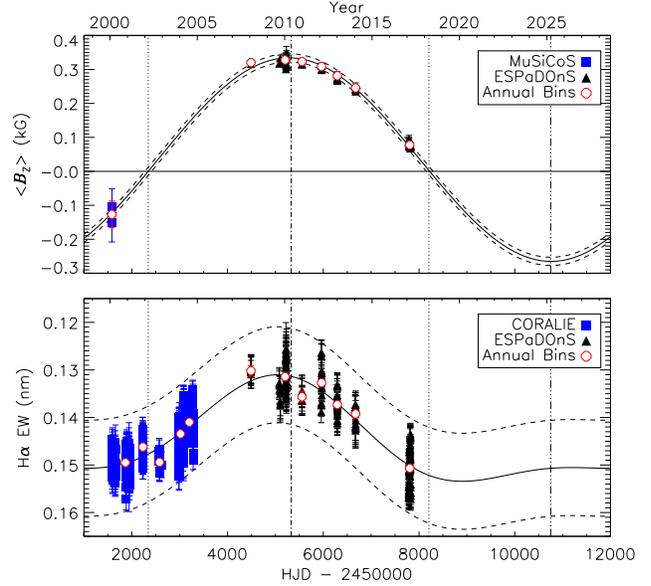}
\caption{{\em Top}: \bz~vs. time. The solid curve shows the least-squares sinusoidal fit to the annual mean measurements using a 30-year period, and the dashed curves indicate the $1\sigma$ uncertainty in the amplitude and mean value of the fit. The horizontal solid line shows \bz~$=0$. Vertical lines indicate the times at which \bz~$=0$ (dotted), \bz$=$\bz$_{\rm max}$ (dot-dashed), and \bz$=$\bz$_{\rm min}$ (dot-dot-dashed). {\em Bottom}: H$\alpha$ EWs. Curves show the least-squares $2^{nd}$-order sinusoidal fit to the annual means, which is performed independently of the fit to \bz. Note that maximum emission corresponds to \bz$_{\rm max}$, and minimum emission corresponds to \bz~$=0$. Also note that the fit predicts a second, weaker emission peak corresponding to \bz$_{\rm min}$.}
\label{bz_time}
\end{figure}

Because \bz~is expected to be modulated by stellar rotation, it can be used to determine the rotation period. The periodograms for \bz~and \nz~are shown in Fig.\ \ref{bz_periods}, where the latter shows the variation arising from noise. There are numerous peaks at periods of a few tens to a few hundreds of days which appear in both the \nz~and the \bz~periodograms. Phasing \bz~with the periods corresponding to these peaks does not produce a coherent variation (e.g., phasing the data with the highest peak in this range, at 177~d, and fitting a first-order sinusoid, yields a reduced $\chi^2$ of 1162). The strongest peak in the \bz~periodogram is at $5100 \pm 300$~d. This peak does not appear in the \nz~periodogram. This is similar to the timespan of the ESPaDOnS dataset, and so the formal uncertainty is certainly under-estimated, with the 5100~d period representing a lower limit. However, the S/N of this peak is 26, above the threshold for statistical significance, indicating that the long-term variation is probably real.

Fig.\ \ref{bz_time} shows the \bz~measurements, both individual (filled symbols) and in annual bins (open circles), as a function of time. There is an obvious long-term modulation, with \bz~steadily declining from a peak of $\sim$330 G in 2010 (HJD 2455200) to value of $\sim$80 G in 2017 (HJD 2457800). The annual mean \bz~measurements are provided in Table \ref{annual_ew_bz_tab}. 

\begin{table}
\centering
\caption[]{Annual weighted mean \bz~measurements and H$\alpha$ EWs. Months/years and HJDs correspond to the mean of the cluster of observations within each bin.}
\begin{tabular}{rrrr}
\hline
\hline
Month/Year & HJD -   & \bz & H$\alpha$ EW       \\
           & 2450000 & (G) & (nm)               \\
\hline
02/2000    & 2451585 & -127$\pm$40 & -- \\
10/2000    & 2451822 & -- & 0.1492$\pm$0.0002 \\
11/2001    & 2452234 & -- & 0.1462$\pm$0.0003 \\
11/2002    & 2452580 & -- & 0.1494$\pm$0.0007\\
12/2003    & 2452989 & -- & 0.1440$\pm$0.0003 \\
05/2004    & 2453156 & -- & 0.1413$\pm$0.0002 \\
01/2008    & 2454489 & 314$\pm$6 & 0.130$\pm$0.001   \\
01/2010    & 2455203 & 328$\pm$2 & 0.1314$\pm$0.0004 \\
12/2010    & 2455556 & 318$\pm$6 & 0.136$\pm$0.001   \\
02/2012    & 2455965 & 305$\pm$3 & 0.1327$\pm$0.0007 \\
01/2013    & 2456294 & 281$\pm$3 & 0.1373$\pm$0.0006 \\
01/2014    & 2456674 & 253$\pm$5 & 0.1392$\pm$0.0008 \\
02/2017    & 2457797 &  78$\pm$2 & 0.1507$\pm$0.0003 \\
\hline
\hline
\end{tabular}
\label{annual_ew_bz_tab}
\end{table}

The MuSiCoS observations, both of which are of negative polarity, indicate (in combination with the ESPaDOnS data) that the rotational period must be longer than 5100~d. The time difference between the MuSiCoS measurements and the maximum \bz~ESPaDOnS observations is $\sim$3600~d. If these two epochs sample the \bz~curve at its positive and negative extrema, then the rotational period must be at least 7200~d ($\sim$20~years), or a half-integer multiple if there is more than one cycle between the observed extrema. The curvature of the ESPaDOnS \bz~suggests that \bz$_{\rm max}$ occurred at $\sim$HJD~2455200, however as it cannot be ruled out that \bz$_{\rm min} < -127 \pm 40$~G, it is possible that $P_{\rm rot} > 7200$~d. Indeed, a longer period seems likely: phasing \bz~with a 20-year period, and fitting a least-squares $1^{st}$-order sinusoid (as expected for a dipolar magnetic field), produces a very poor fit as compared to longer periods. In Fig.\ \ref{bz_time}, the illustrative sinusoudal fit was performed using a 30-year period, which achieves a reasonable fit to the data (the reduced $\chi^2$ is 2.6). Note that with this fit the MuSiCoS data do not define \bz$_{\rm min}$. Longer periods can also be accommodated, however at the expense of $|\langle B_z\rangle_{\rm min}| > |\langle B_z\rangle_{\rm max}|$. We thus adopt $P_{\rm rot} = 30$~yr as the most conservative option allowed by the data. 

\subsection{Comparison to previous results}

\begin{figure}
\centering
\includegraphics[width=8cm]{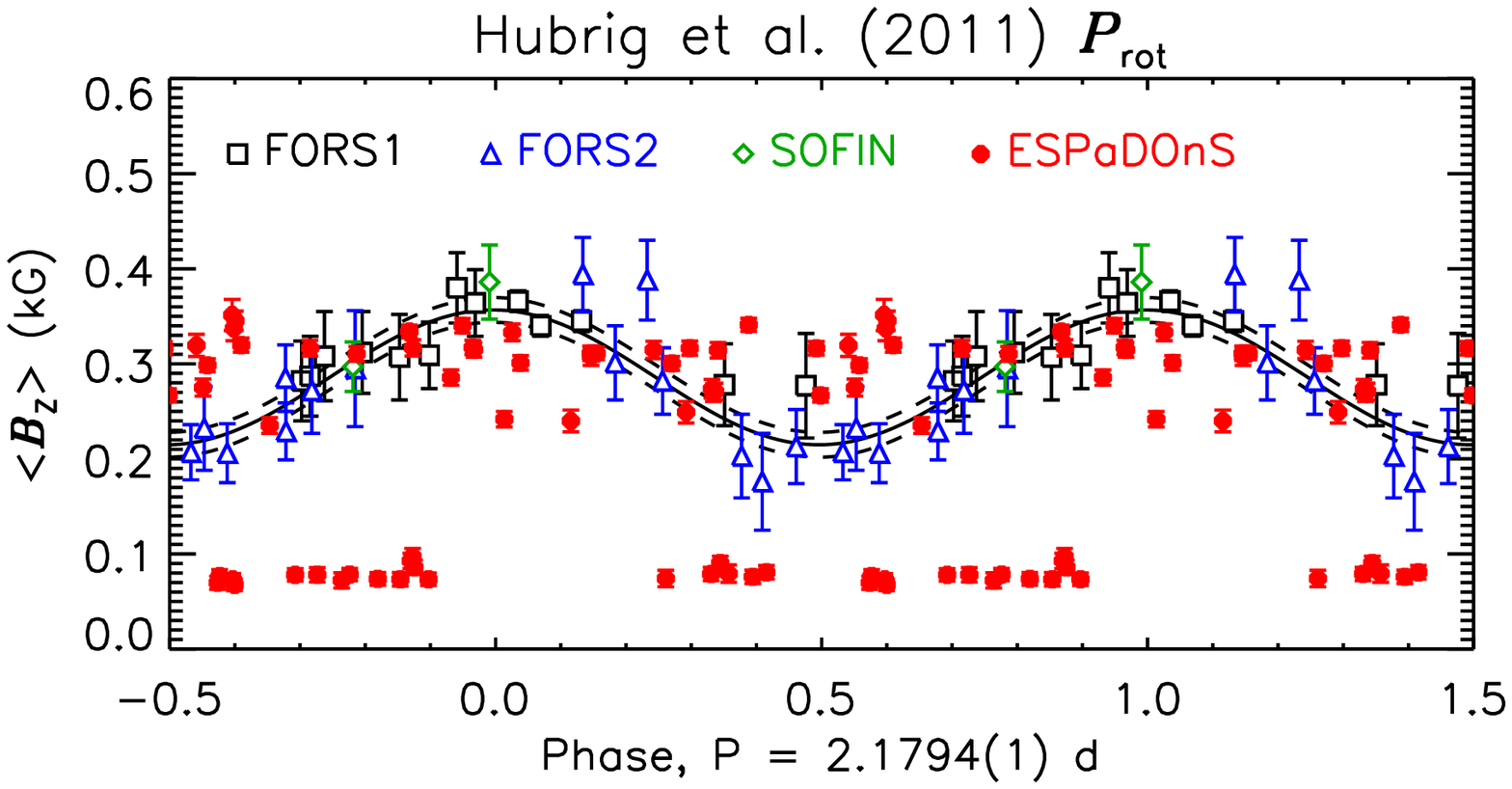}
\includegraphics[width=8cm]{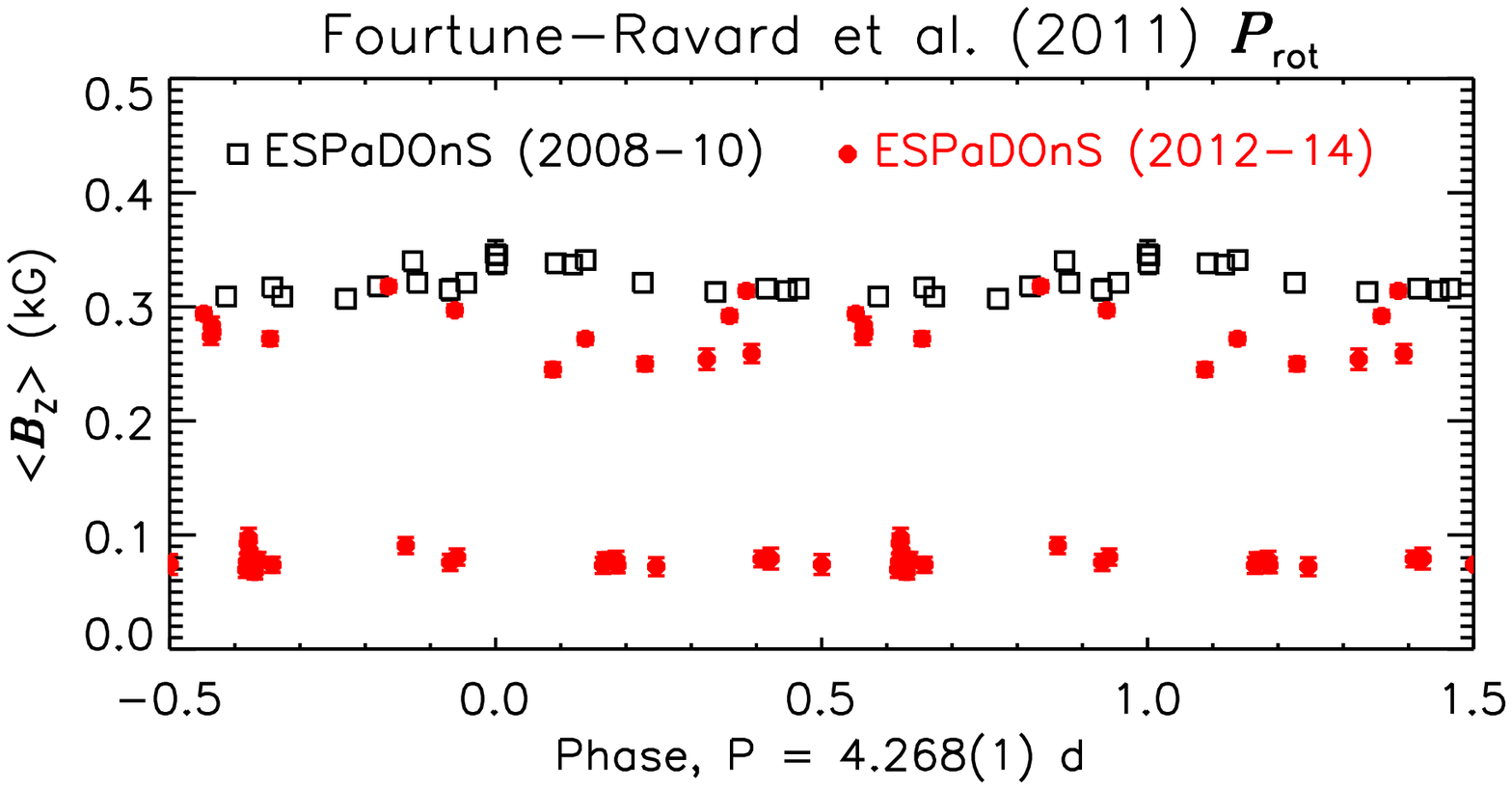}
\caption{{\em Top}: ESPaDOnS, FORS1, FORS2, and SOFIN \bz~measurements phased with the rotational period poposed by Hubrig et al.\ (2011). {\em Bottom}: ESPaDOnS \bz~measurements phased with the rotational period proposed by Fourtune-Ravard et al.\ (2011). Neither of the previously proposed periods provides an acceptable phasing of the data.}
\label{bz_wrongp}
\end{figure}

There are two competing claims for rotational periods in the literature. The first, based on spectropolarimetry collected with FORS1, FORS2, and SOFIN, is approximately 2.18~d \citep{hub2011a}. The second period, based on a preliminary analysis of an earlier, smaller ESPaDOnS dataset, is $\sim 4.27$~d \citep{fr2011}. Period analysis of the ESPaDOnS measurements using Lomb-Scargle statistics rules out both of these periods. The rotational periods provided by \cite{hub2011a} and \cite{fr2011} are respectively indicated with black and red arrows in the top panel of Fig.\ \ref{bz_periods}. There is no significant power in the periodogram at either period. The peak corresponding to the \cite{fr2011} period appears in both the \bz~and \nz~periodograms, suggesting it to be a consequence of noise. Phasing the data with the periods given by \cite{hub2011a} or \cite{fr2011} does not produce a coherent variation, as is demonstrated in Fig.\ \ref{bz_wrongp}. It impossible for a period on the order of days to account for the systematic decline of $\sim$200~G between the earliest data and the most recent data. We also note that the ESPaDOnS dataset presented here includes two epochs with superior time-sampling to that of the FORS1/2 datasets: 14 observations over 10 d in 2010, and 20 observations over 38 d in 2017, as compared to 13 FORS1 observations over 1075 d and 11 FORS2 observations over 60 d. The ESPaDOnS dataset thus enables a much better probe than the FORS1/2 dataset of short-term as well as long-term variability.

All FORS1/2 and SOFIN measurements are of positive polarity \citep{hubrig2006,hub2011a,2015A&A...582A..45F}, in agreement with ESPaDOnS data. Furthermore, the magnitude, $\sim$300 G, is similar. The two SOFIN measurements also agree well with the ESPaDOnS data. The much shorter 2.18~d period determined by \cite{hub2011a} is due to an apparent \bz~variation with a semi-amplitude that is significant at 1.8$\sigma$ in comparison with the median uncertainties in these measurements. \cite{bagn2012} showed that systematic sources of uncertainty such as instrumental flextures must be taken into account in the evaluation of \bz~uncertainties from FORS1 data, that the uncertainties in FORS1 \bz~measurements should thus be about 50\% higher, and therefore that FORS1/2 detections are only reliable at a significance of $>$5$\sigma$, a much higher threshold than is satisfied by the variation the 2.18~d period is based upon. The catalogue of FORS1 \bz~measurements published by \cite{2015AA...583A.115B} additionally reveals differences of up to 150 G for observations of $\xi^1$ CMa between results from different pipelines. Using the published uncertainties, the FORS1/2 measurements presented by \cite{hub2011a} are generally within 2$\sigma$ of the sinusoidal fit to the ESPaDOnS and MuSiCoS data, and only 3 differ by greater than 5$\sigma$. Curiously, the FORS2 data published by \cite{hub2011a} are systematically 80 G lower than the FORS1 measurements. We conclude that the previously published low-resolution magnetic data are not in contradiction with the long-term modulation inferred from high-resolution \bz~measurements.

\section{Emission Lines}\label{emission}


Having rejected the possibility that $\xi^1$ CMa's H$\alpha$ emission originates in the decretion disk of a classical Be companion star (\S~\ref{sec:interferometry}), we proceed under the assumption that it is formed within the stellar magnetosphere. In this section we explore the star's UV and H$\alpha$ emission properties. We evaluate short-term variability with the stellar pulsations, as well as long-term variability consistent with the slow rotation inferred from the magnetic data. 

\subsection{H$\alpha$ emission}\label{subsec:halpha}

\subsubsection{Long-term modulation}\label{subsubsec:halpha_long}

\begin{figure}
\centering
\includegraphics[width=8cm]{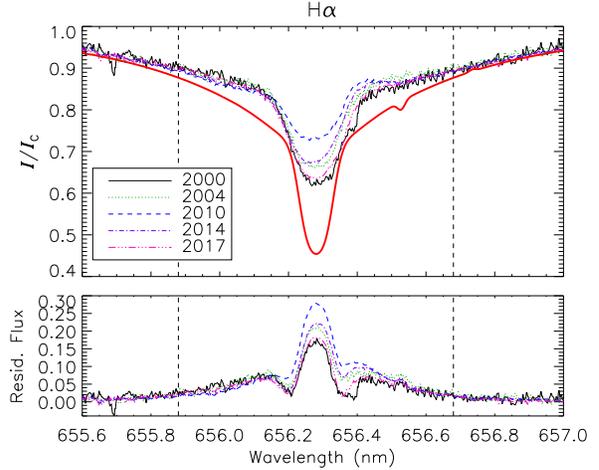}
\caption{H$\alpha$ continuum-normalized flux ({\em top}), and residual flux ({\em bottom}) after subtraction of a photospheric model, at similar pulsation phases ($\phi\sim 0.75$) in successive epochs. The thick red line shows a model line profile disk-integrated with the appropriate pulsation velocity. Vertical dashed lines indicate the integration range for EW measurement.}
\label{halpha_hbeta}
\end{figure}

Fig.\ \ref{halpha_hbeta} shows H$\alpha$ in 2000, 2004, 2010, 2014, and 2017, selected so as to share the same pulsation phase ($\phi = 0.75 \pm 0.03$). A synthetic line profile calculated using the physical parameters obtained in \S~\ref{sec:stellar_pars} is overplotted as a thick line. Comparison of observed to synthetic line profiles demonstrates that emission is present at all epochs, although it is substantially weaker in the earlier CORALIE data. The agreement of H$\beta$ with the model is reasonable at all epochs, although there is also weak emission present in the line core (Fig.\ \ref{logg}). Close analysis of the line core of H$\beta$ shows evidence of evolution, with the same trend as in H$\alpha$, although this evolution is of a much lower amplitude. 

\begin{figure}
\centering
\includegraphics[width=8cm]{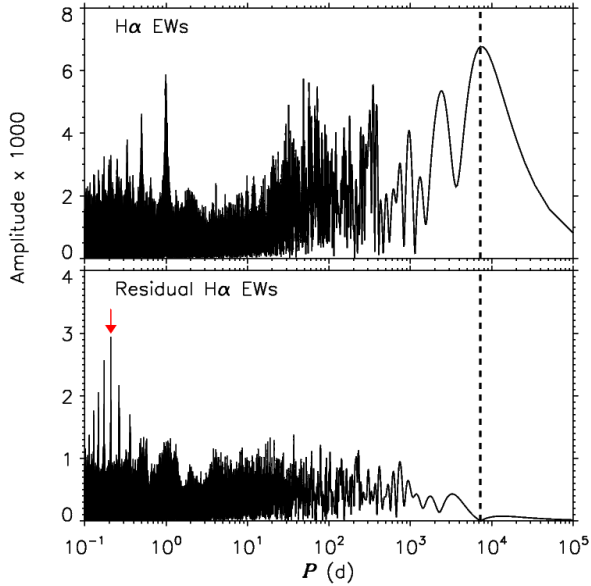}
\caption{Periodogram for H$\alpha$ EWs ({\em top}), and after pre-whitening with the highest-amplitude period ({\em bottom}). The highest-amplitude period is indicated by the dashed line. The red arrow indicates the pulsation period determined from the RVs.}
\label{halpha_prot}
\end{figure}

H$\alpha$ EWs were measured with an integration range of $\pm$0.4 nm of the rest wavelength in order to include only the region with emission (Fig.\ \ref{halpha_hbeta}). The spectra were first shifted to a rest velocity of 0 \kms~by subtracting the RVs measured in \S~\ref{sec:pulsation}. The periodogram for these measurements is shown in the middle panel of Fig.\ \ref{halpha_prot}. It shows a peak at 6700~d or 18~yr. The S/N of this peak is 29, indicating that it is statistically significant. This is consistent with the very long timescale of variation inferred from the ESPaDOnS \bz~periodogram (Fig.\ \ref{bz_periods}), and with the minimum 7200~d period inferred from the positive and negative \bz~extrema in the ESPaDOnS and MuSiCoS data. 

The bottom panel of Fig.\ \ref{bz_time} shows the H$\alpha$ EWs as a function of time. Maximum emission and maximum \bz~occur at approximately the same phase, as do the \bz~minimum and the minimum emission strength. The variation in EW is seen more clearly in the annual mean EWs (open circles), which are tabulated in Table \ref{annual_ew_bz_tab}. The relative phasing of \bz~and EW is consistent with a co-rotating magnetosphere, and supports the accuracy of the adopted period. Since \bz~is both positive and negative, two local emission maxima are expected at the positive and negative extrema of the \bz~curve, as at these rotational phases the magnetospheric plasma is seen closest to face on. The solid curve in the bottom panel of Fig.\ \ref{bz_time} shows a $2^{nd}$-order sinusoidal fit to the annual mean EWs, which indeed yields two local maxima, the strongest corresponding to \bz$=$\bz$_{\rm max}$, and the second maximum predicted to occur at \bz$=$\bz$_{\rm min}$. Note that, due to the incomplete phase coverage (less than half of a rotational cycle), in the event that the H$\alpha$ variation is indeed a double-wave a Lomb-Scargle periodogram should show maximum power at close to half of the rotational period. The periodogram peak at 18-yr would then indicate $P_{\rm rot} = 36$~yr, consistent with the rotational period inferred from \bz. 

\subsubsection{Short-term modulation}\label{subsubsec:halpha_short}

\begin{figure}
\centering
\includegraphics[width=8cm]{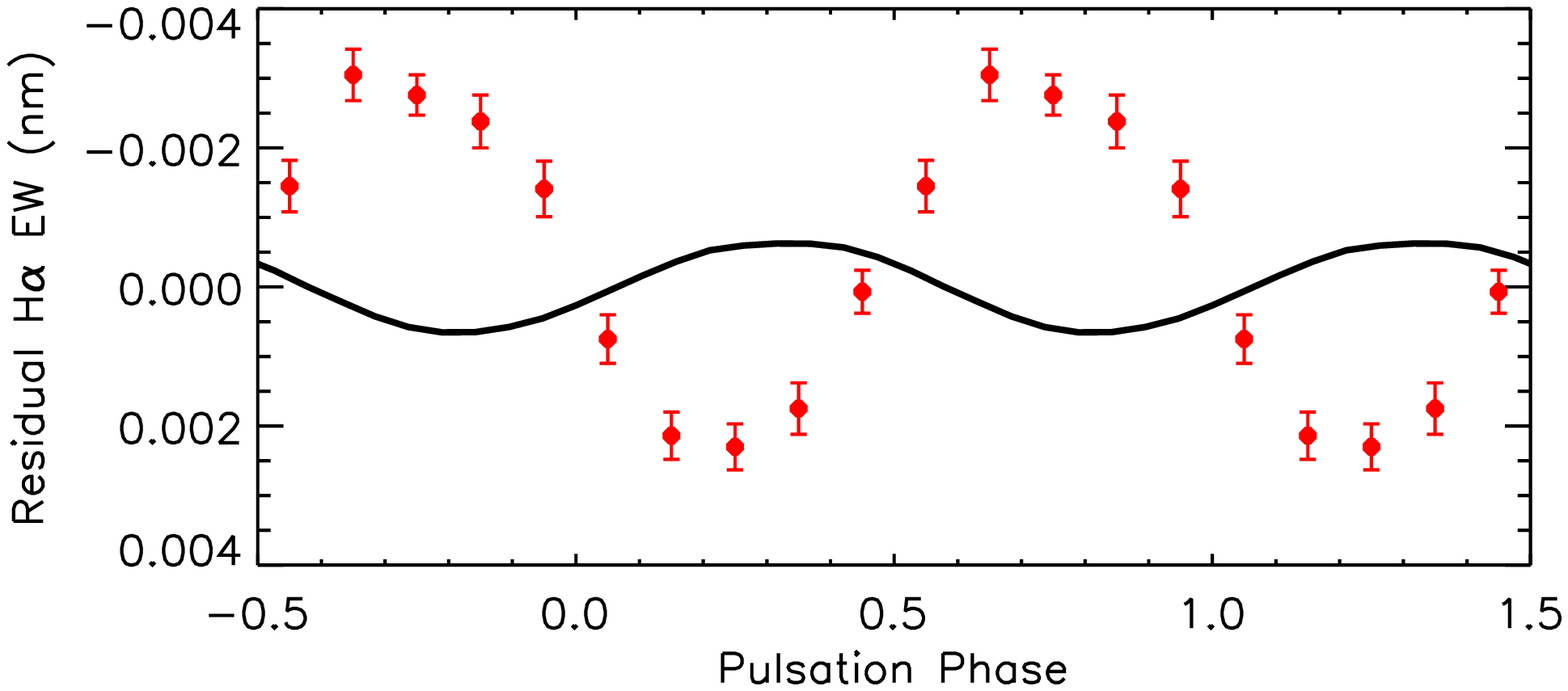}
\includegraphics[width=8cm]{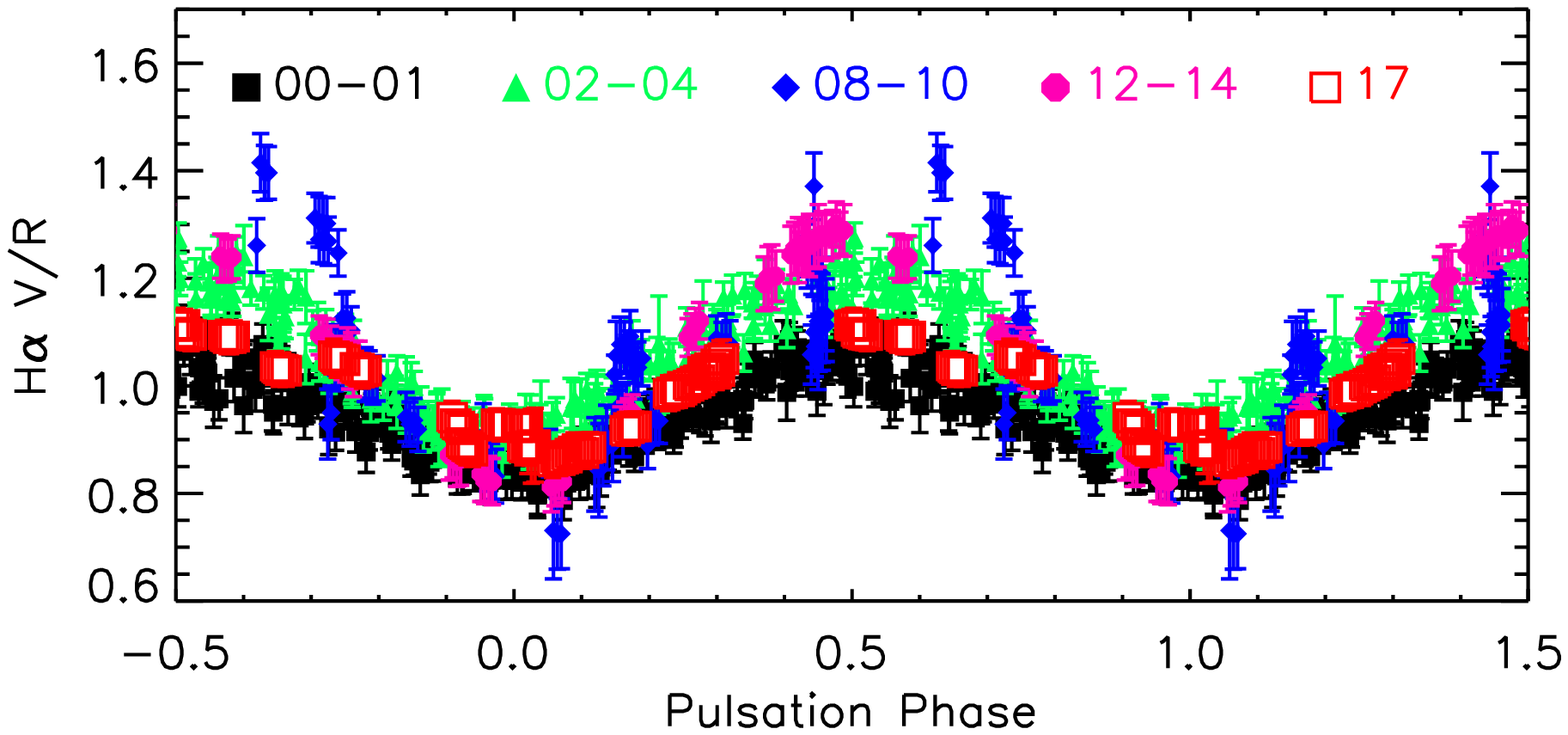}
\caption{{\em Top}: residual H$\alpha$ EWs, after removal of the sinusoidal variation shown in the bottom panel of Fig.\ \ref{bz_time}, phased with the pulsation period and binned with pulsation phase. The thick solid black line indicates the EW variation expected from a photospheric model calculated using the $\pm$300 K \teff~variation determined in \S~\ref{subsec:teffvar}, normalized by subtracting the mean EW of the model. The amplitude of the residual EW variation is 3 times larger than predicted by the photospheric model, and is about 0.5 cycles out of phase. {\em Bottom}: H$\alpha$ V/R variation. Symbols and colors indicate the year of observation, the last two digits of which are given in the legend. The amplitude of the variation is much higher than predicted by the photospheric model, which is essentially flat on this scale and therefore is not shown. The amplitude furthermore changes between epochs, which is not expected for photospheric pulsations given the high stability of pulsation properties in other lines.}
\label{halpha_redblue_ew}
\end{figure}

\begin{figure}
\centering
\includegraphics[width=8.cm]{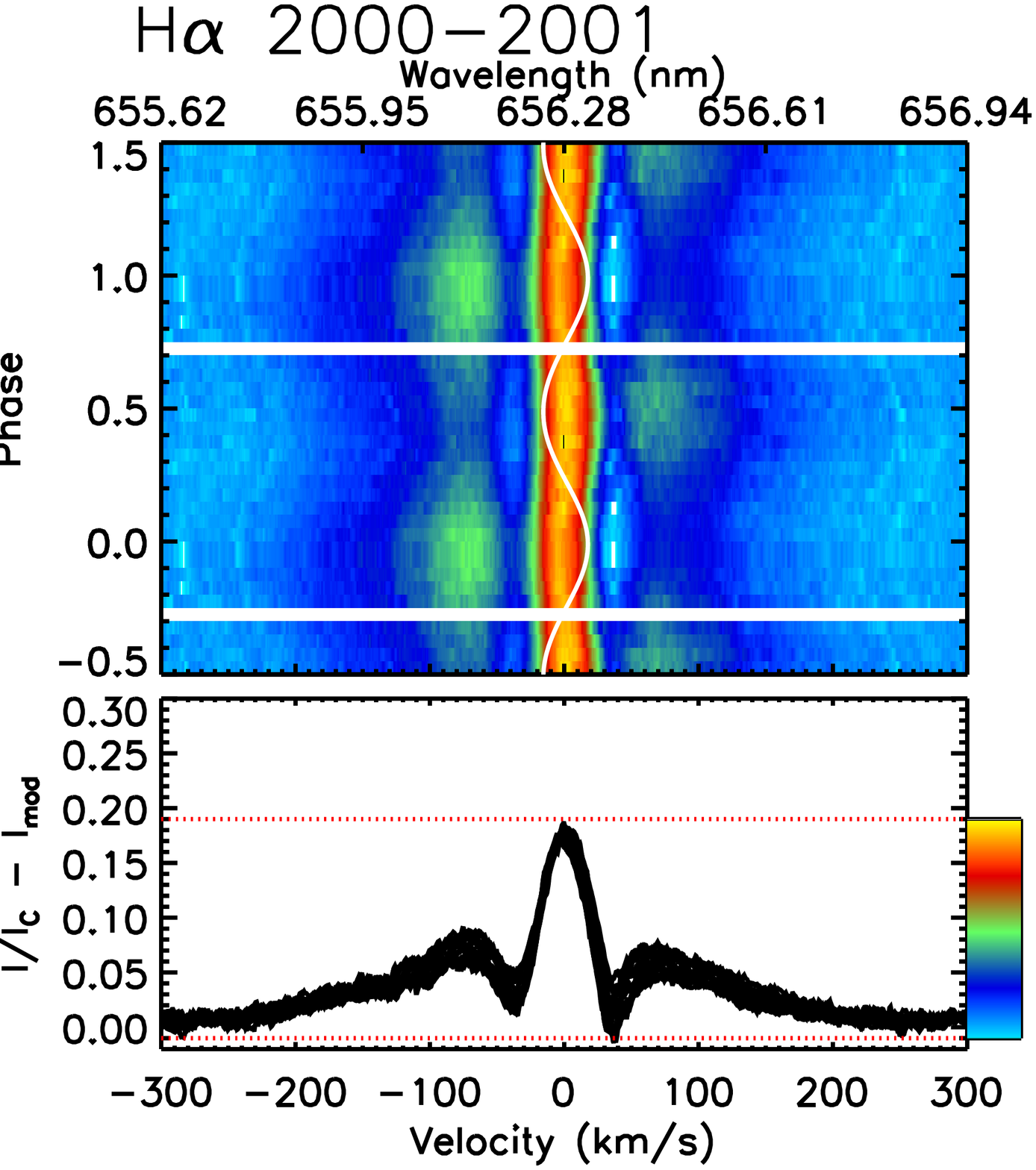} 
\includegraphics[width=8cm]{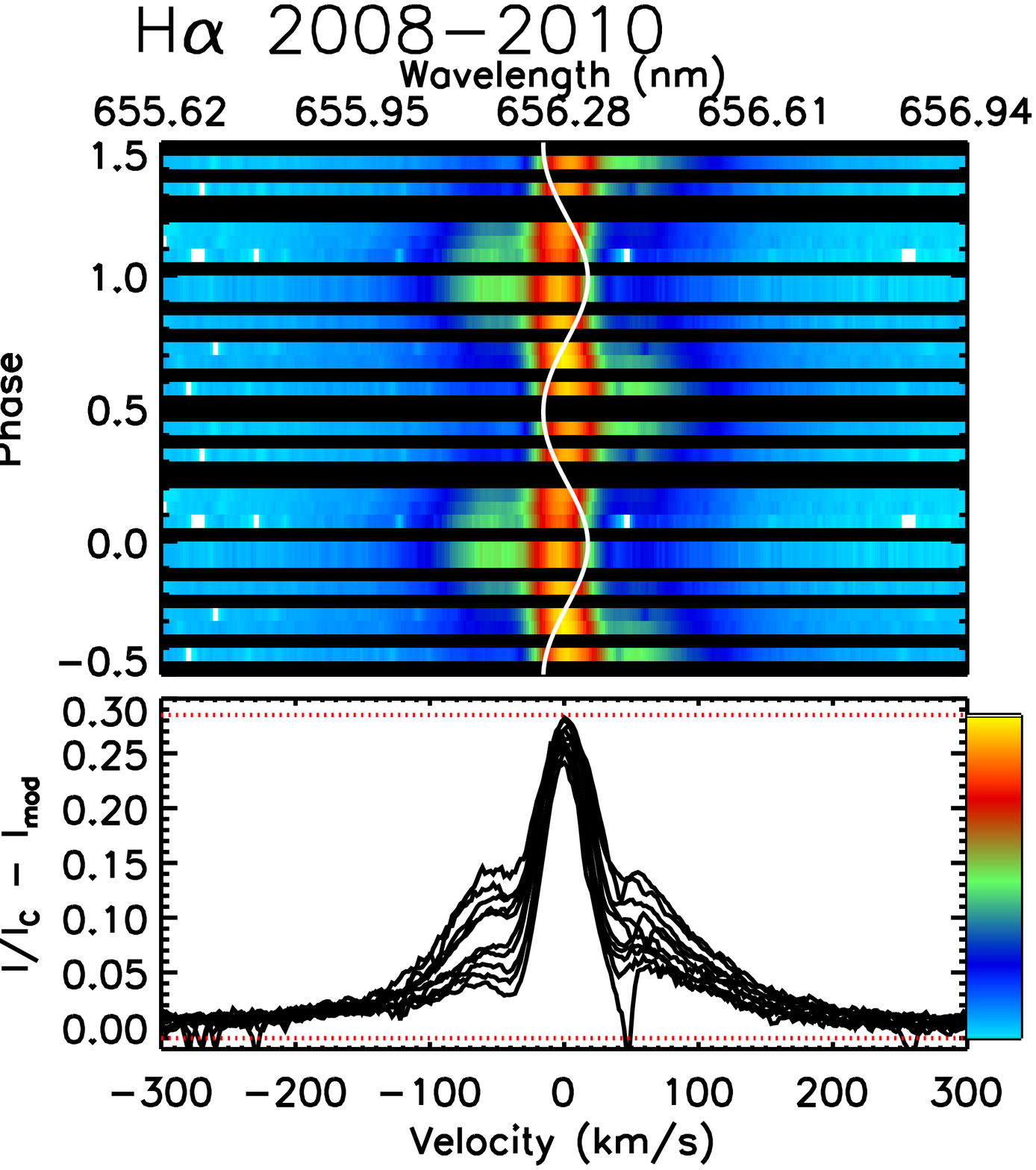} 
\caption{H$\alpha$ dynamic spectra in the epochs of minimum (2000-2001) and maximum (2008-2010) emission. The synthetic spectrum shown in Fig.\ \ref{halpha_hbeta} was used as a reference spectrum. Observed spectra were shifted to 0 \kms~by subtracting their RVs, and then co-added in 20 phase bins. The white curves indicate the RV curve due to pulsations. In all epochs, the emission profile has three components: a central peak near 0~\kms, which shifts slightly about line centre, but stays nearly constant in strength; a blue-shifted peak, which is at maximum strength at phase 0 and disappears by phase 0.5; and a red-shifted peak, which disappears at phase 0 and is at a maximum at phase 0.5. All three components increase in strength from 2000 to 2010. Note the greater amplitude of the variation in the secondary emission lobes in the 2008-2010 data.}
\label{halpha_model_dyn}
\end{figure}

As a first step to investigating whether H$\alpha$ is affected by pulsation, residual EWs were obtained by subtracting the least-squares 2$^{nd}$-order sinusoidal fit to the annual mean EWs. The bottom panel of Fig.\ \ref{halpha_prot} shows the period spectrum of the residual EWs. The strongest peak is at the stellar pulsation period, with a S/N of 12. After prewhitening the EWs with both the rotational and pulsational frequencies, the S/N of the highest peak in the periodogram is close to 4, suggesting that all significant variation is accounted for by these two frequencies.

While the semi-amplitude of the residual EW variation is similar to the median error bar, binning the residual EWs by pulsation phase does not change the semi-amplitude, but increases the significance of the variation to $\sim 8\sigma$ with respect to the mean error bar. The phase-binned residual EWs are shown phased with the pulsation period in the top panel of Fig.\ \ref{halpha_redblue_ew}. There is a coherent variation of the phase-binned residual EWs with the pulsation period, with a semi-amplitude of approximately 0.003 nm. 

One obvious candidate mechanism for producing this pulsational modulation is the change in the EW of the underlying photospheric profile due to the changing \teff, as explored in \S~\ref{subsec:teffvar}. To investigate this hypothesis we calculated synthetic spectra via linear interpolation between the grid of {\sc tlusty} BSTAR2006 models \citep{lanzhubeny2007}. We used the physical parameters from \S~\ref{sec:stellar_pars}, including the $\pm$300 K \teff~variation found in \S~\ref{subsec:teffvar}. Deformation of the line profile due to pulsation was accounted for using the RV curve from \S~\ref{sec:pulsation}, and modelling the (assumed radial) pulsations as described in \S~\ref{subsec:vsini}. Spectra were calculated at 20 pulsation phases, and the EW was measured at each phase using the same integration range as used for the observed data, and after moving the synthetic spectra to zero RV. The black line in the top panel of Fig.\ \ref{halpha_redblue_ew} shows the resulting variation, normalized by subtracting the mean EW across all models. The semi-amplitude of the EW variation expected due to changes in the photospheric profile due to pulsation is about 0.001 nm, much smaller than observed. Furthermore, it is out of phase with the observed variation by about 0.5 pulsation cycles. Note that the maximum and minimum EWs of the predicted variation correspond to the minimum and maximum of the \teff~variation (Fig.\ \ref{phot_mod}), as expected for H lines, which grow weaker with increasing \teff~in this \teff~range. 

Radial pulsation introduces asymmetry into the line profile, which can be quantified by $V/R$, defined as the ratio of the EW in the blue half to the EW in the red half of the line. We measured $V/R$ from EWs calculated from $-$0.4~nm to line centre (defined at the laboratory rest wavelength shifted by the RV measured from metallic lines in \S~\ref{sec:pulsation}), and line centre to $+0.4$~nm. The bottom panel of Fig.\ \ref{halpha_redblue_ew} shows the $V/R$ variation. The semi-amplitude of the variation is about 0.25. This is much higher than the semi-amplitude of 0.0007 predicted by synthetic spectra calculated using a radially pulsating photospheric model; since this is essentially flat on the scale of Fig.\ \ref{halpha_redblue_ew}, the photospheric variation is not shown. The low level of line asymmetry in the synthetic spectra is a consequence of the large Doppler broadening of the H$\alpha$ line, approximately 30~\kms~or about twice the semi-amplitude of the RV curve. For narrower lines, e.g. the C~{\sc ii} 656.3 nm line for which the Doppler velocity is similar to the RV semi-amplitude, the predicted and observed $V/R$ variations are in reasonable agreement. 

The semi-amplitude of the H$\alpha$ $V/R$ variation increases steadily from 0.16 in 2000-2001, to 0.20 in 2002-2004, to 0.35 in 2008-2010, and then declines to 0.24 in 2012-2014 and 0.13 in 2017. This is the same pattern as the change in total emission strength, thus, the amplitude of the V/R variation correlates to the total emission strength and, therefore, to \bz. 

For a more detailed view of the pulsational modulation of H$\alpha$ we calculated dynamic spectra phased with the pulsation period. These are shown in Fig.\ \ref{halpha_model_dyn}, using the synthetic line profile from Fig.\ \ref{halpha_hbeta} as a reference spectrum. Individual line profiles were moved to 0 \kms~by subtracting their RVs, and then binned by pulsation phase using phase bins of 0.05 cycles. The two panels of Fig.\ \ref{halpha_model_dyn} show dynamic spectra in the epochs of minimum emission (2000-2001) and maximum emission (2008-2010). The emission line morphology and pattern of variability is essentially identical in other epochs. In both cases, the emission peaks near the centre of the line. In the CORALIE data, the H$\alpha$ emission peak is $\sim$20\% of the continuum. This rises to about 28\% of the continuum in the ESPaDOnS data. The emission peak anticorrelates slightly with the RV, as shown by the overplotted white lines. The strong red and blue emission variability revealed by the $V/R$ variation in the bottom panel of Fig.\ \ref{halpha_redblue_ew} is due to secondary emission peaks which occur at phases 0.0 and 0.5, respectively blue- and red-shifted with respect to the line centre. As with the central emission peak, these are stronger in the ESPaDOnS data, reflecting the change in $V/R$ amplitude over time. Note that, in the data acquired at earlier epochs, these secondary emission bumps are apparently separated from the main emission peak, while in later epochs they are connected (although this depends on the choice of the reference spectrum).

Both the emission strength (as measured by the residual EWs after pre-whitening with the $2^{nd}$-order fit in Fig.\ \ref{bz_time}) and the line asymmetry (as quantified by $V/R$) vary coherently with the pulsation phase. Synthetic photospheric spectra calculated using a radially pulsating model are unable to reproduce the residual EW variation, which is both 3$\times$ larger than predicted, and 0.5 cycles out of phase. The amplitude of the observed $V/R$ variation is about 300$\times$ larger than predicted by the model, which does not reproduce the prominent blue- and red-shifted secondary emission bumps. The amplitude of $V/R$ is furthermore variable with time, increasing and decreasing in strength in the same fashion as the total emission strength. As the star's pulsation amplitude is extremely regular, a change in the amplitude of $V/R$ with epoch cannot be explained by photospheric pulsation. These discrepancies between model and observation suggest either that the origin of the pulsational modulation of H$\alpha$ is either not photospheric, or that it cannot be explained due to \teff~variation alone. Exotic processes, such as temperature inversions related to shockwaves produced by the star's supersonic pulsations, may be one explanation. Alternatively, the origin of the pulsational modulation may reside within the magnetosphere.

\subsection{UV emission lines}\label{subsec:uv}

\begin{figure}
\centering
\includegraphics[width=8.5cm]{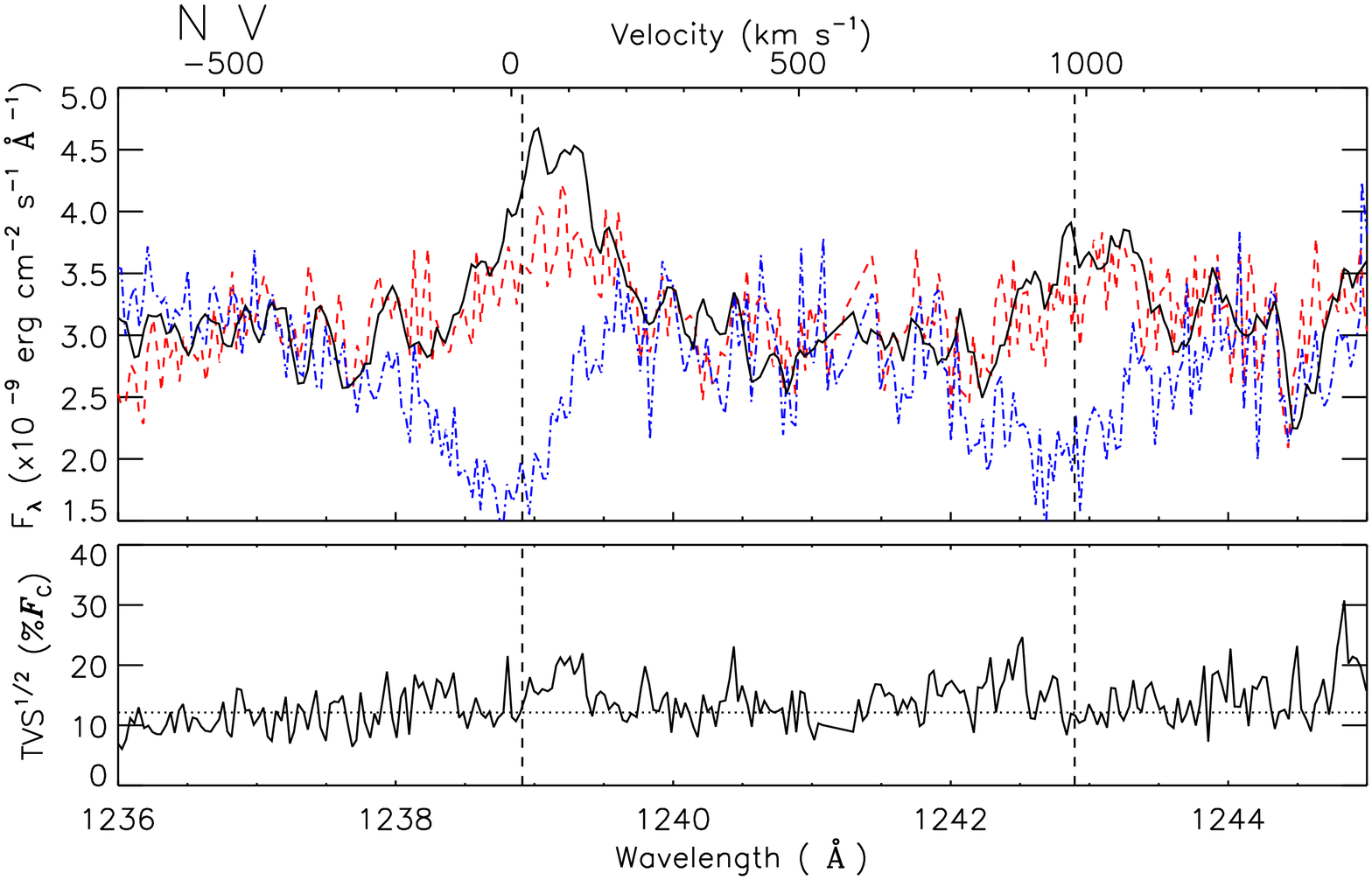} 
\includegraphics[width=8.5cm]{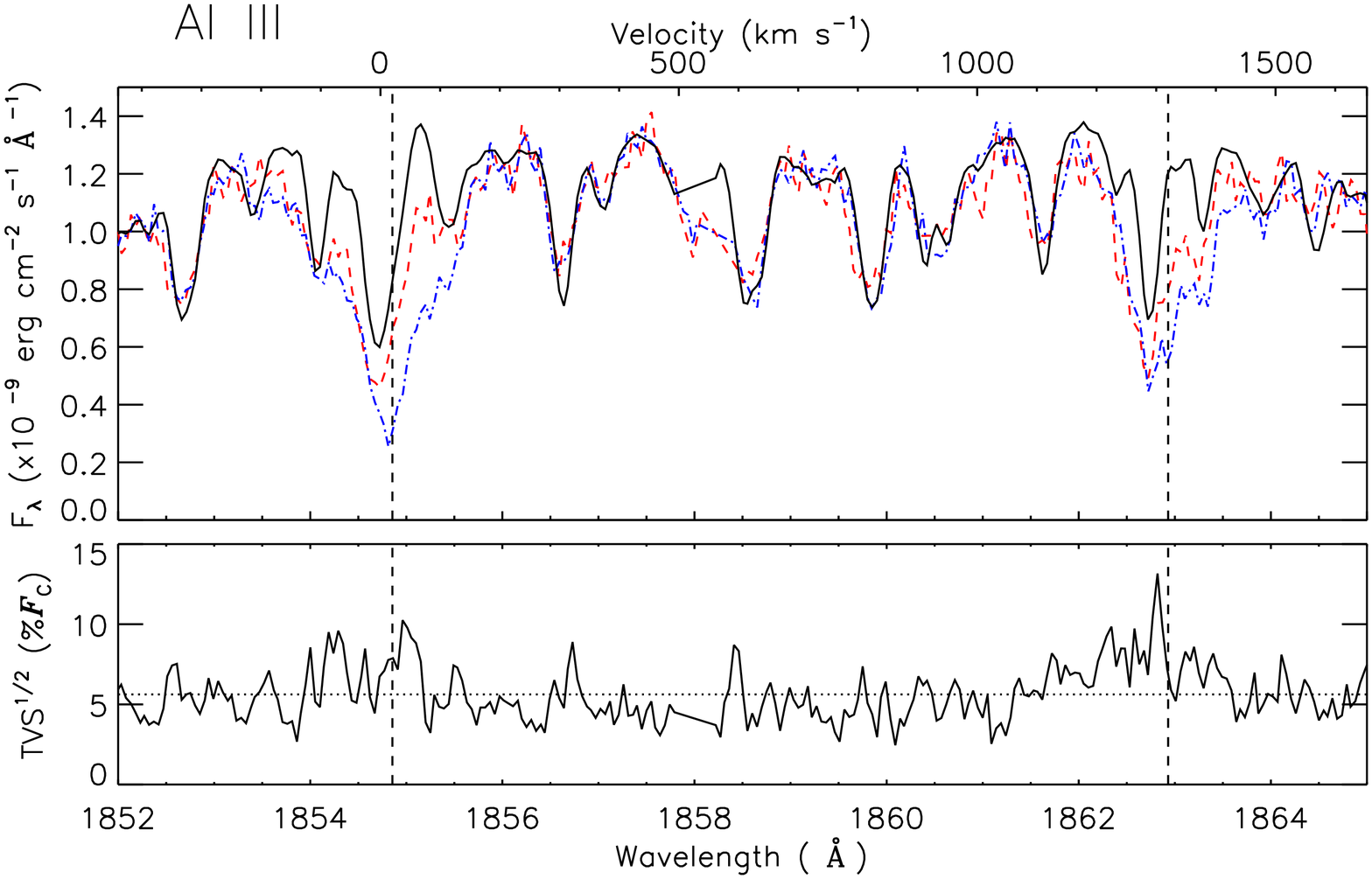} 
\caption{Comparison between the mean N~{\sc v} ({\em top})  and Al~{\sc iii} ({\em bottom}) emission lines of $\xi^1$ CMa (solid, black), and those of $\beta$ Cep at maximum emission (red, dashed) and maximum absorption (blue, dot-dashed). Vertical dashed lines indicate rest wavelengths. $\xi^1$ CMa's emission lines are similar to those of $\beta$ Cep at maximum emission, albeit somewhat stronger. The lower panels show Temporal Variance Spectra (TVS, see text), calculated from all spectra, which characterize the variability of the spectra normalized to the continuum. The horizontal line indicates the continuum variation, in comparison to which there is no statistically significant variability inside the line profile.}
\label{uv_compare}
\end{figure}

Emission is present in four of the UV doublets often used to diagnose the wind properties of early-type stars: N {\sc v} 1239, 1243 \AA; Si {\sc iv} 1394, 1403 \AA; C {\sc iv} 1548, 1551 \AA; and Al {\sc iii} 1854, 1863 \AA. The emission profiles of these doublets are similar to those of the magnetic $\beta$ Cep pulsator $\beta$ Cep at maximum emission. \cite{schnerr2008} performed this comparison for the C~{\sc iv} doublet. Fig.\ \ref{uv_compare} compares the mean line profiles for $\xi^1$ CMa's N~{\sc v} and Al~{\sc iii} lines to those of $\beta$ Cep, and demonstrates that these lines are also similar to those of $\beta$ Cep at maximum emission. Such emission is unique to magnetic stars, and as an indirect diagnostic of stellar magnetism has historically prompted the search for magnetic fields in such stars (e.g., \citealt{1993npsp.conf..186H,henrichs2013}; indeed, the detection of $\xi^1$ CMa's UV emission motivated the collection of the MuSiCoS data). 

The presence of emission in the N {\sc v} line is particularly interesting. This line is not seen in normal early B-type stars \citep{1995yCat.3188....0W}, nor is it present in most normal late O-type stars (e.g.\ \citealt{1998AJ....116..941B}). In their study of IUE data for 4 magnetic B-type stars, including $\beta$ Cep, \cite{smithgroote2001} concluded that this doublet must be formed at $\sim$30 kK, somewhat higher than the photospheric \teff. The presence of N~{\sc v} lines in the UV spectra of relatively cool stars is thought to be a consequence of Auger ionization due to the presence of X-rays \citep{1979ApJ...229..304C}. Since $\xi^1$ CMa has a strong magnetic field, it is expected to be overluminous in X-rays due to magnetically confined wind shocks, and is indeed observed to be overluminous in X-rays to the degree predicted by models \citep{oskinova2011,naze2014}.

$\beta$ Cep's wind lines show clear variability synchronized with its rotation period \citep{henrichs2013}. In contrast, $\xi^1$ CMa's wind lines show only a very low level of variability, more similar to that seen in a normal (magnetically unconfined) stellar wind \citep{schnerr2008}. The bottom panels of Fig.\ \ref{uv_compare} show Temporal Variance Spectra (TVS), which compare the variance within spectral lines to the variance in the continuum \citep{fullerton1996}. To minimize variation due to pulsation, spectra were moved to 0 \kms~by subtracting the RV computed on the basis of the RV curve and ephemeris determined in Section \ref{sec:pulsation}. With the uncertainty in $P_0$ and in $\dot{P}$ (\S~\ref{sec:pulsation}), the uncertainty in pulsation phase is about 0.028, corresponding to a maximum uncertainty in RV of 3~\kms, less than the $\sim$7.8~\kms~velocity pixel of the IUE data. RV correction reduces the TVS pseudocontinuum level by a factor of about 2 to 3.

Comparison of EW measurements of these lines to EWs of synthetic spectra, using the same radially pulsating model described above in \S~\ref{subsec:halpha}, yielded ambiguous results, due to the small number of high-dispersion IUE observations (13 spectra), and formal uncertainties similar to the maximum level of variability, $\sim$0.02 nm. This low level of variability is consistent with the weak variability of the residual H$\alpha$ EWs. 

The lack of variability in $\xi^1$ CMa's wind lines is consistent with a rotational pole aligned with the line of sight, an aligned dipole, a long rotation period, or some combination of these. The similarity to $\beta$ Cep's emission lines at maximum emission suggests that the magnetic pole was close to being aligned with the line of sight when the UV data were acquired. The long-term modulations of both H$\alpha$ and \bz~favour an oblique dipole with a long rotational period, in which case the UV data should have been acquired at a rotational phase corresponding to one of the extrema of the \bz~curve. Phasing the UV data with the same 30~yr rotational period as in Fig.\ \ref{bz_time} yields a phase close to 1.0, i.e. they would indeed have been obtained close to magnetic maximum. Assuming that the data must have been acquired near an extremum of the \bz~curve (i.e.\ at a phase close to 0.5 or 1.0) would require a period of 10, 15, 20, 30, or 60 years, of which only 30 and 60 years are not excluded by the magnetic data. This assumes that the UV data were, in fact, acquired close to an extremum. Given that the UV emission is only slightly stronger than that of $\beta$ Cep, which has a weaker magnetic field and no H$\alpha$ emission, it may be the case that the true maximum UV emission strength is significantly in excess of observations, in which case the UV data cannot be used to infer $P_{\rm rot}$.

\section{Magnetic and magnetospheric parameters}\label{sect:magpars}

In this section, the $\sim 30$~yr rotation period inferred from magnetic and spectroscopic data is used with the star's physical parameters and \bz~measurements to establish constaints on the properties of $\xi^1$ CMa's surface magnetic field, circumstellar magnetosphere, and spindown timescales, using the self-consistent Monte Carlo method described by \cite{my_phd_thesis}. These results are summarized in Table \ref{orm_tab}. 

\begin{table}
\caption{Magnetic, magnetospheric, and rotational properties. The final column gives the section in which the parameter is derived.}
\begin{center}
\begin{tabular}{lrl}
\hline
\hline
Parameter & Value & Ref. \\
\hline
$P_{\rm rot}$ (d) & $>$30 yr & 6.1 \\
$r$ & -0.78$\pm$0.02 & 8.1 \\
$\beta$ ($^\circ$) & 83$^{+5}_{-12}$ & 8.1 \\
$\epsilon$ & 0.36$\pm$0.02 & 4.2 \\
$B_{\rm d}$ (kG) & $>1.1$ & 8.1 \\
\hline
$W$ & $<7\times 10^{-5}$ & 8.2 \\
\rk~($R_*$) & $>580$ & 8.2 \\
$\log{\dot{M}}~(M_\odot~{\rm yr}^{-1})$ & -8.0$\pm$0.1 & 8.2, \cite{vink2001} \\
\vinf~(\kms) & 2070$\pm$60 & 8.2, \cite{vink2001} \\
$\eta_*$  & $>340$ & 8.2 \\
\ra~($R_*$)  & $>4.6$ & 8.2 \\
\hline
$\tau_{\rm J}$  & 4$\pm$1 & 8.3 \\
$t_{\rm S,max}$  & 42$\pm$12 & 8.3 \\
\hline
\hline
\end{tabular}
\end{center}
\label{orm_tab}
\end{table}

\begin{figure}
\centering
\includegraphics[width=9cm]{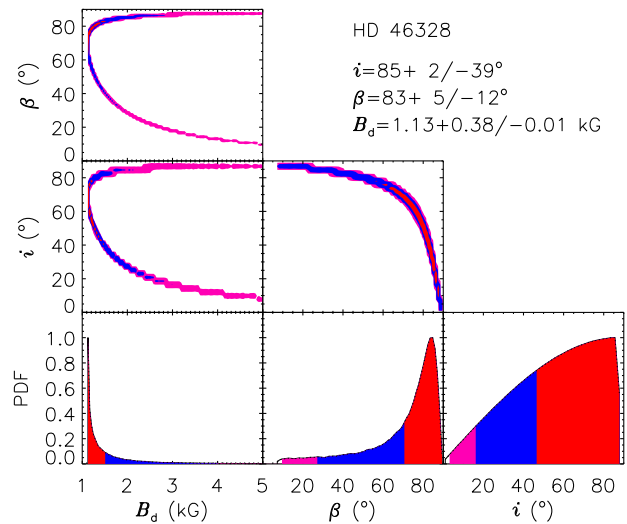}
\caption{Oblique Rotator Model parameters determined via a Monte Carlo approach. Red, blue, and pink show the 1, 2, and 3$\sigma$ contours of the grid. The bottom panels show probability density functions (PDFs) for $B_{\rm d}$, $\beta$, and $i$, where $i$ was drawn from a random distribution over a solid angle of 4$\pi$.}
\label{orm}
\end{figure}

The \bz~variation of a rotating star with a dipolar magnetic field can be reproduced with a three-parameter model: the inclination angle $i$ between the rotational axis and the line of sight, the obliquity angle $\beta$ between the magnetic and rotational axes, and the polar strength of the magnetic dipole at the stellar surface $B_{\rm d}$. The two angular parameters are related via $\tan{\beta} = (1-r)/(1+r)\cot{i}$, where $r$ is the ratio defined by \cite{preston1967} as $r=(|B_0|-B_1)/(|B_0|+B_1)$, with $B_0$ and $B_1$ the mean and semi-amplitude of the sinusoidal fit to \bz~when phased with $P_{\rm rot}$. We determined $r=-0.78\pm 0.02$ from the sinusoidal fit to the annual mean \bz~measurements shown in the top panel of Fig.\ \ref{bz_time}. In consequence these results rely on the 30-year period. 

The degeneracy between $i$ and $\beta$ is usually broken by determining $i$ independently, e.g.\ from \vsini, $P_{\rm rot}$, and $R_*$. Using this method \cite{hub2011a} found $i \sim 3^\circ$, however, this was based on their rotation period of 2.18~d, which was shown in \S~\ref{subsec:bz} to be incorrect. Indeed, the equatorial rotational velocity implied by a 30-year period, $v_{\rm eq}<0.04$ \kms, is much less than the upper limit on \vsini~$<8$ \kms~found in \S~\ref{subsec:vsini} (and indeed, much less than the 1.8 \kms~velocity resolution of ESPaDOnS data). Since $i$ cannot be constrained, we applied a probabilistic prior, requiring that $i$ be drawn from a random distribution over 4$\pi$ steradians such that $P(i)=1-\cos{i}$ (e.g., \citealt{landstreetmathys2000}). The corresponding probability density function (PDF) is shown in the bottom right panel of Fig.\ \ref{orm}. The bottom middle panel shows the resulting PDF for $\beta$, which peaks at $83^\circ$. Despite the inverse relationship of $i$ and $\beta$ obtained for $r=-0.78$ (middle panel of Fig.\ \ref{orm}), and the bias towards large $i$ in the prior, large $\beta$ are favoured overall. We note that we obtain a similar $\beta$ to that given by \cite{hub2011a}, however this is simply a coincidence as their value was obtained with an incorrect rotation period and a very different value of the Preston $r$ parameter (0.56, as compared to -0.78).

For each ($i$,~$\beta$) pair, $B_{\rm d}$ was determined using Eqn.\ 1 from \cite{preston1967}. This also requires knowledge of the limb darkening coefficient $\epsilon$, which we obtained from the tables calculated by \cite{diazcordoves1995} as $\epsilon=0.36\pm0.02$ (\S \ref{subsec:vsini}). The resulting PDF peaks sharply at the minimum value permitted by Preston's equation, $B_{\rm d}=1.1$ kG. As demonstrated in the middle and upper left panels of Fig.\ \ref{orm}, $B_{\rm d} < 2$ kG over the range $20^\circ<i<85^\circ$, with a low-probability tail extending out to several kG for very large and very small $i$ and $\beta$. This is much lower than the value of $\sim$5.3 kG given by \cite{hub2011a}, which arose from their very small $i$.

If instead the ESPaDOnS and MuSiCoS measurements define the extrema of \bz, we obtain $r=-0.33 \pm 0.02$. This does not affected $B_{\rm d}$, although a slightly smaller $\beta \sim 75^\circ$ is favoured.

Magnetic wind confinement is governed by the balance of kinetic energy density in the radiative wind to the magnetic energy density. This is expressed by the dimensionless wind magnetic confinement parameter $\eta_*$ \citep{ud2002}. If $\eta_* > 1$, the star possesses a magnetosphere. 

\cite{oskinova2011} analyzed IUE observations of $\xi^1$ CMa using the Potsdam Wolf-Rayet (PoWR) code in order to determine the wind parameters, obtaining $\log{(\dot{M}/M_\odot~{\rm yr^{-1}})}=-10$ and \vinf$=700$ \kms. However, magnetic confinement reduces the net mass-loss rate and, more seriously, strongly affects line diagnostics due to the departure from spherical symmetry in the circumstellar environment. Indeed, \citeauthor{oskinova2011} were unable to achieve a simultaneous fit to the Si~{\sc iv} and C~{\sc iv} doublets. Comparison of magnetohydrodynamic (MHD) simulations and spherically symmetric models to the magnetospheric emission of Of?p stars has demonstrated that MHD simulations yield a superior fit, and require mass-loss rates comparable to those obtained from the \cite{vink2001} recipe, whereas spherically symmetric models in general require much lower mass-loss rates to achieve a relatively poor match to the observations \citep{2012MNRAS.426.2208G,2012MNRAS.422.2314M,2013MNRAS.431.2253M}. In addition to this, magnetic wind confinement is not expected to modify the {\em surface} mass flux, which is the relevant quantity in calculating the strength of magnetic confinement \citep{ud2002}. Thus, $\eta_*$ should be determined using the wind parameters as they would be in the absence of a magnetic field, rather than those measured via spectral modelling (e.g., \citealt{petit2013}). Using the mass-loss recipe of \cite{vink2001} with \teff, $\log{L}$, and $M_*$ from Table \ref{params}, assuming the metallicity $Z/Z_\odot=1$, and determining the wind terminal velocity \vinf$=2070\pm60$ \kms~via scaling the star's escape velocity by 2.6 as suggested by Vink et al., yields a mass-loss rate $\log{(\dot{M}/M_\odot~{\rm yr^{-1}})}=-8.0\pm0.1$. From Eqn.\ 7 of \cite{ud2002}, $\eta_* > 340$, so the wind is magnetically confined. 

The physical extent of the magnetosphere is given by the Alfv\'en radius \ra, defined as the maximum extent of closed magnetic field lines in the circumstellar environment. \ra~can be calculated heuristically from $\eta_*$ (Eqn.\ 7 in \citealt{ud2008}): we find \ra~$>4.6~R_*$. Unless $i$ is particularly large or small, in which case $B_{\rm d}$ is significantly in excess of 2 kG, the magnetic and magnetospheric parameters will be close to the derived lower limits. \ra~is almost certainly below 20 $R_*$.  

$\xi^1$ CMa has the highest X-ray luminosity and the hardest X-ray spectrum of any of the magnetic $\beta$ Cep stars \citep{cassinelli1994, oskinova2011}. The X-ray luminosity, corrected for interstellar absorption corresponding to $E(B-V) = 0.015$, is $2.4 \times10^{31} {\rm erg s^{-1}}$ \citep{2014ApJS..215...10N}. Using 2D MHD simulations \cite{ud2014} calculated X-ray emission from magnetically confined wind shocks and developed an X-ray Analytic Dynamical Magnetosphere (XADM) scaling for \lx~with \bd, $R_*$, \mdot, and \vinf. Comparison to available X-ray data for magnetic early-type stars has indicated that \lx~predicted by XADM should be scaled by $\sim$5-20\%~to match the observed \lx~\citep{2014ApJS..215...10N}. This efficiency factor accounts for dynamical infall of the plasma. XADM successfully predicts the X-ray luminosity of magnetic, hot stars across 3 decades in $\log{L}$, 2 decades in $B_{\rm d}$, and 5 decades in \lx~\citep{2014ApJS..215...10N}, i.e. the range of the model's successful application is much larger than the uncertainty introduced by the efficiency factor. Assuming an efficiency of 10\%, $B_{\rm d}=1.1$ kG, and taking into account the uncertainties in the stellar parameters in Table \ref{params}, XADM predicts \lx$=31.5 \pm 0.1$, in excellent agreement with the observed X-ray luminosity \lx$=31.47$. Adopting the higher value of $B_{\rm d}=5.3$~kG suggested by \cite{hub2011a} yields \lx~$=31.7$, slightly higher than observed (although this can be reconciled by lowering the efficiency factor to 5\%). If the lower \mdot~and \vinf~determined by \cite{oskinova2011} are used instead, XADM predicts \lx~$=28.9$ with an efficiency factor of 100\%, 2.6 dex lower than observed. Since the efficiency factor can only lower \lx, it is impossible for the XADM model to match $\xi^1$ CMa's observed X-ray luminosity with a mass-loss rate significantly lower than the \cite{vink2001} prediction. 

\cite{petit2013} divided magnetic, massive stars into two classes, those with dynamical magnetospheres (DMs) only, and those also possessing centrifugal magnetospheres (CMs). CMs appear when \rk~$<$~\ra, where \rk~is the Kepler radius, defined as the radius at which the centrifugal force due to corotation compensates for the gravitational force \citep{town2005c, ud2008}. Solving for \rk~using Eqn.\ 12 from \cite{town2005c} with $M_*$ from Table \ref{params} and $P_{\rm rot}>30$~yr yields \rk~$>580~R_*$. As \rk~$\gg$~\ra, the magnetosphere does not include a CM. The Kepler radius is related to the dimensionless rotation parameter $W\equiv v_{\rm eq}/v_{\rm orb}=R_{\rm K}^{-3/2}$, where $v_{\rm orb}$ is the velocity required to maintain a Keplerian orbit at the stellar surface \citep{ud2008}. Critical rotation corresponds to $W=1$, and no rotation to $W=0$. For $\xi^1$ CMa, $W<7 \times 10^{-5}$.

Magnetic wind confinement leads to rapid spindown due to angular momentum loss via the extended moment arm of the magnetized wind \citep{wd1967, ud2009}. The rotation period will decrease exponetially, with a characteristic angular momentum loss timescale $\tau_{\rm J}$ of \citep{ud2009}:

\begin{equation}\label{tauj}
\tau_{\rm J} = \frac{3}{2}f \tau_{\rm M}\left(\frac{R_*}{R_{\rm A}}\right)^2,
\end{equation}

\noindent where $\tau_{\rm M} \equiv M_*/\dot{M}$ is the mass-loss timescale, and $f$ is the moment of inertia factor, which can be evaluated from the star's radius of gyration $r_{\rm gyr}$ as $f = r_{\rm gyr}^2$. Consulting the internal structure models calculated by \cite{claret2004}, $f \simeq 0.06$ for a star of $\xi^1$ CMa's mass and age. Solving Eqn.\ \ref{tauj} then yields $\tau_{\rm J}=4\pm1$ Myr. The rotation parameter at a time $t$ after the birth of the star is

\begin{equation}\label{pt}
W(t) = W_0 e^{-t/\tau_{\rm J}},
\end{equation}

\noindent where $W_0$ is the initial rotation parameter. Assuming $W_0=1$ yields the maximum spindown {\em age} $t_{\rm S,max}=42\pm12$ Myr \citep{petit2013}. This is about 4 times longer than the age inferred from evolutionary tracks. Solving Eqn.\ \ref{pt} for $W_0$ yields $W_0 < 0.02$. Thus, either magnetic braking must have been much more rapid than predicted, or almost all of the star's angular momentum loss must have occurred before it began its main sequence evolution, i.e.\ the star was already a slow rotator at the ZAMS.

Given the important role played by the mass-loss rate in determining \ra~and $\tau_{\rm J}$, it is of interest to explore the sensitivity of these results to different mass-loss prescriptions. \cite{2012A&A...537A..37M} found that they were unable to reproduce the observed mass-loss rates of stars with $\log{L} < 5.2$, suggesting that \mdot~may be lower than predicted by the \cite{vink2001} recipe. We first note that, since $\tau_{\rm J}$ increases with decreasing $\dot{M}$, a lower mass-loss rate cannot resolve the discrepancy between $t_{\rm S,max}$ and the age inferred from the HRD. Second, we note that satisfying the condition for a CM (\ra$>$\rk) requires $\log{(\dot{M}/M_\odot~{\rm yr^{-1}})} \le -16.5$. If the lower \mdot~and \vinf~found by \cite{oskinova2011} are used to calculate the magnetospheric and spindown parameters, none of the above conclusions are fundamentally changed (\ra~$\sim 14~R_* \ll R_{\rm K}$ , and $t_{\rm S,max} = 250^{+80}_{-100}$~Myr). \cite{2010A&A...512A..33L} computed mass-loss rates for late O-type stars in the weak-wind domain using the theory of reversing layers \citep{2007A&A...468..649L}, and for a star with $\xi^1$ CMa's \teff~and $\log{g}$ predict $\log{(\dot{M}/M_\odot~{\rm yr^{-1}})} \sim -8.8$. \cite{krticka2014} provided an alternate calculation for \mdot, intended specifically for B-type stars, and for $\xi^1$ CMa predicted $\log{(\dot{M}/M_\odot~{\rm yr^{-1}})} = -8.9$ (see Table 5 in \citealt{krticka2014}), similar to the \cite{2010A&A...512A..33L} prediction. Thus, while there are systematic differences between theoretical mass-loss rates, these are much smaller than would be required to change the basic conclusions that the star lacks a CM and has a spindown age much less than its evolutionary age. 


\section{Discussion}


\subsection{Impact of pulsation on the magnetometry}\label{subsec:pulsation_impact}

\begin{figure}
\centering
\includegraphics[width=8.5cm]{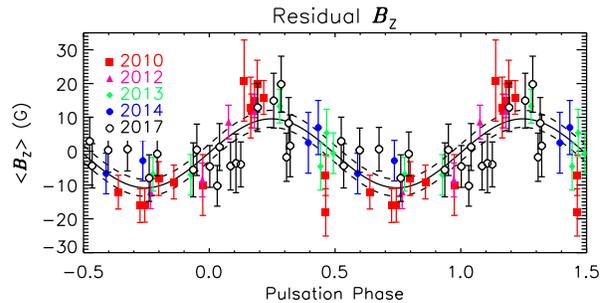}
\caption{Residual ESPaDOnS \bz~measurements, after subtraction of the mean \bz~at each epoch, phased with the pulsation period. The legend indicates the year in which each dataset was acquired. Only epochs with at least 4 observations are shown.}
\label{resid_bz}
\end{figure}

\begin{figure}
\centering
\includegraphics[width=8.5cm]{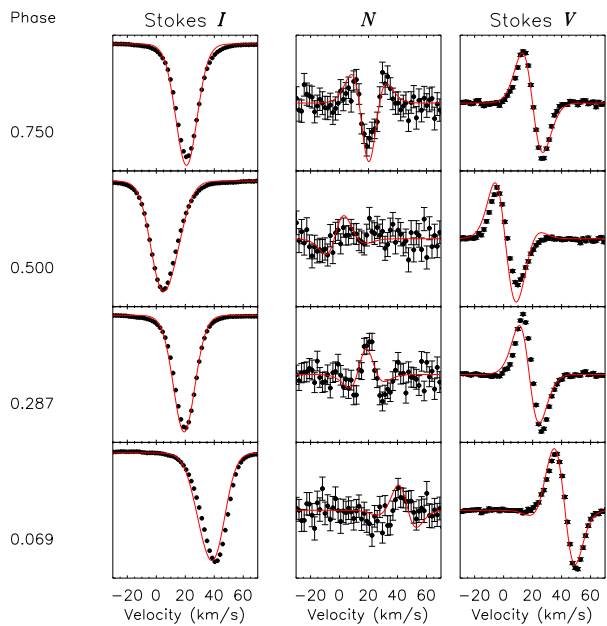}
\caption{Stokes $I$ (left), null $N$ (middle), and Stokes $V$ (right) LSD profiles. Observed profiles are shown by black circles, synthetic profiles by red lines.}
\label{null_model}
\end{figure}

While \bz~exhibits a clear long-term modulation, it also shows evidence for short-term variability. In particular in 2010 (HJD$\sim$2455200), the most densely time-sampled epoch, there is substantial apparent scatter in the measurements: the mean error bar is 5 G, but the standard deviation of \bz~in this epoch is 15 G. 

We calculated residual \bz~measurements by subtracting the mean \bz~at each annual epoch in order to remove the long-term trend. These are shown phased with the pulsation period in Fig.\ \ref{resid_bz}. The reduced $\chi^2$~of a least-squares sinusoidal fit to all residual \bz~measurements is 1.2, as compared to 2.8 for the null hypothesis of no variation. In 2010 there is an apparently coherent variation with a semi-amplitude of 16$\pm$2~G, consistent with the standard deviation in \bz~of 15~G, and 3 times larger than the median error bar of 5 G. The reduced $\chi^2$ of the least-squares sinusoidal fit to the 2010 residual \bz~measurements is 0.9, as compared to 6.1 for the null hypothesis. In 2017, the epoch with the largest number of observations and the most systematic sampling of the phase curve, the same fit yields a semi-amplitude of 7$\pm$2~G, which is identical to both the standard deviation of, and the median uncertainty in, \bz~during this epoch. The reduced $\chi^2$ of a sinusoidal fit and the null hypothesis to the 2017 data is close to 1 in both cases, indicating that in 2017 there is no evidence for a coherent variation with pulsation phase. In 2013, the only other epoch with sufficient observations to constrain a sinusoidal fit, the semi-amplitude is 10$\pm$5~G, the reduced $\chi^2$ of the fit is 1.3, and the reduced $\chi^2$ of the null hypothesis is 2.3, i.e. 2013 yields results intermediate between 2010 and 2017. This suggests that the observed variation of \bz~may be explained as the superposition of a short-term variation with the stellar pulsation period on top of the long-term variation due to rotation, with the short-term variation declining in amplitude with \bz. 

We next turn our attention to the LSD profiles. It is clear from the numerous `definite' detections in LSD $N$ profiles that pulsation has affected our data to some degree. Signatures in $N$ profiles are sometimes seen in stars in which the stellar lines vary during acquisition of the polarized sub-exposures used to construct the final Stokes $V$ spectrum. Following the analysis performed by \cite{neiner2012a}, we modelled the $N$ profiles by creating disk-integrated Stokes $I$ and $V$ profiles at the radial velocities corresponding to the pulsation phases of each sub-exposure, and then combining the individual model profiles in such a way as to simulate the double-ratio method's combination of $I \pm V$ beams to yield Stokes $I$, Stokes $V$, and $N$ \citep{d1997}. Each model profile was created with \vsini$~=0$~\kms, a macroturbulent velocity of 8 \kms, the measured RVs, a projection factor of 1.45, and a constant dipolar magnetic field with $i = \beta = 90^\circ$ and \bd$=1.1$ kG, with the positive magnetic pole at the centre of the stellar disk. The model profiles were normalized to the observed Stokes $I$ EW.

Illustrative results are shown for Stokes $I$, $N$, and Stokes $V$ in Fig.\ \ref{null_model}. The model reproduces the variation in $N$ reasonably well. We conclude from this that the $N$ profile signatures are principally a consequence of line profile variability between sub-exposures. 

There is no change in \bz~measured from model LSD profiles created with or without introducing RV shifts between sub-exposures. Furthermore, \nz~remains null, as expected. This result is in agreement with the observational results of \cite{neiner2012a}, who found that while shifting sub-exposures by their pulsation velocities greatly reduced the $N$ profile signatures, \bz~remained unchanged within error bars. 

If the modulation of \bz~with pulsation phase is not a consequence of an instrumental effect, it might be due to a real change in the intrinsic surface magnetic field strength of the star. An obvious mechanism that might produce such an effect is conservation of magnetic flux throughout radial pulsation cycles. Magnetic flux is conserved as $1/R_*^2$, and the semi-amplitude of the change in stellar radius is $\Delta r/R_* \sim 1-1.5\%$, thus \bz~could vary by about $\pm$2--3\% or $\pm 7-10$ G at \bz$_{\rm max}=328$ G, which is close to the observed semi-amplitude in 2010 of 16$\pm$2 G. In this scenario, maximum \bz~should occur at phase 0.25, when the star is at its most contracted, and minimum \bz~at phase 0.75, when the star is at its most extended; this is indeed what is observed. Thus, a modulation of \bz~arising due to an actual change in the surface magnetic field strength with the pulsation period is consistent with both the magnitude and the phasing of the observed effect. The apparent proportionality of the semi-amplitude of the pulsational modulation of \bz~to the magnitude of \bz~is also consistent with this hypothesis, as a change in \bz~due to a change in $B_{\rm d}$ should be largest when there is the least amount of magnetic flux cancellation across the stellar disk (i.e. at magnetic maximum). 

While our model does not predict a change in \bz~due to RV shifts between subexposures, we cannot entirely rule out the possibility that the modulation shown in Fig.\ \ref{resid_bz} is due to line profile variability. Whatever its origin, the influence of pulsation on \bz~is much smaller than the long-term modulation of \bz, thus the fundamental conclusion that $P_{\rm rot} > 30$~yr should not be affected by pulsation.

\subsection{Does extremely slow rotation in an evolved magnetic B-type star make sense?}

Due to magnetic braking evolved magnetic hot stars are expected to rotate more slowly than either similar, younger magnetic stars or non-magnetic stars of comparable age and mass. However, as was shown in \S~\ref{sect:magpars}, the spindown age inferred for $\xi^1$ CMa is difficult to reconcile with the age inferred from its position on the HRD. The \cite{ekstrom2012} evolutionary models imply an age of 11$\pm$0.7 Myr, while its spindown timescale is $\tau_{\rm J}=4\pm1$ Myr. Thus, at most two or three e-foldings of the initial rotational period can have occurred since the ZAMS. Assuming initially critical rotation, $\xi^1$ CMa should now have a rotational period of about 6 d; conversely, calculating from its actual rotation period, its initial rotation fraction $W_0$ must have already been very close to 0, implying a rotational period on the ZAMS already on the order of years. 

\begin{figure}
\centering
\includegraphics[width=\hsize]{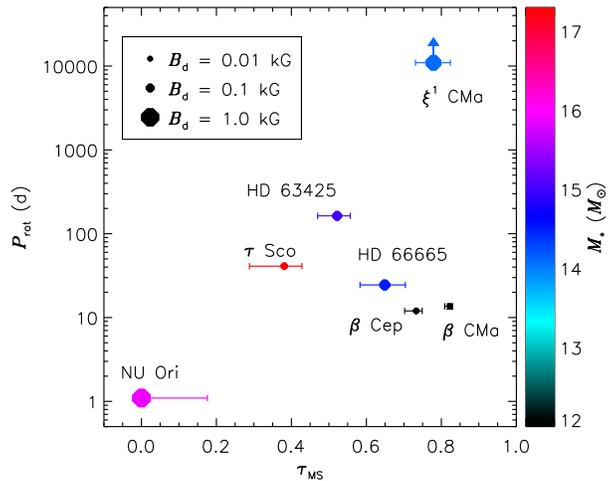}
\caption{$P_{\rm rot}$ as a function of fractional main sequence age $\tau_{\rm MS}$ for magnetic early B-type stars with masses similar to $\xi^1$ CMa, with colour corresponding to stellar mass and symbol size to $B_{\rm d}$. Uncertainties in $P_{\rm rot}$ are negligible on the scale of the diagram.}
\label{bd_prot}
\end{figure}

\cite{petit2013} calculated spindown ages for all stars in Fig.\ \ref{hrd} except $\beta$ CMa and $\epsilon$ CMa, finding $40 < t_{\rm S,max} < 130$ Myr for these stars, i.e.\ their maximum spindown ages are 4 to 10 times greater than their main-sequence ages, which for stars in this mass range are typically 8 to 15 Myr. We note that the rotational periods of $\tau$ Sco and $\beta$ Cep are both known to high precision \citep{2006MNRAS.370..629D,henrichs2013}. For $\tau$ Sco, $t_{\rm S,max}=126^{+16}_{-32}$~Myr, while its position on the HRD indicates $t=3.3^{+0.5}_{-1.1}$~Myr, a discrepancy of 1.6 dex; similarly for $\beta$ Cep, $t_{\rm S,max}=204^{+9}_{-19}$~Myr, differing by 1.2 dex with its age of $t=13.2^{+0.4}_{-0.7}$~Myr \citep{my_phd_thesis}. 

These discrepancies between the stellar age and the maximum spindown age cannot be reconciled by utilizing a different set of evolutionary models: while the absolute ages inferred from different prescriptions vary depending on assumptions about rotation, mixing, or mass-loss, these uncertainties in the models are typically on the order of 1 to 2 Myr for a 15 \msun~star, much less than the $\sim$1 dex difference between gyrochronological and evolutionary ages. It should be noted that the \cite{ekstrom2012} models do not include magnetic fields. Fully self-consistent evolutionary models of magnetic stars are not yet available; however, it seems unlikely that such models will resolve the discrepancy via modification of the age inferred from a star's position on the HRD. 

\cite{2016A&A...592A..84F} also found that spindown timescales are typically much longer than stellar ages for magnetic, massive stars, which they interpreted as evidence for rapid magnetic flux decay. At first order, the surface magnetic field should weaken as $1/R_*^2$ due to conservation of fossil magnetic flux. As an evolved star, $\xi^1$ CMa's radius has increased by a factor of 1.8$\pm$0.1 since the ZAMS, thus its minimum surface magnetic field strength on the ZAMS should have been $B_{\rm d,ZAMS} > 3.2 $ kG. Recalculating $\tau_{\rm J}$ on the ZAMS using Vink mass-loss rates suggests that $\tau_{\rm J}$ should not have been very different from its present value, as the lower mass-loss rate on the ZAMS compensates for the stronger magnetic field and greater Alfv\'en radius. Thus, magnetic flux conservation alone cannot account for $\xi^1$ CMa's slow rotation, and some other mechanism or combination of mechanisms -- magnetic flux decay, rapid angular momentum loss on the PMS, or modification of the internal structure of the star due to the magnetic field -- is necessary. While it is of course possible that the theoretical mass-loss rates are simply incorrect, resolving the discrepancy by changing \mdot~alone would require an increase of $\sim$1 dex, well outside the range of uncertainty of current models which typically differ a factor of 2 to 3, and are in addition usually {\em lower} than the \cite{vink2001} mass-loss rates (e.g., \citealt{2014A&A...568A..59S,2017A&A...598A...4K}).

The irreconcilability of gyrochronological and evolutionary timescales amongst magnetic early B-type stars is worthy of future investigation. Here, it is pointed out primarily to emphasize that the difficulty in explaining $\xi^1$ CMa's extremely slow rotation should not count as an argument against a long rotational period, as this problem is shared by all similar stars.

Fig.\ \ref{bd_prot} shows $P_{\rm rot}$ as a function of fractional main sequence age $\tau_{\rm MS}$ for those stars from Fig.\ \ref{hrd} for which both $P_{\rm rot}$ and $B_{\rm d}$ are known. The majority of the dipolar field strengths and rotational periods were obtained from the catalogue published by \cite{petit2013}, and references therein. $\beta$ CMa's magnetic, rotational, and stellar properties were obtained from \cite{2015AA...574A..20F}. The rotational periods of NU Ori and HD 63425 were determined using ESPaDOnS data \citep{my_phd_thesis}. Masses and fractional main sequence ages were determined via linear interpolation between the \cite{ekstrom2012} evolutionary models on the basis of the stars' positions on Fig.\ \ref{hrd}. For HD 63425 and HD 66665, we used the \teff-$\log{g}$ diagram to infer stellar parameters, as the uncertainties in $\log{L}$ are very large for these stars. 

In addition to having by far the longest rotational period of any magnetic B-type star, $\xi^1$ CMa has a stronger magnetic field than any star in Fig.\ \ref{bd_prot} but NU Ori. It is also amongst the most evolved stars in the sample: the only stars of a comparable fractional age are $\beta$ Cep and $\beta$ CMa, both of which are less massive ($\sim$12 \msun), and have much weaker magnetic fields ($\sim$0.1 kG). Since mass-loss rates increase with $M_*$, more massive stars should spin down more rapidly than less massive stars. With the exception of NU Ori, stars more massive than $\xi^1$ CMa all have rotational periods of 10-100 d, although they are less evolved. NU Ori, which has a stronger magnetic field, a much shorter rotational period, and a slightly higher stellar mass than $\xi^1$ CMa, is also the youngest star on the diagram, likely accounting for its rapid rotation. This comparison shows that, viewed in the context of stars with similar magnetic and stellar parameters, $\xi^1$ CMa's very slow rotation is not anomalous given its age. 

\subsection{A B-type star with an optically detectable dynamical magnetosphere?}

\begin{figure}
\centering
\includegraphics[width=\hsize]{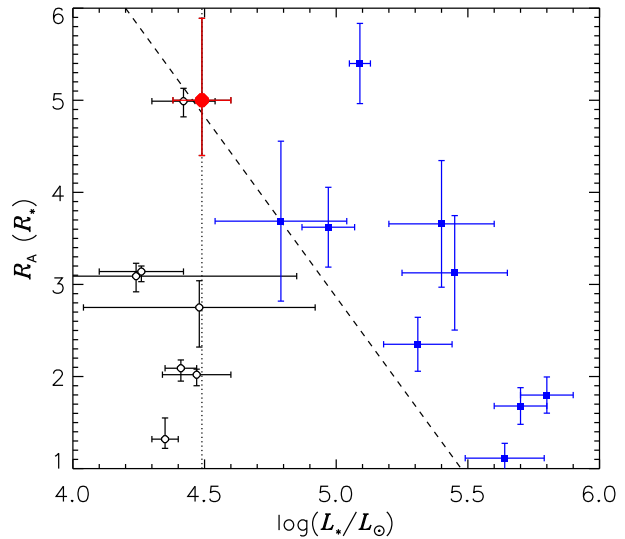}
\caption{\ra~as a function of $\log{L}$ for magnetic early B-type stars (open black circles) and the magnetic O-type stars (solid blue squares). $\xi^1$ CMa is shown as a large, filled red circle. With the exception of $\xi^1$ CMa, only the O-type stars display emission in H$\alpha$. The ad-hoc dashed line suggests a possible relationship between the onset of emission in DMs, and \ra~and $\log{L}$.}
\label{ra_lum}
\end{figure}

In DMs, plasma deposited in the magnetosphere by the stellar wind falls back due to gravity, while in CMs centrifugal support due to the magnetically enforced corotation of the plasma with the star prevents infall above \rk. \cite{petit2013} noted that stars without CMs only display emission if their mass-loss rates are high enough to replenish the magnetosphere on dynamical timescales, thus enabling the circumstellar environment to remain optically thick despite continuous depletion as material returns to the photosphere. As a consequence, the only stars with optical emission lines originating in their DMs are the magnetic O-type stars, whereas the only magnetic B-type stars listed by \citeauthor{petit2013} as hosting H$\alpha$ emission possess CMs. While they were aware of $\xi^1$ CMa's H$\alpha$ emission, \citeauthor{petit2013} assumed the 4.26~d rotation period found by \cite{fr2011}, and thus that the emission originated in the star's CM. However, as \rk~$\gg$~\ra, $\xi^1$ CMa does not have a CM.

The H$\alpha$ emission profile shown in Fig.\ \ref{halpha_model_dyn} is furthermore entirely distinct from the double-horned profile characteristic of a CM, which generally produces two emission peaks located at approximately $\pm$\rk~(e.g.\ \citealt{leone2010,oks2012,grun2012,rivi2013}). The strong central peak is much more similar to the typical profile of a magnetic O-type star with a DM (e.g.\ \citealt{sund2012,ud2013,2015MNRAS.447.2551W}), although the emission peaks of such stars tend to be slightly red-shifted, whereas $\xi^1$ CMa's is closer to line centre. $\xi^1$ CMa's ultraviolet emission line morphology is also consistent with an origin in a DM, at least insofar as it is very similar to the emission lines of $\beta$ Cep, which does not have a CM. 

This leads to the question of how a B-type star can have an optically detectable DM. The quantity of plasma in a DM is a function of two parameters: the size of the magnetosphere, and the mass-loss rate which feeds it. As is clear from Fig.\ \ref{hrd}, $\xi^1$ CMa is one of the most luminous magnetic B-type stars known: while $\tau$ Sco and NU Ori have similar $\log{L}$, no stars are known to have a higher luminosity. In Fig.\ \ref{ra_lum} \ra~is plotted as a function of $\log{L}$ for the stars in Fig.\ \ref{hrd}. Not only does $\xi^1$ CMa have a higher luminosity than most of the B-type stars, it also has one of the largest Alfv\'en radii. 

Also plotted on Fig.\ \ref{ra_lum} are the magnetic O-type stars from the \cite{petit2013} catalogue. All of these stars have emission, despite the very small \ra~of the stars with higher luminosities. The dashed diagonal line indicates a possible division between stars with and without emission line DMs. Under the assumption that a star with a smaller \ra~requires a higher mass-loss rate in order to show emission, stars above the line should show emission, while those below should not. Alternatively, the presence or absence of emission could be related purely to the mass-loss rate, as indicated by the vertical dotted line: in this case, stars with $\log{L/L_\odot} \gtrsim 4.5$ should show emission. $\xi^1$ CMa's position in either case is consistent with the presence of H$\alpha$ emission. The presence or absence of magnetospheric emission in magnetic OB stars with $4.5 \le \log{L/L_\odot} \le 5.0$, but \ra~$\ge 3~R_*$, would help to distinguish between these scenarios.

Only one other magnetic B-type star, NU Ori, is in the same part of the diagram as $\xi^1$ CMa. As noted above, NU Ori is a rapid rotator, and possesses a CM. This star shows neither H$\alpha$ nor UV emission \citep{petit2013}. The absence of emission, given its high luminosity and comparable magnetic confinement strength to $\xi^1$ CMa, is a distinct puzzle deserving of future investigation.

\subsection{A pulsating magnetosphere?}

Due to the limited number and quality of IUE observations, the evidence for pulsational modulation of UV resonance doublets is inconclusive. Evidence for pulsational modulation of the star's H$\alpha$ emission is much stronger. The residual H$\alpha$ EWs, after subtraction of a sinusoidal fit to the annual mean EWs phased with the presumed rotation period, yield a clear peak in the periodogram at the pulsation period. When phased with the pulsation period, the amplitude of the residual EW variation is 3 times higher than the amplitude expected from a photospheric modulation of the absorption line's EW, and is furthermore 0.5 pulsation cycles out of phase with the predicted photospheric variation. Finally, H$\alpha$ shows a much stronger V/R variation than is predicted by the photospheric model. 

Stochastic changes in the EW of the H$\alpha$ line have been observed for some magnetic O-type stars with dynamical magnetospheres, e.g.\ the well-studied object $\theta^1$ Ori C \citep{2008AA...487..323S}. 3D MHD simulations are able to reproduce this short-term variability as a consequence of turbulent plasma infall \citep{ud2013}. As \mdot~is constant for such stars, their short-term variability is wholly a consequence of the complex behaviour of MHD plasmas. Since $\xi^1$ CMa's mass-loss rate should change with the pulsation phase, it is conceivable that this could force a periodicity in the outflow and infall of the magnetically confined plasma. The observation of prominent blue- and red-shifted secondary peaks in the H$\alpha$ emission structure occuring at distinct pulsation phases, together with the stronger modulation in H$\alpha$ $V/R$ as compared to residual H$\alpha$ EWs, further suggests that, throughout the course of a pulsation cycle, there are larger local changes in the distribution of plasma within the magnetosphere, as compared to the total mass of magnetically confined plasma. 

The majority of the H$\alpha$ emission is expected to be produced in the innermost regions \citep{ud2013}. The wind flow timescale, calculated using a standard $\beta = 1$ velocity law\footnote{While confinement in a dipolar magnetic field does modify the wind-flow timescale, in practice this is a second-order effect \citep{ud2014}.}, ranges from $\sim$0.13 d at a distance of 2 $R_*$, to $\sim$0.19 d at the minimum Alfv\'en radius of 4.6 $R_*$. The free-fall timescale is somewhat longer, ranging from 0.2 to 1.2 d over this same distance. These timescales are similar to the pulsation period, suggesting that periodic density waves arising due to the varying mass-loss rate might lead to a periodicity in the plasma infall. The strongest blue-shifted secondary emission peak occurs at about phase 1.0, and the strongest secondary red-shifted emission peak at phase 0.5. If the blue- and red-shifted emission peaks are respectively consequences of outflowing and infalling plasma, and assuming the majority of the emission is formed relatively close to the star, a lag of approximately 0.7 cycles between the peak mass-loss rate (at phase 0.3) and the strongest blue-shifted emission (at phase 1.0) would make sense.

While no other magnetic, massive star's H$\alpha$ emission has ever been found to be modulated by pulsation as well as rotation, \cite{2014NatCo...5E4024O} reported that $\xi^1$ CMa's X-ray light curve exhibits a coherent variation with the pulsation period, with both hard and soft X-ray emission peaking $\sim$0.1 cycles after the maximum of the visible light curve. 

Since $\xi^1$ CMa is a pulsating star, \mdot~is not constant. The changes in \teff~and $\log{L_*}$ should lead to a variation in \mdot~with a semi-amplitude of 0.05 dex. Changes in \vinf~should be negligible. Calculation of $\eta_*$ and \ra~as a function of pulsation phase suggests the overall changes in these parameters to be small ($\sim$20\% in $\eta_*$ and 4\% in \ra). Similar X-ray pulsations have also been reported for the $\beta$ Cep star $\beta$ Cru \citep{2008MNRAS.386.1855C}. As $\beta$ Cru does not have a detected magnetic field (J.\ Grunhut, priv.\ comm.), this argues against a purely magnetic explanation for the X-ray variability. 

\cite{2014NatCo...5E4024O} considered a variation in \mdot~as a possible mechanism behind the star's X-ray pulsations, and rejected it based upon the absence of X-ray pulsations in other $\beta$ Cep stars, either magnetic or not. However, they did not consider their results in the context of the XADM model. The XADM model predicts a modulation in \lx~of approximately $\pm10$\% to arise from a variation in \mdot~of $\pm$0.05 dex, with the maximum and minimum of the predicted X-ray light curve corresponding to the maximum and minimum of the visible light curve. This is the same amplitude as that reported by \cite{2014NatCo...5E4024O}. 

As a sanity check, we performed similar calculations for $\beta$ Cep and $\beta$ Cen, the only other magnetic $\beta$ Cep stars with X-ray light curves. $\beta$ Cep is slightly cooler (\teff$\sim$26 kK), less strongly magnetized ($B_{\rm d}\sim260$~G; \citealt{henrichs2013}), and has somewhat lower amplitude pulsations leading to a smaller change in luminosity ($\pm$0.015 mag and $\pm$0.006~dex, respectively, as evaluated by phasing the star's Hipparcos observations with its primary pulsation period of 0.19~d; \citealt{2000ApJ...531L.143S}) and, hence, a smaller change in the mass-loss rate ($\pm$0.03 dex). The XADM model predicts that $\beta$ Cep should have X-ray pulsations with a semi-amplitude of $\le 6$\%, close to the upper limit determined by \cite{2009A&A...495..217F}. For $\beta$ Cen, it is not clear which of the many pulsation frequencies identified by \cite{2016A&A...588A..55P} belongs to the magnetic secondary. However, even the highest amplitude frequency ($\pm$0.0015 mag) leads to a luminosity modulation of just 0.006 dex, hence a change in \mdot~of less than $\pm$0.01 dex, and, with $B_{\rm d}=200$~G \citep{my_phd_thesis}, a variation in the X-ray luminosity of $<3$\%, below the upper limit established by \cite{2005A&A...437..599R}. 

It is probably not the case that simple calculation of global parameters at each pulsation phase provides an accurate picture. Strong local density variations due to the changing \mdot~may lead to correspondingly strong local changes in the magnetic confinement strength. Magnetohydynamic simulations will be necessary to properly explore the impact of stellar pulsation on magnetospheric dynamics. 

\section{Conclusions and Future Work}

The principal result of this paper is that both magnetic and spectroscopic data indicate that the rotation period of $\xi^1$ CMa is very long, on the order of decades. Rotational periods on the order of days, as previously suggested in the literature, are conclusively ruled out by the ESPaDOnS \bz~measurements alone, which require a period of at least 5100~d. Inclusion of the MuSiCoS results requires that $P_{\rm rot}$ be about 30 years, a conclusion supported by the modulation of the H$\alpha$ emission strength seen in the CORALIE and ESPaDOnS datasets. A period of approximately 30 years is consistent with the timing of the star's strong UV emission, which should correspond to an extremum of the \bz~curve. Given that phase coverage of the rotational cycle is incomplete to an unknown degree, the constraints that can be placed upon the geometry and strength of the magnetic field are necessarily somewhat loose. Further spectropolarimetric observation over decades is essential to characterization of the magnetic field. If the 30-year period is correct, magnetic crossover (\bz$=0$) will occur in 2018, and magnetic minimum in 2025. 

We have conducted the first detailed examination of the star's weak H$\alpha$ emission, which we detect in all available spectroscopic data. There is no evidence in the interferometric data of a binary companion, and combined constraints from interferometry and RV residuals rule out a binary companion of sufficient luminosity to host the weak emission. Furthermore, there is no indication of a Keplerian disk in the high-resolution AMBER spectro-interferometry. Thus, the interferometric data does not support the hypothesis that the emission originates in the Keplerian disk of a classical Be companion star, indicating that this emission is produced in $\xi^1$ CMa's circumstellar environment. This makes $\xi^1$ CMa the first magnetic $\beta$ Cep star with a magnetosphere detectable in visible light. Given that its extremely slow rotation implies a dynamical rather than a centrifugal magnetosphere, $\xi^1$ CMa is also the coolest star with a dynamical magnetosphere detectable in H$\alpha$.

Finally, in addition to the already reported modulation of X-ray flux with the pulsation period, the H$\alpha$ emission also appears to correlate with the pulsations in a fashion that seems likely to be intrinsic to the circumstellar plasma. The best evidence for this is in the strong H$\alpha$ $V/R$ variations, which are much greater than expected from a radially pulsating photospheric model. The total H$\alpha$ EW in any given epoch also varies in antiphase with the expected EW variation due to photospheric \teff~variation, and furthermore has a larger amplitude than predicted by a photospheric model. However, the EW is only weakly variable compared to $V/R$, suggesting that the distribution of plasma within the magnetosphere, rather than the total quantity of magnetically confined material, is the primary origin of the variability. Given the complex velocity fields within dynamical magnetospheres, magnetohydrodynamic (MHD) simulations will be necessary to understand the interplay between pulsation, mass-loss, and magnetic confinement. As the magnetic wind confinement parameter $\eta_*$ is on the order of 10$^2$, modeling $\xi^1$ CMa's magnetosphere should be within the capabilities of the current generation of 2D MHD codes.

We also find a weak modulation of \bz~with the pulsation period. Further modeling is needed to understand the origin of this phenomenon. Signatures in the diagnostic null profile are accurately reproduced as a consequence of RV variations between subexposures, however, this model does not predict any modulation of \bz. Polarized radiative transfer in a moving atmosphere will be essential to answering the question of whether these modulations are an artefact of the measurement methodology, or intrinsic to the stellar magnetic field. 
\\
\\
{\small {\em Acknowledgements} The authors acknowledge the assistance of Conny Aerts, who provided the CORALIE dataset. This work has made use of the VALD database, operated at Uppsala University, the Institute of Astronomy RAS in Moscow, and the University of Vienna. MS acknowledges the financial support provided by the European Southern Observatory studentship program in Santiago, Chile. MS and GAW acknowledge support from the Natural Science and Engineering Research Council of Canada (NSERC). CN and the MiMeS collaboration acknowledge financial support from the Programme National de Physique Stellaire (PNPS) of INSU/CNRS. We acknowledge the Canadian Astronomy Data Centre (CADC). The IUE data presented in this paper were obtained from the Multimission Archive at the Space Telescope Science Institute (MAST). STScI is operated by the Association of Universities for Research in Astronomy, Inc., under NASA contract NAS5-26555. Support for MAST for non-HST data is provided by the NASA Office of Space Science via grant NAG5-7584 and by other grants and contracts. This project used the facilities of SIMBAD and Hipparcos. This research has made use of the Jean-Marie Mariotti {\sc litpro} service co-developed by CRAL, IPAG, and LAGRANGE.

\bibliography{bib_dat.bib}{}

\newpage
\appendix


\section{Reliability of MuSiCoS data}\label{append:musicos}

\begin{figure}
\centering
\includegraphics[width=9cm]{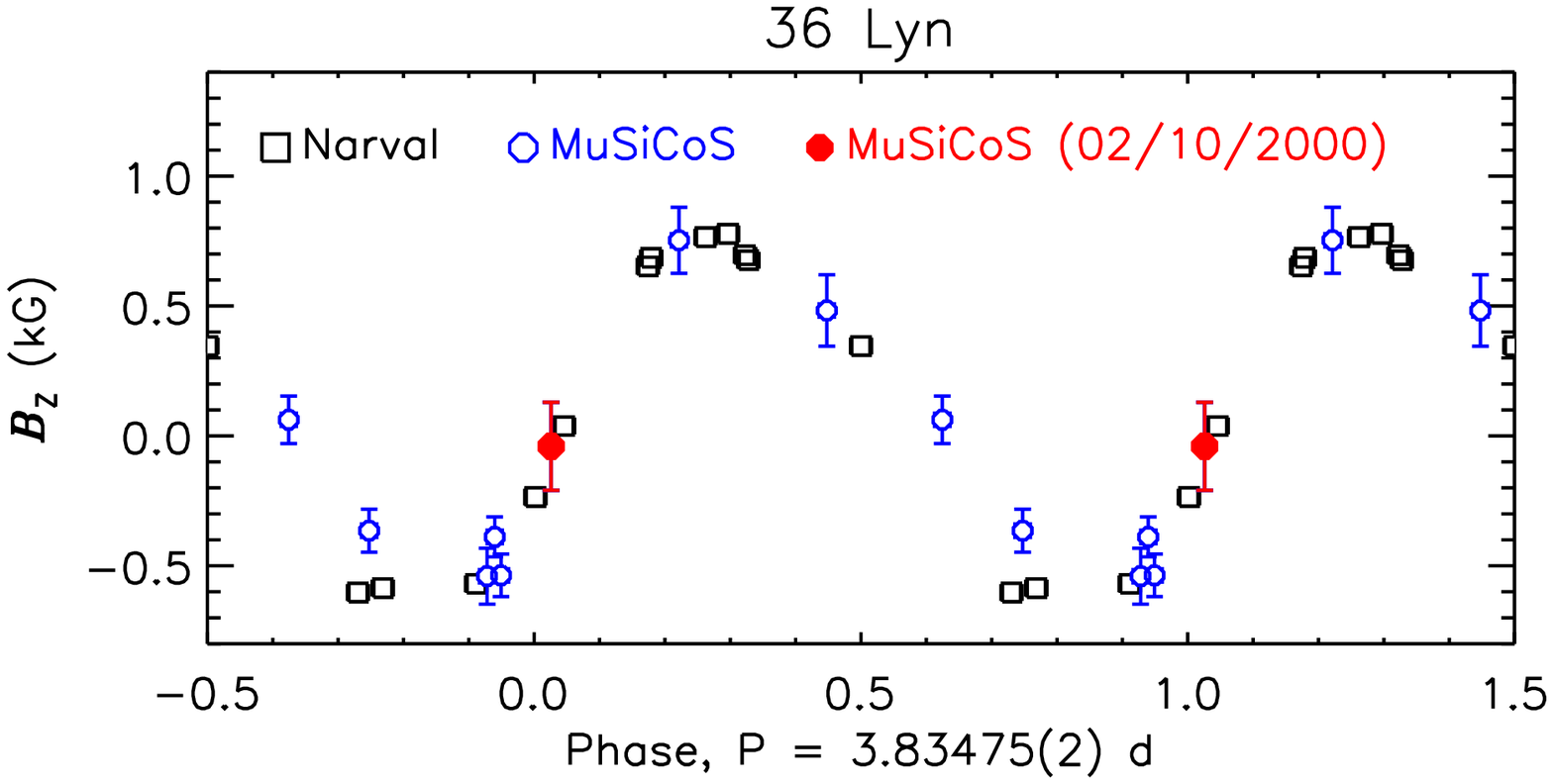}
\caption{\bz~measurements of 36 Lyn obtained with Narval and MuSiCoS, phased with the ephemeris determined from the MuSiCoS data by \protect\cite{2006A&A...458..569W}. The MuSiCoS measurement obtained during the same observing run as the $\xi^1$ CMa observations is indicated by the filled red circle.}
\label{36lyn_bz}
\end{figure}

\begin{figure}
\centering
\includegraphics[width=8cm]{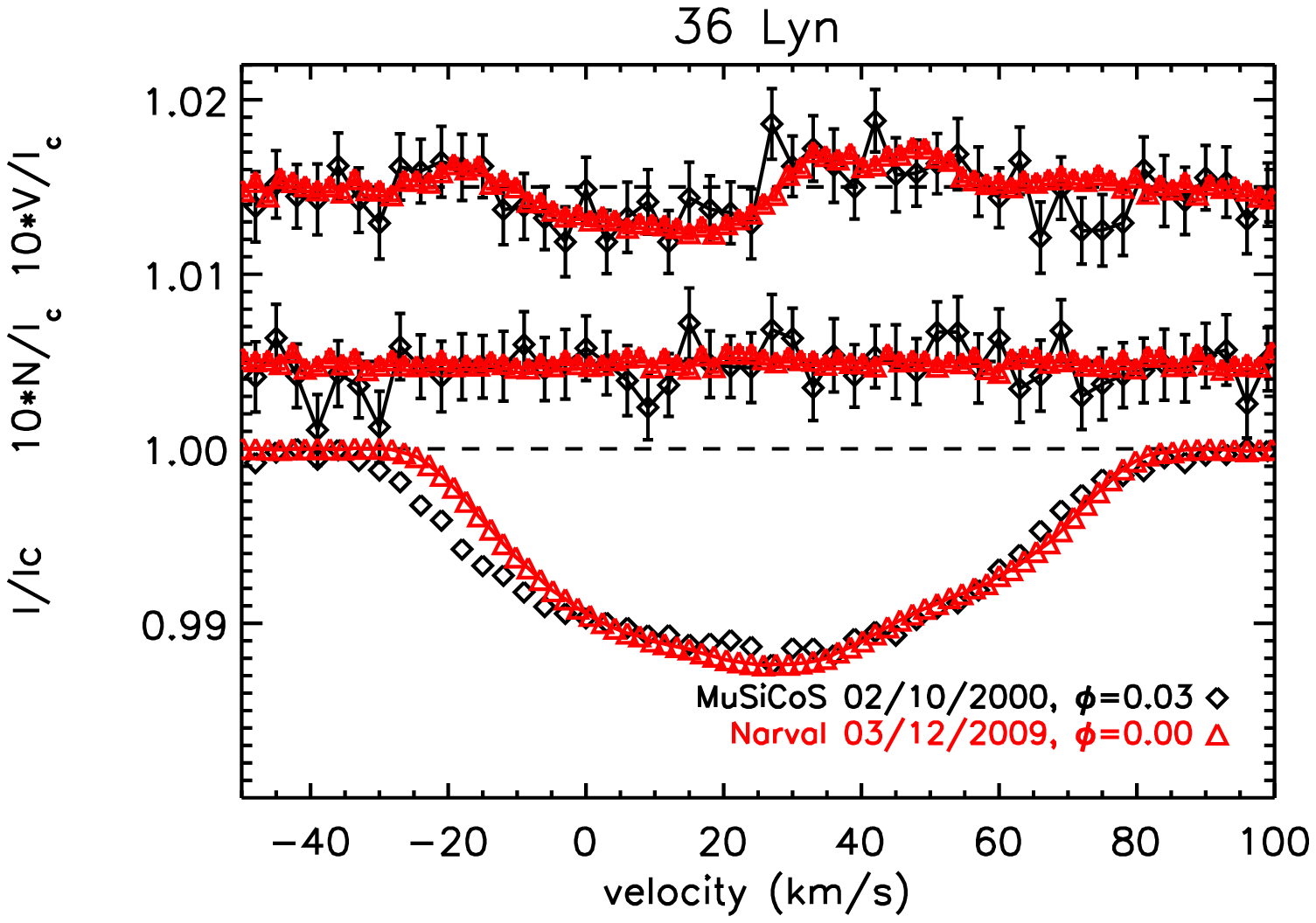}
\caption{Comparison between the MuSiCoS LSD profile of the magnetic Bp star 36 Lyn obtained during the same observing run as the $\xi^1$ CMa data, and a Narval LSD profile of 36 Lyn at a similar rotational phase.}
\label{36lyn_nar_mus_lsd}
\end{figure}

\cite{2015A&A...582A..45F} objected to a long rotation period, first suggested by \cite{2015IAUS..307..399S}, on the grounds that the period is largely constrained by the MuSiCoS measurements, and further suggesting that as these data were collected at high airmass the measurements were unreliable. As we have shown, it is not the case that a long rotation period is implied primarily by the MuSiCoS data: to the contrary, a gradual decline in \bz~is clearly evident in the ESPaDOnS data alone, and the conclusion that the star is slowly rotating is also supported by the long-term modulation of the H$\alpha$ emission strength. However, the MuSiCoS measurements are an important constraint, and in this appendix we address the question of their reliability. While we are aware of no reason why the sign of the Stokes $V$ signature should be flipped due to the star being observed at high airmass, the fibers were plugged in differently each observing run, which could result in an apparent polarity change. To check this possibility, we acquired the MuSiCoS dataset of the Bp star 36 Lyn reported by \cite{2006A&A...458..569W}, of which one measurement was acquired on 02/10/2000, during the same observing run as the $\xi^1$ CMa data. We compare these data with Narval observations of 36 Lyn. Narval, a clone of ESPaDOnS, obtains essentially identical results to its sibling instrument \citep{2016MNRAS.456....2W}. LSD profiles were extracted from the two datasets using the same method described in \S~\ref{subsec:lsd}, with a line mask appropriate to 36 Lyn's \teff. 

Fig.\ \ref{36lyn_bz} shows the resulting \bz~measurements, phased with the rotational ephemeris determined by \cite{2006A&A...458..569W}. The two datasets agree well at all phases. While the measurement obtained on 02/10/2000 is in its expected position, \bz$=-40\pm170$~G is close to 0, thus a sign flip would not change its position on the \bz~curve. As an additional check, Fig.\ \ref{36lyn_nar_mus_lsd} compares the LSD profile of this MuSiCoS observation to the Narval LSD profile with the closest rotational phase. The Stokes $V$ profiles of the two observations are a good match to one another. Moreover, a change in polarity of the MuSiCoS signature would render it in conflict with the Narval profiles. We conclude that there is no compelling reason to doubt the reliability of MuSiCoS, which obtained results fully consistent with those of modern high-resolution spectropolarimeters, and thus that the negative polarity of $\xi^1$ CMa's Stokes $V$ profile during this epoch is real.

\renewcommand{\thetable}{\Alph{section}\arabic{table}}

\newpage

\section{Log of magnetic measurements}

\begin{table*}
\centering
\caption[Log of ESPaDOnS Observations]{Log of spectropolarimetric observations. $\phi_{\rm P}$ gives the pulsation phase, calculated using the ephemeris in Eqn.\ \ref{ephem}. Exposure times $t_{\rm exp}$ correspond to the exposure time of a single subexposure: the full exposure time is 4$\times$ this value. Peak signal-to-noise ratios (S/N) are per spectral pixel. Detection flags (DFs) are given for Stokes $V$ and $N$, where DD is a definite detection, MD is a marginal detection, and ND is a non-detection, as evaluated using false alarm probabilities (FAPs, see text).}
\begin{tabular}{llcrrrrrrrr}
\hline
\hline
HJD - & Cal. & $\phi_{\rm P}$ & $t_{\rm exp}$ & Peak & \bz & $\sigma_{\rm B}$ & DF$_V$ & \nz & $\sigma_{\rm N}$ & DF$_N$ \\
2450000 & Date &  & (s) & S/N & (G) & (G) & &  (G) & (G) & \\
\hline
\multicolumn{11}{c}{MuSiCoS} \\
1580.40470 & 2000-02-05 & 0.41264 & 649 &   91 & -- &  -- & ND &   -- &  57 & -- \\
1584.39190 & 2000-02-09 & 0.43770 & 648 &  917 & -151 &  57 & ND &   17 &  57 & ND \\
1586.41240 & 2000-02-11 & 0.07858 & 649 &  335 & -105 &  54 & ND &  -21 &  54 & ND \\
\hline
\multicolumn{11}{c}{ESPaDOnS} \\
4488.98104 & 2008-01-23 & 0.75730 &  75 &  829 &  316 &   8 & DD &    8 &   8 & DD \\
\hline
\hline
\end{tabular}
\label{esp_obs_log}
\end{table*}

\begin{table*}
\contcaption{Log of spectropolarimetric observations.}
\label{esp_obs_log:continued}
\begin{tabular}{llcrrrrrrrr}
\hline
\hline
HJD - & Cal. & $\phi_{\rm P}$ & $t_{\rm exp}$ & Peak & \bz & $\sigma_{\rm B}$ & DF$_V$ & \nz & $\sigma_{\rm N}$ & DF$_N$ \\
2450000 & Date &  & (s) & S/N & (G) & (G) & &  (G) & (G) & \\
\hline
7788.88397 & 2017-02-04 & 0.25608 &  72 & 1008 &   92 &   8 & DD &   -2 &   8 & ND \\
7788.89021 & 2017-02-04 & 0.28586 &  72 &  983 &   97 &   8 & DD &    7 &   8 & ND \\  
7788.89646 & 2017-02-04 & 0.31568 &  72 & 1068 &   85 &   7 & DD &   11 &   7 & ND \\
7789.91848 & 2017-02-05 & 0.19224 &  72 & 1194 &   90 &   7 & DD &   -8 &   7 & DD \\
7795.75648 & 2017-02-11 & 0.04821 &  72 &  485 &   58 &  47 & ND &   -1 &  47 & ND \\ 
7795.76132 & 2017-02-11 & 0.07131 &  72 &  219 &   62 &  61 & ND &   -7 &  61 & ND \\
\hline
\hline
\end{tabular}
\end{table*}

\section{Radial Velocity and H$\alpha$ EW measurements}

\begin{table}
\caption{RVs and H$\alpha$ EWs from ESPaDOnS sub-exposures.}
\label{esp_rv_ew}
\begin{tabular}{l r r r}
\hline
\hline
HJD - & $\phi_{\rm P}$ &RV & H$\alpha$ EW \\
 2450000 & & (\kms) & (nm) \\
\hline
4488.97344 & 0.71204 & 19.2$\pm$0.7 & 0.130$\pm$0.003 \\
4488.97480 & 0.71853 & 19.8$\pm$0.8 & 0.130$\pm$0.003 \\
4488.97615 & 0.72497 & 20.4$\pm$0.8 & 0.130$\pm$0.003 \\
4488.97926 & 0.73981 & 22.0$\pm$0.8 & 0.130$\pm$0.003 \\
4488.98061 & 0.74626 & 22.6$\pm$0.8 & 0.130$\pm$0.003 \\
4488.98196 & 0.75270 & 23.2$\pm$0.8 & 0.130$\pm$0.004 \\
\hline
\hline
\end{tabular}
\end{table}
 
\begin{table}
\contcaption{RVs and H$\alpha$ EWs from ESPaDOnS sub-exposures.}
\label{esp_rv_ew:continued}
\begin{tabular}{l r r r}
\hline
\hline
HJD - & $\phi_{\rm P}$ &RV & H$\alpha$ EW \\
 2450000 & & (\kms) & (nm) \\
\hline
5228.93590 & 0.44974 &  6.8$\pm$0.5 & 0.132$\pm$0.004 \\
5228.93711 & 0.45551 &  6.7$\pm$0.5 & 0.131$\pm$0.004 \\
5228.93834 & 0.46138 &  6.4$\pm$0.5 & 0.131$\pm$0.004 \\
5229.70502 & 0.11960 & 33.3$\pm$1.1 & 0.130$\pm$0.008 \\
5229.70624 & 0.12542 & 33.0$\pm$1.1 & 0.132$\pm$0.008 \\
5229.70745 & 0.13120 & 32.5$\pm$1.1 & 0.132$\pm$0.005 \\
\hline
\hline
\end{tabular}
\end{table}
 
\begin{table}
\contcaption{RVs and H$\alpha$ EWs from ESPaDOnS sub-exposures.}
\label{esp_rv_ew:continued}
\begin{tabular}{l r r r}
\hline
\hline
HJD - & $\phi_{\rm P}$ &RV & H$\alpha$ EW \\
 2450000 & & (\kms) & (nm) \\
\hline
6294.85303 & 0.47647 &  6.7$\pm$0.5 & 0.138$\pm$0.003 \\
6294.85420 & 0.48205 &  5.8$\pm$0.5 & 0.137$\pm$0.003 \\
6294.85539 & 0.48773 &  5.5$\pm$0.5 & 0.138$\pm$0.003 \\
6668.99874 & 0.71196 & 20.0$\pm$0.7 & 0.138$\pm$0.003 \\
6668.99992 & 0.71759 & 20.6$\pm$0.8 & 0.138$\pm$0.003 \\
6669.00109 & 0.72317 & 21.2$\pm$0.8 & 0.137$\pm$0.003 \\
\hline
\hline
\end{tabular}
\end{table}
 
\begin{table}
\contcaption{RVs and H$\alpha$ EWs from ESPaDOnS sub-exposures.}
\label{esp_rv_ew:continued}
\begin{tabular}{l r r r}
\hline
\hline
HJD - & $\phi_{\rm P}$ &RV & H$\alpha$ EW \\
 2450000 & & (\kms) & (nm) \\
\hline
7800.78423 & 0.01819 & 40.4$\pm$1.2 & 0.149$\pm$0.002 \\
7800.78533 & 0.02344 & 40.1$\pm$1.2 & 0.149$\pm$0.002 \\
7800.78643 & 0.02868 & 39.9$\pm$1.2 & 0.149$\pm$0.002 \\
7800.84199 & 0.29379 & 16.2$\pm$0.7 & 0.152$\pm$0.003 \\
7800.84309 & 0.29904 & 15.8$\pm$0.7 & 0.151$\pm$0.003 \\
7800.84418 & 0.30424 & 15.3$\pm$0.6 & 0.152$\pm$0.003 \\
\hline
\end{tabular}
\end{table}

\onecolumn
\begin{table}
\caption{Log of CORALIE observations, RV measurements, and H$\alpha$ EW measurements. Peak signal-to-noise ratios (SNRs) are per spectral pixel.}
\label{cor_obs_log}
\begin{tabular}{l l r r r r r}
\hline
\hline
HJD - & Cal. & $\phi_{\rm P}$ & $t_{\rm exp}$ & Peak & RV & H$\alpha$ EW \\
 2450000 & Date & & (s) & SNR & (\kms) & (nm) \\
\hline
1591.54854 & 2000-02-17 & 0.58586 &  300 &  132 &  10.24$\pm$0.7 &  0.136$\pm$0.003 \\ 
1591.55522 & 2000-02-17 & 0.61772 &  300 &  139 &  12.30$\pm$1.1 &  0.137$\pm$0.003 \\ 
1591.60926 & 2000-02-17 & 0.87556 &  550 &  193 &  34.52$\pm$0.9 &  0.137$\pm$0.003 \\ 
1591.61956 & 2000-02-17 & 0.92471 &  550 &  192 &  38.18$\pm$0.9 &  0.138$\pm$0.003 \\ 
1591.68529 & 2000-02-17 & 0.23838 & 1000 &  217 &  22.89$\pm$0.9 &  0.143$\pm$0.003 \\ 
1591.70651 & 2000-02-17 & 0.33960 & 1500 &  258 &  13.14$\pm$0.5 &  0.143$\pm$0.003 \\
\hline
\hline
\end{tabular}
\end{table}
 
\begin{table}
\contcaption{Log of CORALIE observations.}
\label{cor_obs_log:continued}
\begin{tabular}{l l r r r r r}
\hline
\hline
HJD - & Cal. & $\phi_{\rm P}$ & $t_{\rm exp}$ & Peak & RV & H$\alpha$ EW \\
 2450000 & Date & & (s) & SNR & (\kms) & (nm) \\
\hline
1882.68429 & 2000-12-04 & 0.74907 &  400 &  210 &  23.72$\pm$0.6 &  0.141$\pm$0.002 \\ 
1882.70838 & 2000-12-04 & 0.86400 &  300 &  182 &  34.49$\pm$0.8 &  0.141$\pm$0.002 \\ 
1882.71922 & 2000-12-04 & 0.91575 &  350 &  200 &  37.89$\pm$0.7 &  0.142$\pm$0.002 \\ 
1882.73929 & 2000-12-04 & 0.01151 &  350 &  199 &  40.41$\pm$1.0 &  0.144$\pm$0.003 \\ 
1882.74643 & 2000-12-04 & 0.04559 &  350 &  198 &  39.45$\pm$0.8 &  0.144$\pm$0.003 \\ 
1882.75347 & 2000-12-04 & 0.07917 &  350 &  209 &  37.77$\pm$0.7 &  0.144$\pm$0.002 \\
\hline
\hline
\end{tabular}
\end{table}
 
\begin{table}
\contcaption{Log of CORALIE observations.}
\label{cor_obs_log:continued}
\begin{tabular}{l l r r r r r}
\hline
\hline
HJD - & Cal. & $\phi_{\rm P}$ & $t_{\rm exp}$ & Peak & RV & H$\alpha$ EW \\
 2450000 & Date & & (s) & SNR & (\kms) & (nm) \\
\hline
1947.65653 & 2001-02-07 & 0.76606 &  500 &  227 &  25.36$\pm$0.9 &  0.139$\pm$0.003 \\ 
1947.71283 & 2001-02-07 & 0.03467 &  500 &  229 &  39.80$\pm$0.6 &  0.141$\pm$0.003 \\ 
1947.73233 & 2001-02-07 & 0.12775 &  447 &  204 &  33.81$\pm$0.6 &  0.142$\pm$0.003 \\ 
1948.55915 & 2001-02-08 & 0.07290 &  500 &  213 &  38.22$\pm$0.7 &  0.143$\pm$0.003 \\ 
1948.61481 & 2001-02-08 & 0.33853 &  600 &  266 &  13.53$\pm$0.8 &  0.143$\pm$0.002 \\ 
1948.65575 & 2001-02-08 & 0.53385 &  500 &  248 &   7.85$\pm$0.8 &  0.140$\pm$0.002 \\
\hline
\hline
\end{tabular}
\end{table}
 
\begin{table}
\contcaption{Log of CORALIE observations.}
\label{cor_obs_log:continued}
\begin{tabular}{l l r r r r r}
\hline
\hline
HJD - & Cal. & $\phi_{\rm P}$ & $t_{\rm exp}$ & Peak & RV & H$\alpha$ EW \\
 2450000 & Date & & (s) & SNR & (\kms) & (nm) \\
\hline
2232.77493 & 2001-11-19 & 0.21658 &  197 &  197 &  25.12$\pm$0.6 &  0.137$\pm$0.002 \\ 
2232.79373 & 2001-11-19 & 0.30630 &  250 &  221 &  16.57$\pm$0.5 &  0.137$\pm$0.002 \\ 
2232.80927 & 2001-11-19 & 0.38047 &  234 &  196 &  10.37$\pm$0.7 &  0.140$\pm$0.002 \\ 
2232.82424 & 2001-11-19 & 0.45188 &  250 &  205 &   7.38$\pm$0.6 &  0.134$\pm$0.002 \\ 
2233.66354 & 2001-11-20 & 0.45662 &  250 &  188 &   7.22$\pm$0.6 &  0.133$\pm$0.002 \\ 
2233.72912 & 2001-11-20 & 0.76954 &  250 &  205 &  26.01$\pm$0.7 &  0.134$\pm$0.002 \\
\hline
\hline
\end{tabular}
\end{table}
 
\begin{table}
\contcaption{Log of CORALIE observations.}
\label{cor_obs_log:continued}
\begin{tabular}{l l r r r r r}
\hline
\hline
HJD - & Cal. & $\phi_{\rm P}$ & $t_{\rm exp}$ & Peak & RV & H$\alpha$ EW \\
 2450000 & Date & & (s) & SNR & (\kms) & (nm) \\
\hline
2985.57321 & 2003-12-12 & 0.21228 &  412 &  162 &  24.42$\pm$0.9 &  0.144$\pm$0.003 \\ 
2985.58629 & 2003-12-12 & 0.27471 &  453 &  185 &  18.84$\pm$0.4 &  0.144$\pm$0.003 \\ 
2985.59988 & 2003-12-12 & 0.33955 &  499 &  198 &  12.78$\pm$0.8 &  0.138$\pm$0.003 \\ 
2985.61331 & 2003-12-12 & 0.40361 &  483 &  215 &   8.51$\pm$0.6 &  0.138$\pm$0.002 \\ 
2985.63357 & 2003-12-12 & 0.50032 &  439 &  211 &   6.54$\pm$0.7 &  0.136$\pm$0.003 \\ 
2985.66530 & 2003-12-12 & 0.65170 &  439 &  188 &  14.55$\pm$0.7 &  0.135$\pm$0.003 \\
\hline
\hline
\end{tabular}
\end{table}
 
\begin{table}
\contcaption{Log of CORALIE observations.}
\label{cor_obs_log:continued}
\begin{tabular}{l l r r r r r}
\hline
\hline
HJD - & Cal. & $\phi_{\rm P}$ & $t_{\rm exp}$ & Peak & RV & H$\alpha$ EW \\
 2450000 & Date & & (s) & SNR & (\kms) & (nm) \\
\hline
3072.55234 & 2004-03-08 & 0.23514 &  250 &  126 &  22.63$\pm$0.7 &  0.139$\pm$0.003 \\ 
3072.60451 & 2004-03-08 & 0.48402 &  276 &  110 &   6.20$\pm$0.9 &  0.135$\pm$0.004 \\ 
3072.62053 & 2004-03-08 & 0.56051 &  300 &  108 &   7.99$\pm$0.9 &  0.136$\pm$0.004 \\ 
3072.64107 & 2004-03-08 & 0.65847 &  400 &  110 &  15.14$\pm$0.6 &  0.132$\pm$0.004 \\ 
3073.51755 & 2004-03-09 & 0.84063 &  250 &   76 &  32.86$\pm$1.5 &  0.135$\pm$0.005 \\ 
3073.55929 & 2004-03-09 & 0.03982 &  270 &   76 &  39.66$\pm$1.7 &  0.132$\pm$0.005 \\
\hline
\hline
\end{tabular}
\end{table}
 
\begin{table}
\contcaption{Log of CORALIE observations.}
\label{cor_obs_log:continued}
\begin{tabular}{l l r r r r r}
\hline
\hline
HJD - & Cal. & $\phi_{\rm P}$ & $t_{\rm exp}$ & Peak & RV & H$\alpha$ EW \\
 2450000 & Date & & (s) & SNR & (\kms) & (nm) \\
\hline
3270.82444 & 2004-09-22 & 0.29461 &  802 &  253 &  16.75$\pm$0.6 &  0.133$\pm$0.002 \\ 
3270.88071 & 2004-09-22 & 0.56311 &  769 &  245 &   8.26$\pm$0.7 &  0.128$\pm$0.002 \\ 
3270.90008 & 2004-09-22 & 0.65554 &  696 &  234 &  14.66$\pm$0.6 &  0.127$\pm$0.002 \\ 
3271.81924 & 2004-09-23 & 0.04132 &  800 &  210 &  39.07$\pm$1.0 &  0.131$\pm$0.002 \\ 
3271.84105 & 2004-09-23 & 0.14537 &  800 &  224 &  30.93$\pm$0.7 &  0.135$\pm$0.002 \\ 
3272.83696 & 2004-09-24 & 0.89739 &  800 &  238 &  37.17$\pm$0.6 &  0.130$\pm$0.002 \\
\hline
\end{tabular}
\end{table}
\twocolumn

\section{Log of interferometric observations}

\begin{table}
\centering
\caption[Log of VLTI Observations]{Log of VLTI observations. Type is photometry (H or K band) or high-resolution (HR).}
\begin{tabular}{lllr}
\hline
\hline
Cal. & Type & VLTI \\
Date & & Config. \\
\hline
\\
\multicolumn{3}{c}{AMBER} \\
2014-03-08 & H & A1-K0-G1 \\
2014-03-08 & K & A1-K0-G1 \\
2014-03-08 & HR & A1-K0-G1 \\
\\
\hline
\\
\multicolumn{3}{c}{PIONIER} \\
2014-03-23 & H & A1-G1-J3-K0 \\
\hline
\hline

\end{tabular}
\label{vlti_obs_log}
\end{table}

\end{document}